# Future Supply of Medical Radioisotopes for the UK Report 2014

***Report prepared by:***
***British Nuclear Medicine Society and Science & Technology Facilities Council.***
***December 2014***

# Preface

Technetium-99m ($^{99m}Tc$) is the principal radioisotope used in medical diagnostics worldwide. Current estimates are that $^{99m}Tc$ is used in 30 million procedures per year globally and accounts for 80 to 85% of all diagnostic investigations using Nuclear Medicine techniques. Its 6-hour physical half-life and the 140 keV photopeak makes it ideally suited to medical imaging using conventional gamma cameras. $^{99m}Tc$ is derived from its parent element molybdenum-99 ($^{99}Mo$) that has a physical half-life of 66 hours. At present $^{99}Mo$ is derived almost exclusively from the fission of uranium-235 targets (using primarily highly-enriched uranium) irradiated in a small number of research nuclear reactors.

A global shortage of $^{99}Mo$ in 2008/09 exposed vulnerabilities in the supply chain of medical radioisotopes. In response, and at the request of member states, the Organization of Economic Co-operation and Development (OECD) Nuclear Energy Agency (NEA) assembled a response team and in April 2009 formed a High-Level Group on the security of supply of Medical Radioisotopes (HLG-MR). The HLG-MR terms of reference are: to review the total $^{99}Mo$ supply chain from uranium procurement for targets to patient delivery; to identify weak points and issues in the supply chain in the short, medium and long-term; to recommend options to address vulnerabilities to help ensure stable and secure supply of radioisotopes.

The UK has no research nuclear reactors and relies on the importation of $^{99}Mo$ and other medical radioisotopes (e.g. Iodine-131) from overseas (excluding PET radioisotopes). The UK is therefore vulnerable not only to global shortages, but to problems with shipping and importation of the products. In this context Professor Erika Denton UK national Clinical Director for Diagnostics requested that the British Nuclear Medicine Society lead a working group with stakeholders including representatives from the Science & Technology Facilities Council (STFC) to prepare a report. The group had a first meeting on 10 April 2013 followed by a working group meeting with presentations on 9$^{th}$ September 2013 where the scope of the work required to produce a report was agreed.

The objectives of the report are: to describe the status of the use of medical radioisotopes in the UK; to anticipate the potential impact of shortages for the UK; to assess potential alternative avenues of medical radioisotope production for the UK market; and to explore ways of mitigating the impact of medical radioisotopes on patient care pathways. The report incorporates details of a visit to the Cyclotron Facilities at Edmonton, Alberta and at TRIUMF, Vancouver BC in Canada by members of the report team.

Brian Neilly, December 2014.

## Authors

**Brian Neilly**, (Report Lead). Consultant Physician, Nuclear & Respiratory Medicine, Glasgow Royal Infirmary, Glasgow G4 0SF, Past-President BNMS, Chair Intercollegiate Committee Nuclear Medicine.

**Sarah Allen**, Lead Clinical Scientist, Nuclear Medicine, Guys and St Thomas' NHS Foundation Trust, London SE1 9RT.

**Jim Ballinger**, Chief Radiopharmaceutical Scientist, Nuclear Medicine, Guys and St Thomas' NHS Foundation Trust, London SE1 9RT.

**John Buscombe**, Consultant Physician, Addenbrookes Hospital, Cambridge CB2 0QQ.

**Rob Clarke**, Acting Head of Experimental Science Group, Central Laser Facility, STFC Rutherford Appleton Laboratory, Harwell Science and Innovation Campus, Didcot OX11 0QX.

**Beverley Ellis**, Head of Radiopharmacy Services at Central Manchester and Manchester Children's University Hospitals NHS Trust, M13 9WL.

**Glenn Flux**, Head of Radioisotope Physics at the Royal Marsden NHS Trust and Institute of Cancer Research, Sutton SM2 5PT.

**Louise Fraser:** Specialist Scientist - Nuclear Medicine, Centre for Radiation, Chemical and Environmental Hazards, Public Health England, Chilton OX11 0RQ.

**Adrian Hall**, Head of Radiopharmacy at the Royal Marsden NHS Trust, Sutton SM2 5PT.

**Hywel Owen**, Lecturer in Accelerator Physics, School of Physics and Astronomy, University of Manchester, Manchester M13 9PL.

**Audrey Paterson,** Professor and Past-Director of Professional Policy**,** The Society and College of Radiographers, London SE1 2EW.

**Alan Perkins**, Professor of Radiological and Imaging Sciences, Medical Physics and Clinical Engineering, School of Medicine, Queen’s Medical Centre, Nottingham NG7 2UH.

**Andrew Scarsbrook**, Consultant Radiologist & Nuclear Medicine Physician, Leeds Teaching Hospitals NHS Trust, Honorary Clinical Associate Professor, University of Leeds, Department of Nuclear Medicine, St James's University Hospital, Leeds LS9 7TF.

## Stakeholders

British Nuclear Medicine Society
Public Health England (Formerly Health Protection Agency)
Royal College of Physicians
Royal College of Radiologists
Royal College of Radiologists Patient Representative
Science & Technology Facilities Council
Society and College of Radiographers

The initial meeting of the Medical Radioisotope Development Group took place in London on April 9th 2013 and was attended by the following:

| **Stakeholder** | **Representative(s)** |
|---|---|
| Department of Health | Erika Denton, NCD Diagnostics<br>Phillip Webster |
| British Nuclear Medicine Society | Brian Neilly, President BNMS |
| Clinical Science | Alan Perkins |
| Public Health England | Louise Fraser |
| Molecular Radiotherapy | Glenn Flux |
| Patient representative | Chris Wiltsher |
| Royal College of Physicians | John Buscombe |
| Royal College of Radiologists | Andrew Scarsbrook<br>Pete Cavanagh |
| Radiopharmacy | Beverley Ellis, Jim Ballinger |
| Science & Technology Facilities Council | Barbara Camanzi |
| Society and College of Radiographers | Audrey Paterson |
| Specialty Advisor Accelerator Science | Hywel Owen |
| Specialty Advisor Accelerator Research | Susan Smith |
| Specialty Advisor Nuclear Physics | Ian Lazarus/John Simpson |
| Specialty Advisor Laser Science | Rob Clarke |

## Acknowledgements

BNMS would like to thank Barbara Camanzi of the Science & Technology Facilities Council (STFC) for her help and encouragement with the project. Funding for the two meetings of the medical radioisotopes group was provided by the BNMS and by the STFC.

Funding for the delegation to Canada (Chapter 8) was provided by the Global Partnership Fund of the Foreign and Commonwealth Office and by the STFC. BNMS wish to thank Aatif Baskanderi, Science, Innovation and Energy Officer of the British Consulate-General, Calgary, Alberta, Canada for his help securing the arrangements for the visit of the UK delegation to Edmonton and Vancouver.

The BNMS also would like to thank the staff at the Medical Isotope and Cyclotron Facility, Edmonton and at TRIUMF, Vancouver for their time, consideration and openness during the delegation visit. Thanks also to Charlotte Weston for arranging the travel itinerary and the accommodation in Canada.

Thanks to Bernard Ponsard for his input to Chapter 1 and his permission to reproduce Figure 1.4.

# Table of Contents

## Glossary of Terms

| | |
|---|---|
| AECL | Atomic Energy Canada Ltd. |
| ANSTO | Australian Nuclear Science & Technology Organization. |
| ARSAC | Administration of Radioactive Substances Advisory Committee. |
| Becquerel (Bq) | The international (SI) unit to measure radioactivity. One Becquerel is equal to one nuclear decay per second. Activity is typically expressed as mega ($10^6$, MBq), giga ($10^9$, GBq) or terabecquerel ($10^{12}$, TBq) units of radioactivity. |
| BNMS | British Nuclear Medicine Society. |
| BP | British Pharmacopoeia. |
| BR-2 | Belgian Reactor 2. |
| CT | Computed Tomography, a technique that uses a computer to generate multiple X-ray images in three orthogonal planes. |
| Curie (Ci) | A unit to measure radioactivity. One Curie is 3.7 x $10^{10}$ Becquerel (37 GBq). Often expressed in millicuries where 1 mCi is 1/1000 of a Ci. |
| Cyclotron | A particle accelerator that uses electromagnetic fields to accelerate a charge towards a target to produce radionuclides, commonly $^{18}$F for PET imaging. |
| Electron volt (eV) | A unit of energy equal to approximately $1.6\times10^{-19}$ joules, typically expressed as kilo ($10^3$, keV) or mega ($10^6$, MeV) electron volts. |
| $^{18}$ FDG | $^{18}$F-Fluorodeoxyglucose, the principal radiopharmaceutical used in PET imaging. |
| FCR | Full Cost Recovery, term used to describe the objective to ensure that all of the costs associated with irradiation (operational and capital costs) are recovered. |
| Gamma Camera | Equipment used to image gamma radiation emitted from a patient previously administered with a radiopharmaceutical. |
| Generator | A mechanism that permits elution of short-lived radionuclides from longer-lived precursors e.g. $^{99}$Mo/$^{99m}$Tc and $^{81}$Rb/$^{81m}$Kr. |
| GMC | General Medical Council. |
| GMP | Good Manufacturing Practice. |
| HEU | Highly enriched uranium containing more than 20% $^{235}$Uranium. |
| HFR | High Flux Reactor, Petten, Netherlands. |
| HLG-MR | High Level Group-Medical Radioisotopes, the NEA-OECD group set up to oversee the global response to the molybdenum shortages. |
| Jules Horowitz | New French Nuclear Reactor to replace OSIRIS. |
| LEU | Low enriched uranium (≤ 20%). |
| Linac | A linear particle accelerator is a type of particle accelerator commonly used in radiotherapy to accelerate electrons into a target to produce a beam of high energy X-rays. |
| MDT | Multidisciplinary Team Meeting. |
| MHRA | Medicine and Healthcare Products Regulatory Agency. |
| MRI | Magnetic Resonance Imaging, an imaging technique that uses magnetic and radiofrequency fields to reconstruct high-resolution images of the body. |
| MRT | Molecular Radiotherapy, the branch of nuclear medicine that uses the properties of radionuclides to treat disease. |
| MSC | Modernising Scientific Careers. |
| $^{99}$Mo | Molybdenum-99, a radioisotope of Molybdenum and the precursor of $^{99m}$Tc. |
| MWth | $10^6$ Watts (thermal). |
| NRU | National Research Universal Reactor at Chalk River, Canada. |
| NEA | Nuclear Energy Authority. |
| OECD | Organisation for Economic Co-operation & Development. |
| OPAL | Open Pool Australia Light water reactor. |
| ORC | Outage Reserve Capacity, the system in place to ensure that processors hold sufficient product in reserve to allow for unplanned or extended shutdowns of reactors. |
| OSIRIS | Experimental Nuclear Reactor at Saclay, France. |
| PALLAS | Research Nuclear Reactor at Petten to replace HFR. |
| PET | Positron Emission Tomography, a diagnostic imaging test based on the detection of radiation from positrons emitted from short-lived radioisotopes (e.g. $^{18}F$-FDG) injected into a patient. |
| PET-CT | Positron Emission Tomography with Computed Tomography. |
| PET-MR | Positron Emission Tomography with Magnetic Resonance Imaging. |
| Ph. Eur. | European Pharmacopeia. |
| Radiopharmaceutical | A pharmaceutical or a drug labelled with a radionuclide used in nuclear medicine imaging and therapy. |
| Radiotherapeutics | Term used to describe the use of beta or alpha-emitting radionuclides in therapy of disease especially cancer. |
| RCP | Royal College of Physicians. |
| RCR | Royal College of Radiologists. |
| SAFARI-1 | South African Nuclear Research Reactor. |
| Sievert | The international (SI) unit of effective radiation dose, commonly expressed in millisieverts (mSv, $10^{-3}$ Sievert). |
| SCoR | Society and College of Radiographers. |
| SPECT | Single Photon Emission Tomography, an imaging technique that uses images from a gamma camera to reconstruct a series of trans-axial slices that are combined to produce a three-dimensional dataset. |
| SPECT-CT | Single Photon Emission Tomography combined with Computed Tomography. |
| STFC | Science & Technology Facilities Council. |
| STP | Scientist Training Programme. |
| $^{99m}Tc$ | Technetium 99m, the principal radioisotope used in medical imaging. |
| US | Ultrasound scan. |
| USP | US Pharmacopoeia. |

## Executive Summary

**Molybdenum: Global Supply Perspective**

- Technetium-99m ($^{99m}$Tc) is the daughter product of molybdenum-99 and is the principal radioisotope used for medical diagnostic imaging, accounting for almost 85% of the global radiopharmaceutical market estimated to be in excess of 3 billion USD.
- $^{99m}$Tc is a versatile element that supports a spectrum of different radiopharmaceuticals targeting multiple disease sites, in particular bone, cardiac, renal, lung and endocrine disorders. Its relatively short physical half-life results in a low radiation dose burden for the patient.
- At present approximately 600,000 imaging procedures using $^{99m}$Tc are undertaken each year in the UK. The global estimate is around 40 million procedures per annum.
- There is good evidence that the demand for $^{99m}$Tc for imaging diagnostics will continue in the short to medium term. Growth predictions are difficult to assess but demand will continue to grow at a slowing rate (estimated at 0.5% per annum until about 2030). Estimates are that $^{99m}$Tc will be needed for at least the next 15-25 years.
- As new technologies for molecular imaging (e.g. PET) develop, $^{99m}$Tc's share of the medical diagnostics market is expected to fall but in context of an increasing annual budget to over 4.1billion US dollars, the demand for $^{99m}$Tc will remain high. The future demand for $^{99m}$Tc will be driven by the availability of new diagnostic ligands.
- There are systematic problems in the global supply chain for medical radioisotopes. These problems are complex, rendering the supply system fragile and subject to unpredictable outages.
- $^{99}$Mo is at present produced exclusively in a small number of research nuclear reactors most of which are >40 years old. Of these reactors OSIRIS (France) and NRU (Canada) will cease production permanently at the end of 2015 and 2016 respectively. Loss of these reactors will remove approximately 30% of the global production of molybdenum-99 from the supply chain by the end of 2016.
- The production of $^{99}$Mo until now has been achieved by nuclear reactor-based fission of highly enriched (weapons-grade) uranium. For non-proliferation reasons, $^{99}$Mo derived from HEU will no longer be accepted beyond 2020.
- The economic model of medical radioisotope production is poorly developed and is a disincentive to investment and to the development of any outage reserve capacity (ORC).
- The OECD NEA HLG-MR has determined that full cost recovery (FCR) and a mechanism for ORC is vital to sustain future global supply of $^{99}$Mo from reactors. As a result of the implementation of FCR and ORC, it is inevitable that the cost of $^{99m}Tc$ will increase. The magnitude and timing of this increase in cost of $^{99m}Tc$ is uncertain.

- Individual countries and companies are exploring options for a future supply of medical radioisotopes. The UK at present has limited capacity for the production of radioisotopes for medical use restricted to a small number of cyclotrons that manufacture mainly $^{18}F$- for PET imaging with FDG.
- There are potential reactor-based and non-reactor solutions for the secure supply of $^{99}Mo$ and $^{99m}Tc$ in the medium to long-term future however, there are many assumptions as to the availability of new reactor projects as outlined by the latest HLG-MR report (July 2014).
- The purpose of the present report is to assess and forecast the UK's requirement for medical radioisotopes and the options that are available for the development of secure and sustainable source and distribution of radioisotopes for medical use.

**Nuclear Medicine: UK perspective**

- Nuclear Medicine as currently practiced in the UK is a multidisciplinary specialty employing a workforce of more than 2,500 individuals
- There has been a significant increase over time in the NHS of the availability of gamma cameras with SPECT and SPECT-CT capability. These cameras are optimal for imaging using $^{99m}Tc$-based radiotracers.
- The demand for conventional (non-PET) imaging and non-imaging diagnostics in the UK is in the region of 600,000 per annum. Bone, cardiac and renal investigations make up more than half of the requests for these investigations.
- There is an increasing availability of PET-CT across the UK for a growing number of indications. In most of these centres there is limited capacity for further growth due to high use for FDG-PET-CT for oncological indications. Non-oncological and non-FDG PET indications presently make up only 10% of the total.
- The cyclotrons available in the UK for the production of $^{18}FDG$ and other PET-tracers are not suitable for the large-scale production of $^{99m}Tc$. Higher-energy cyclotrons would have to be installed if commercial quantities of $^{99m}Tc$ are to be produced.
- Nuclear medicine services in the UK receive generators from only three commercial companies. The majority of radiopharmacy services in the UK have been affected by generator shortages.
- Most nuclear medicine services in the UK have developed contingencies to deal with periods of shortage. These include re-scheduling the patient, employing alternative radiopharmaceuticals, or using lower administered activities, none of which are optimal. In a small proportion of cases this change has affected future patient referrals.

- There is evidence of unnecessary waste of radioactivity arising from the current generator delivery schedules to departments. Delivery and elution schedules for $^{99m}Tc$ generators are highly complex and software options exist to facilitate the correct choice. This could be improved through partnerships with the commercial generator suppliers.

**Alternative Imaging Strategies UK**

- The 2010 ARSAC report proposed 8 recommendations as to how to mitigate the impact of periods of $^{99m}Tc$ shortage.
- The response of the nuclear medicine community to dose saving and to alternatives is maturing.
- Most alternative radionuclides to $^{99m}Tc$ will be more expensive and some deliver much higher radiation doses to patients and staff.
- Non-radionuclide alternatives are available in certain circumstances but may not achieve the same diagnostic accuracy, may not be useable in all patients and would displace work on other imaging machines such as CT and MR which themselves may be limited in capacity.
- It is apparent that patients will be best served by having access to a plentiful supply of $^{99m}Tc$ for the most effective diagnostic tests.
- Patients will be poorly-served by not having a cheap, plentiful supply of $^{99m}Tc$
- The present distribution of Cyclotrons within the UK is patchy at and while most areas have ready access to FDG, those without Cyclotrons cannot take advantage of the use of the shorter-lived Cyclotron-manufactured radioisotopes

**Non-Reactor Production of $^{99m}Tc$**

- Sourcing secure supplies of $^{99}Mo$ or $^{99m}Tc$ from accelerator-based (linear accelerators, cyclotrons or other technologies) is an option that the UK should consider.
- The model of cyclotron-based manufacture of $^{99m}Tc$ from $^{100}Mo$ is one that is being developed and evaluated in Canada and the production of commercial quantities has been shown to be feasible.
- When compared to other technical approaches, it is concluded that direct cyclotron production of $^{99m}Tc$ is most promising model for the UK; compared to other approaches, cyclotron production is thought to be the most mature and it also lends itself better to co-production with other radioisotopes such as $^{18}F$.

### Cyclotron-Produced $^{99m}$Tc

- International efforts have demonstrated that production of $^{99m}$Tc by cyclotron is a feasible technology, though it requires infrastructure and incurs ongoing costs of daily production and transport. The delegation to Canada established the following:
- It has been demonstrated on a small scale that standard $^{99m}$Tc based radiopharmaceuticals can be prepared using cyclotron-produced $^{99m}$Tc and that their chemical and biological properties appear equivalent.
- Two production methodologies with the potential to supplement UK $^{99m}$Tc supplies on a commercial basis have been identified.
- The production $^{99m}$Tc on high-powered cyclotrons has been accomplished with the potential for routine production of TBq amounts of radioactivity.
- $^{99m}$Tc cyclotron production requires high-energy cyclotrons and cannot be achieved using the existing medical cyclotron facilities in the UK currently used for the production of positron-emitting radiopharmaceuticals.
- The key innovation and intellectual property concerning cyclotron production of $^{99m}$Tc rests with the beam targetry and target plate technology.
- Efficient methods for the radiochemical separation and recovery of $Na^{99m}TcO_4$ from $^{100}$Mo targets have been developed.
- The radionuclidic purity of cyclotron-produced $^{99m}$Tc is well understood. Small amounts of $^{93}$Tc, $^{94m}$Tc, $^{94}$Tc, $^{95}$Tc, and $^{96}$Tc impurities may be present in variable amounts and with optimisation of cyclotron energy these radionuclidic contaminants may contribute an additional 10% in effective radiation dose to the patient.
- Initial work on radiolabelling standard kit formulations has been carried out and products have met the current quality control specifications.
- The position relating to the regulatory approval of cyclotron produced $^{99m}$Tc radiopharmaceuticals is unclear and further information should be sought from both the Canadian regulatory authorities and the MHRA.
- Economic assessment have shown that cyclotron production of $^{99m}$Tc could be achieved at a cost of <$1Ca/mCi.
- The radioactive waste implications of routine cyclotron production of $^{99m}$Tc have not been quantified.
- The reliability of routine large-scale production of $Na^{99m}Tc0_4$ has still to be demonstrated.
- Provision of $^{99m}$Tc-based radiopharmaceuticals across the UK using $^{99m}$Tc produced in a small number of cyclotrons will likely involve a combination of central and local radiopharmacies in order to optimize service provision. There should be public sector oversight of this process.

**UK Nuclear Medicine Workforce**

- Accurate information on the size of the nuclear medicine workforce in the UK is lacking. This is due to a number of factors including the nomenclature in use and the diverse ways of recording such data by the professional bodies. An initiative is required to collect and collate this data for the entire UK (not just for NHS England) for future workforce planning.
- Despite the lack of accurate data, there is good evidence of a workforce shortage at all levels within the multi-disciplinary team. This is evident from surveys done by professional bodies, from difficulties with recruitment, from limited access to training pathways and from projected retirement of the present workforce.
- The nuclear medicine workforce managed the challenge of past shortages of $^{99m}$Tc, but such a response is not sustainable in the long-term. The same consideration applies to the wider imaging workforce outside of nuclear medicine, especially where alternative imaging studies are involved.
- Technological innovations in the field of nuclear medicine have driven the need for further workforce training but to date initiatives to update and make available training programmes are at best patchy.
- Workforce issues and training shortages within the nuclear medicine community in the UK cannot be underestimated – a full impact assessment must be undertaken as part of any move away from the traditional supply chain for $^{99m}$Tc radiopharmaceuticals.
- Further initiatives such as 7-day working, if applied to Nuclear Medicine, will put further pressure on an already stretched workforce and to be effective would have to be fully costed and funded.

## Recommendations

1 The critical period of $^{99m}$Tc shortage is 2016 to 2020. It is uncertain to what extent the global initiatives described in the report will offset the impact of the decommissioning of the French and Canadian reactors in 2015 and 2016 respectively but it is recommended that end-users are fully briefed with as much warning as possible to allow contingency arrangements to be implemented.

2 The recommendations of the 2010 ARSAC Report should be implemented in full. To date the recommendations of the report have been deployed in part. In particular it is recommended that where possible, use should be made of resolution recovery software to permit a reduction in the activity of administered $^{99m}$Tc and that software applications are used to improve selection of $^{99m}$Tc generator and delivery schedules for more efficient use of the available resource.

3 The UK as a signatory to the Joint Declaration On The Security Of Supply Of Medical Radioisotopes (http://www.oecd-nea.org/press/2014/2014-06.html) should recognise in practice the full implications of the six policy principles of the OECD-HLG MR mandate including a commitment to full cost recovery and outage reserve capacity. It is recommended that end-users, hospital administrators, and commissioning bodies be informed that this will mean a rise in the cost per unit of $^{99m}$Tc.

4 For strategic planning of services involving medical radionuclides in the UK, whether workforce or service provision-related, it is vital that that there is accurate information to inform decisions. At present workforce and service-provision databases are patchy and incomplete, often providing conflicting information. It is recommended that the various stakeholders involved in nuclear medicine in the UK are encouraged to collaborate to refine databases to ensure accurate information in relation to workforce, imaging and non-imaging diagnostics, and therapy etc.

5 The magnitude of the workforce issues and training shortages within the nuclear medicine community in the UK cannot be underestimated. It is recommended that a full impact assessment is undertaken as part of any move away from the traditional supply chain of $^{99m}$Tc radiopharmaceuticals. This should include where appropriate an assessment of the use of alternative imaging (e.g. PET, CT, MR, or US) or therapy pathways, and a business case to estimate the costs of alternative strategies.

6 It is likely that nuclear reactor fission-derived $^{99}$Mo will continue to provide the backbone of the future UK supply of $^{99m}$Tc through the existing generator distribution network during the critical period. However, in context of the uncertainty about the future global supply of $^{99}$Mo, it is recommended that the UK should diversify its strategy of reliance on reactor-based $^{99m}$Mo and support the development of novel technologies for the non-reactor production of $^{99m}$Tc either directly or via its $^{99}$Mo precursor. Based on an assessment of the relative maturity of the different options and the possible co-use for purposes such as manufacture of other radioisotopes, it is concluded that the most promising technology for the provision of $^{99m}Tc$ in the UK is its direct production using proton cyclotron bombardment at moderate energies between 18 and 24 MeV.

7 $^{99m}Tc$ production methodologies using cyclotrons with the potential to supplement UK demand have been identified, but would require the installation of high-powered cyclotrons that do not presently exist in the UK. Pioneering work in Canada has established the methodology and the means for radiochemical separation and recovery of $^{99m}Tc$ from $^{100}Mo$ targets. Radiolabelling of kits and satisfactory quality control has been achieved but uncertainty remains around regulatory approval of cyclotron-produced $^{99m}Tc$, the continued use of existing licensed kits, the clinical utility of the products and routine large-scale production and distribution has yet to be demonstrated. Given the time scale involved, it is recommended that UK seeks to work with industrial partner(s) to explore the possibility of commissioning cyclotron(s) with the capacity to supplement the supply of $^{99m}Tc$ tracers and PET radiopharmaceuticals.

8 Provision of $^{99m}Tc$ based radiopharmaceuticals across the UK using $^{99m}Tc$ produced in a small number of cyclotrons will likely involve a combination of central and local radiopharmacies in order to optimize service provision. It is recommended that there is public sector oversight of this process.

9 The visit to Canada opened up a number of opportunities for collaboration and research in the area of medical radioisotopes. It is recommended that these opportunities for collaboration and research be followed up. This may also mean co-operation through international bodies such as the European Union and the International Atomic Energy Agency.

10 It is recommended that a national strategy for the use of radiotherapeutics for cancer treatment should be developed to address the supply of radiotherapeutics, projected costs of drugs and resources, the clinical introduction of new radioactive drugs, national equality of access to treatments and resource planning. Consideration should be given to centres of excellence supporting satellites.

# Chapter 1: The Molybdenum Supply Chain

**Brian Neilly and Alan Perkins**

## 1.1 Background

The global shortage of $^{99}$Mo/$^{99m}$Tc in 2008/2009 was a wake up call and a reminder of the vulnerability inherent in the global supply chain for $^{99m}$Tc [1], the principal radioisotope used for medical imaging worldwide. At the time notifications from the generator manufacturers to the end-user were not unusual (Figure 1.1).

| UPDATED MAY 2010 | | | | | | | JUNE 2010 | | | | | | |
|---|---|---|---|---|---|---|---|---|---|---|---|---|---|
| S | M | T | W | T | F | S | S | M | T | W | T | F | S |
| | | | | | | 1 | | | X | X | 3 | X | X |
| 2 | 3 | X | X | 6 | X | 8 | 6 | 7 | 8 | 9 | 10 | 11 | 12 |
| 9 | 10 | 11 | X | X | X | 15 | 13 | 14 | 15 | 16 | 17 | 18 | 19 |
| 16 | X | X | X | 20 | 21 | 22 | 20 | 21 | 22 | 23 | 24 | 25 | 26 |
| 23 | X | X | X | X | 28 | 29 | 27 | 28 | 29 | 30 | | | |
| 30 | 31 | | | | | | | | | | | | |

| | |
|---|---|
| | Generator standing orders met with some extra |
| | Majority of generator standing orders met but no extra |
| | Generator standing order shortage resulting in size reductions, Tc 99m shortage |
| | Significant shortage to generator standing orders, severe Tc 99m shortage |
| X | No Mo99 supply expected. Generator production canceled. |

**Figure 1.1** Example of a timetable of availability of generator from manufacturers to end-users

The OECD-HLG in response to the shortage of medical radioisotope commissioned two mandates (2009-11 and 2011-13) and has embarked on a third mandate (2013-15). Arising from these initiatives the OECD NEA has published a series of papers addressing comprehensively the issues posed by the global shortage of $^{99}$Mo [1-15]. The HLG developed six policy principles that have been agreed to by all major $^{99}$Mo-producing countries as follows:

1. All $^{99m}$Tc supply chain participants should implement full cost recovery;
2. Reserve production capacity should be sourced and paid for by the supply chain;
3. Governments should establish a proper environment for efficient and safe market operations, without intervening directly;

4. Governments should facilitate the conversion to low-enriched uranium (LEU) by reactors and processors;
5. International collaboration should continue through a policy and information-sharing forum;
6. Periodically review the supply chain's progress towards economic sustainability and security of supply.

## 1.2 Global Demand for $^{99}$Mo/$^{99m}$Tc

The OECD NEA undertook an assessment of the long-term global demand for $^{99m}$Tc [1] using an online survey that obtained 713 responses from 52 countries. This survey confirmed that, notwithstanding the reduced demand for $^{99}$Mo after the 2009 shortfall, the forecast is that demand for $^{99}$Mo will continue to grow at approximately 2% per annum until 2020 and then level off to a growth rate of 1% per annum until 2030 (Figure 1.2)

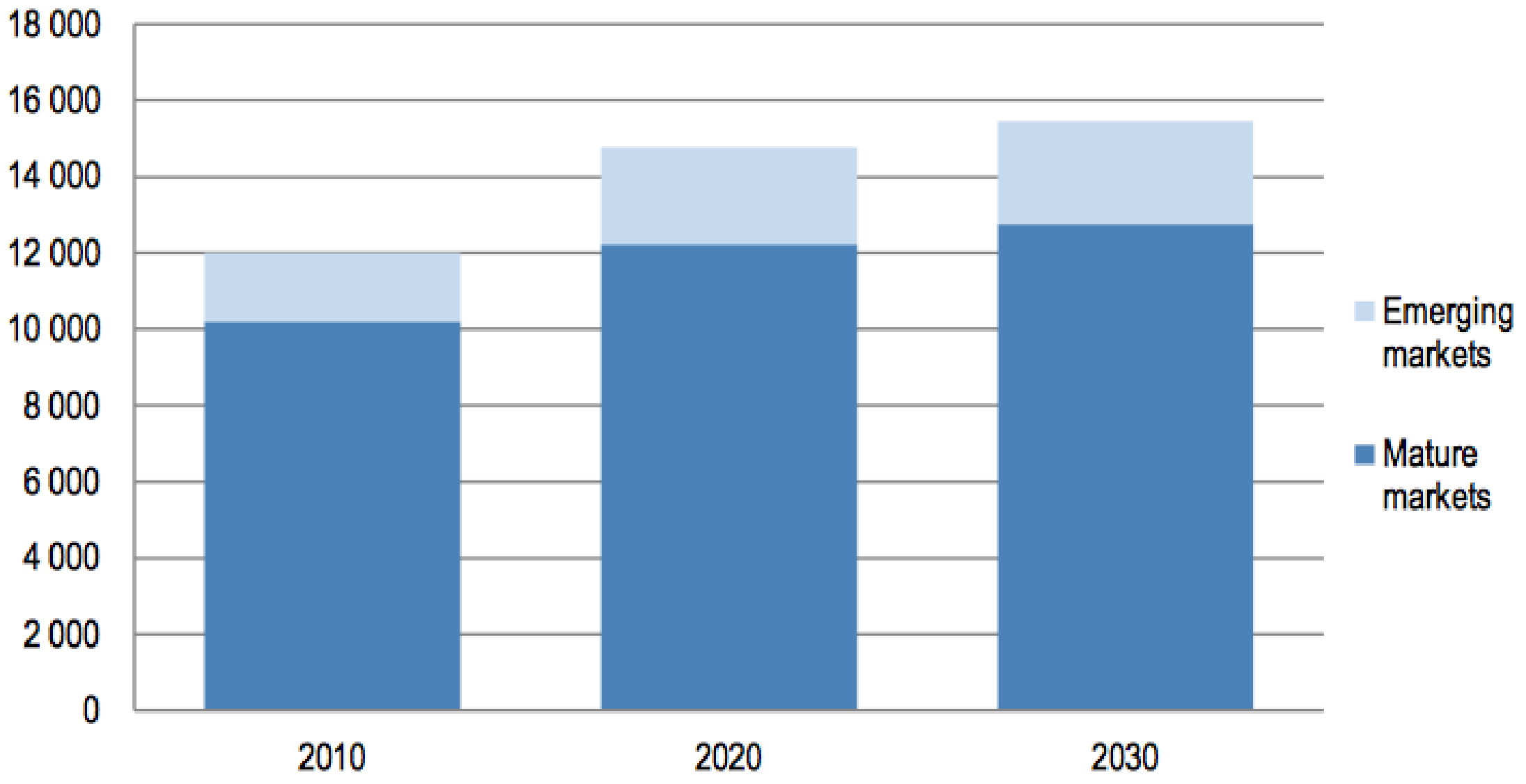

**Figure 1.2** Derived from OECD NEA Report in 'The Supply of Medical radioisotopes: an assessment of Long-term Global Demand for Technetium-99m' [1]. Quantities are demand for $^{99m}$Mo activity in 6-day Curies per week

In 2010 the global demand for $^{99}$Mo was estimated at 12,000 6-day Curies per week (i.e. more than 500,000 per year). The 6-day Curie (6-day Curies) is the accepted term used to describe molybdenum activity and is the measurement of the amount of $^{99}$Mo activity remaining six days after it leaves the processing facility (end of processing (EOP)). Of this global demand for $^{99}$Mo, the regional demand is dominated by North America; 53% of $^{99}$Mo goes to the North American Market, 23% to the European Market, 20% to Asia and 4% to the rest of the World [16].

The 2014 report from the OECD Nuclear Energy Authority (14) revised the estimate of global demand for $^{99}$Mo down from 12,000 to 10,000 6-day Curies, an estimate compiled from data returned from vendors and users. This downturn in global demand is thought to have arisen from the impact of contingency measures adopted by users in response to the shortages in 2009/2010 and that then persisted. These include improved efficiency of use and distribution of the radioisotope, and the use of some alternative diagnostic pathways.

A number of trends were noted in the survey that may influence the future demand for $^{99m}$Tc:

- Growing population and urbanisation will drive growth in access to imaging;
- Ageing population and changing prevalence of medical conditions including malignancy will influence demand;
- Increased availability of SPECT and SPECT-CT cameras;
- Expected shift from SPECT to PET - but it is anticipated that the impact of this will be low at least in the short-term;
- Impact of new PET tracers may have an influence but because of the cost of PET imaging the relative impact will be low in the short to medium-term.

While there is less certainty about the predicted accuracy of the estimates by 2030, most authorities agree that there will be a continued growth in demand for $^{99}$Mo in the medium term before then. Estimates drawn from a number of sources [2, 17-19] reveal that in the short to medium term, $^{99m}$Tc is expected to remain the major market player although the relative proportion of the radiopharmaceutical market occupied by $^{99m}$Tc is expected to decrease (from 85% to 72%). The absolute size of the market is expected to increase in the short to medium term (Figure 1.3).

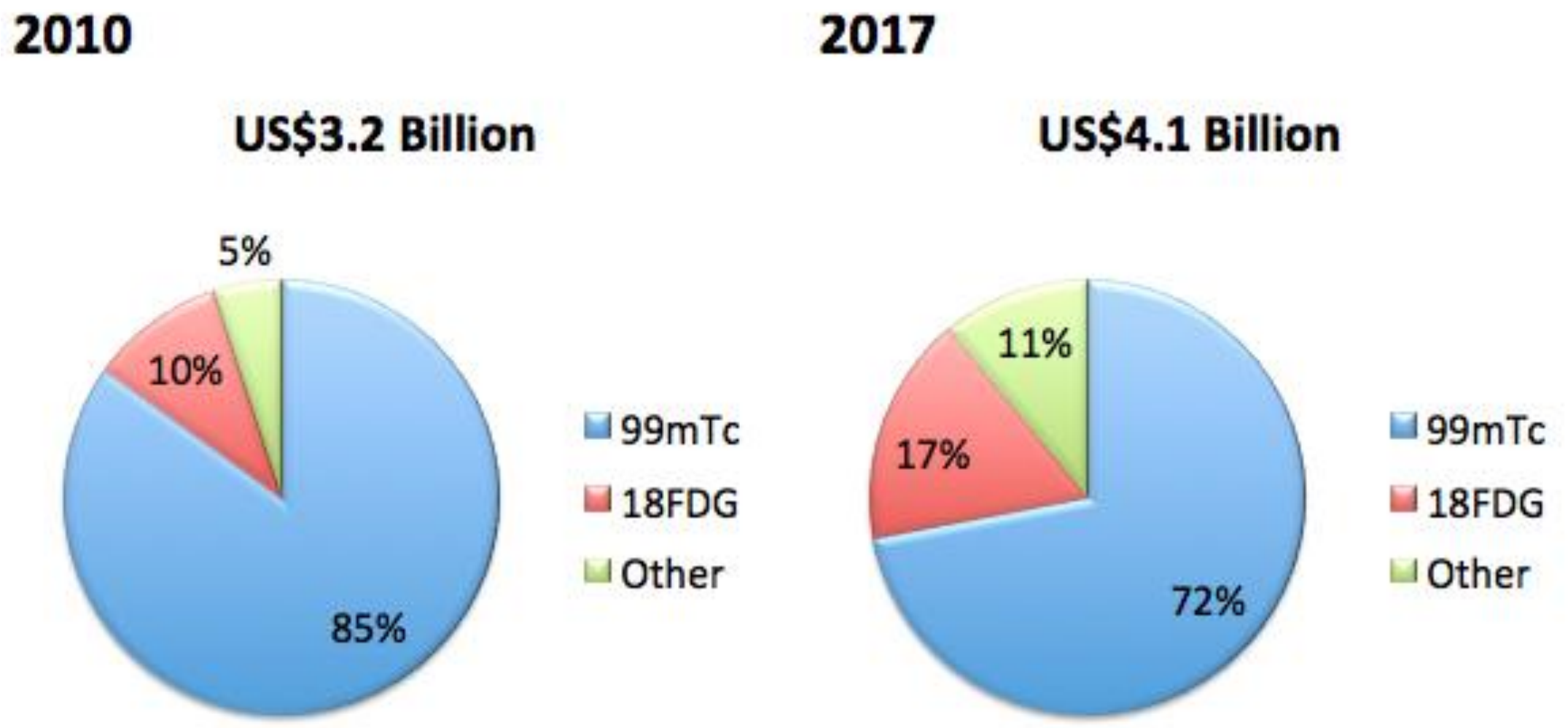

**Figure 1.3** Anticipated change in the global radiopharmaceutical market. Estimates vary according to source [2, 17,18] but are of a comparable amount. Another report estimates a rise in the global radiopharmaceutical market from $3.8 billion in 2012 to $5.5 billion in 2017 [19]

## 1.3 The Molybdenum 99 Supply Chain

The shortage in availability of $^{99m}Tc$ for diagnostic imaging is due to the complexity of the supply chain that involves many steps, each with the capacity for fragility (see Figure 1.4). The following section highlights the various steps in the supply chain involved in the manufacture and distribution of $^{99}Mo$. This involves a number of steps as follows:

- Uranium target manufacturer;
- Target irradiation in a suitable nuclear reactor;
- Processing to dissolve the irradiated targets and chemically extract $^{99}Mo$;
- $^{99}Mo/^{99m}Tc$ generator manufacturer;
- Elution of $^{99m}Tc$ from the generators and prepare radiopharmaceuticals in a radiopharmacy;
- Transport of radiopharmaceuticals to the end user clinics.

At each point in the supply chain there are issues that have to be addressed. These issues are set out below.

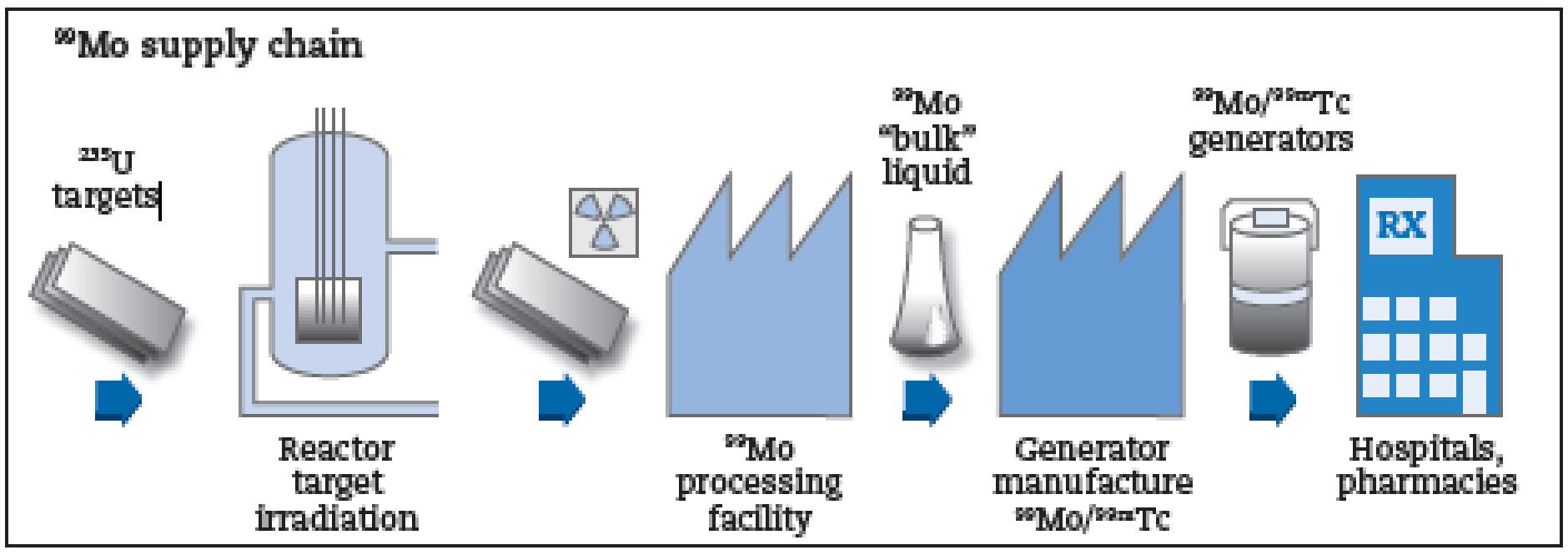

**Figure 1.4** Simplified depiction of supply chain in the conventional production of $^{99m}Tc$ using reactor fission (reproduced from OECD NEA Document 'The Supply of Radioisotopes: The Path to Reliability, 2011" [6])

### *1.3.1 Uranium Targets*

At present the majority of nuclear reactors use highly enriched uranium (HEU) targets for the manufacture of $^{99}Mo$ and also in the reactor fuel itself. Of those listed (Table 1.1) only OPAL, RA3 and Safari-1 use low enriched uranium (LEU) targets. To meet present and future non-proliferation demands, all current $^{99}Mo$-producing nations have agreed to convert from using HEU to LEU targets for $^{99}Mo$ production. The American Medical Radioisotopes Production Act 2012 extends previous non-proliferation efforts and includes provisions to restrict the export of HEU from the United States for the purposes of medical isotope production by 2020. It is anticipated that LEU conversion will have a potential impact on the global supply chain both in terms of capacity and costs. In recognition of this the HLG-MR undertook and published a report on the market impacts of converting to LEU Targets for medical isotope production (15). The advantages of LEU over HEU include: greater proliferation resistance; in consequence an easier availability of target material; and better compliance for target transportation and processing. There are also some downsides, including the fact that around 5 times more LEU must be used than HEU for a given production of $^{99}$Mo, with approximately 25 times more $^{239}$Pu is produced when using LEU targets. However, it is anticipated that the operating costs when producing LEU-derived $^{99}$Mo will be more or less the same as for HEU. The main burden for the producers will be the time and costs associated with the conversion and the increased volumes of radioactive waste. The HLG-MR report recognized that LEU conversion would reduce available irradiation and processing capacity in the interim, but was unlikely to cause long-term shortages. Overall, LEU-based $^{99}$Mo is likely to be more expensive that HEU-based $^{99}$Mo. Nevertheless, all major producers have committed to converting to LEU targets by 2016 and LEU-based $^{99}$Mo is gaining acceptance in markets traditionally supplied with HEU-based material. For example, to encourage use of non-HEU derived $^{99m}$Tc in the US, an additional $10 payment is mandated for each patient radiopharmaceutical administration that is derived from LEU sources.

### *1.3.2 Nuclear Reactor Supply:*

At present a small number of research nuclear reactors are involved in the large-scale global supply of $^{99}$Mo on an industrial scale (see Table 1.1). These reactors are **NRU** (Canada), **BR-2** (Belgium), **HFR** (Netherlands), **OSIRIS** (France), **SAFARI** (South Africa), **MARIA** (Poland), **LVR-15** (Czech Republic), **OPAL** (Australia), **RA-3** (Argentina) and **FRM-II** (Germany). As can be seen from the Table 1.1, most of the world's $^{99}$Mo comes from only five research reactors, namely Canada's NRU, the Netherlands's HFG, Belgium's BR-2, France's Osiris and South Africa's Safari-1 reactor. The majority of these reactors are >40 years old and, in keeping with this and poor investment, are subject to unexpected and prolonged shutdowns. In some cases there will be permanent shutdowns. A brief summary of reactor operations is given below.

**Table 1.1** Current reactor production of $^{99}$Mo

| Reactor | Country | Target | Normal Operating Days | Normal capacity per week (6-d Ci) | Potential annual production (6-d Ci) | Estimated date to cease production |
|---|---|---|---|---|---|---|
| **BR-2** | Belgium | HEU | 140 | 7,800 | 156,000 | 2026 |
| **HFR** | Netherlands | HEU | 280 | 4,680 | 187,200 | 2024 |
| **LVR-15** | Czech Rep | HEU | 210 | 2,800 | 84,000 | 2028 |
| **MARIA** | Poland | HEU | 210 | 2,200 | 42,900 | 2030 |
| NRU | Canada | HEU | 280 | 4680 | 187,200 | **2016** |
| **OPAL** | Australia | LEU | 300 | 1,000 | 42,900 | 2055 |
| OSIRIS | France | HEU | 182 | 2,400 | 62,400 | **2015** |
| **RA-3** | Argentina | LEU | 336 | 400 | 19,200 | 2027 |
| **SAFARI** | South Africa | HEU-LEU | 305 | 3,000 | 130,700 | 2025 |

* Based on operating days and normal available capacity – not necessarily what is actually produced currently. (Reproduced from (OECD –NEA Report 2014[14])

**BR2, Belgium:** BR2 undertook additional irradiation cycles during April and May 2013 to compensate for the reactor problems causing outage of HFR (Petten). The authorities requested stress tests of BR2 in 2013 following the Fukushima incident; implementation of this work is scheduled during 2013-2016. The next operating license period will be for the period 2016-2026; a safety review will be due by 2016. A major maintenance and design update - including the replacement of the beryllium matrix - will be undertaken with a shut down of up to 16 months commencing March 2015. Restart of the reactor is scheduled for July 2016. The operator SCK.CEN plans to construct and operate a new research reactor (MYRRHA), an Accelerator Driven System that aims to be operational from 2024. This new facility is intended for nuclear technology research and isotope production as a replacement for the BR2 reactor.

**HFR Petten, Netherlands:** HFR was out of action for eight months during 2012-13 (tritium contamination was found in groundwater due to cooling pipe leakage). The reactor was restarted in June 2013 with new operating systems in place. It is intended that HFR will remain operational until a new Dutch reactor is built at the Petten site. The operator NRG plans to construct and operate a new reactor (**PALLAS)** that aims to be operational from 2024. This new facility is intended for isotope production and nuclear technology research. The Dutch government have pledged (in October 2014) a loan of €30 million to keep the HFR reactor operational until 2024.

**LVR-15, Czech Republic:** LVR-15 has responded to the global shortage and increased production capacity of $^{99}$Mo. This has helped offset the shortages in the market from the other sites although is only likely to be operational in the short to medium term.

**Maria, Poland:** As per LVR-15 above.

**NRU, Canada:** Prior to 2008 NRU was the major world producer of $^{99}$Mo, responsible for 40-50% of global demand through Nordion Inc. (Ottawa, Canada). Due to licensing problems preventing operation, in 2008 Atomic Energy Canada Ltd (AECL) was forced to cancel the replacement reactors for NRU, MAPLE-I and MAPLE-II (Multipurpose Applied Physics Lattice Experiment). The MAPLE reactors had the potential to supply the entire global needs of $^{99}$Mo as well as supplying $^{131}$I, $^{125}$I and $^{133}$Xe. Short-term provision of $^{99}$Mo will be maintained at NRU until November 2016 after which the facility will cease operation permanently.

**OPAL, Australia**: A new research reactor, known as OPAL (Open Pool Australia Light water reactor), was opened in 2007. OPAL is a 20 MW open-pool design using both LEU fuel and targets. At present OPAL supplies mainly the Australian domestic market, however further investment has enabled additional processing and waste capacity to increase production to also supply into the Asian and US markets.

**OSIRIS, France:** In 2011 the French regulator Autorité de Sûreté Nucléaire (ASN) determined that the OSIRIS Reactor would continue operation until 2015 but would close no later than the end of 2015 in compliance with the ASN decision of 2008. The French are currently building a new 100 MW materials-testing reactor (Jules Horowitz) at Cadarache in southern France. This reactor is not expected to be operational until at least 2021.

**SAFARI-1, South Afric**a: Conversion of the 20 MW SAFARI-1 reactor at NESCA's Pelindaba facility to LEU fuel was completed in 2009 with the introduction of LEU-based targets for isotope production. The first radioisotopes produced by SAFARI-1 were shipped to international customers in July 2010.

### *1.3.3 Processors*

It can be seen from Tables 1.1 and 1.2 and from Figure 1.5 that the global supply of $^{99}$Mo is dependent on a small number of research reactors and processing facilities. Currently, six processors are involved in the production of $^{99}$Mo on a globally significant scale. These include AECL/Nordion (Canada), ANSTO (Australia) and MALLINCKRODT (Netherlands). Regional limitations on processing capacity can limit the full use of reactors during outage situations. It is understood that AECL/Nordion will cease processing after the closure of NRU Canada.

**Table 1.2** Current processors of $^{99}$Mo

| Processor | Country | Target | Capacity per week (6-d Ci) | Annual Capacity (6-d Ci)* | Expected date of conversion to LEU targets |
|---|---|---|---|---|---|
| **AECL/NORDION** | Canada | HEU | 4,680 | 187,200 | Not expected |
| **ANSTO HEALTH** | Australia | LEU | 1,000 | 52,000 | Started LEU |
| **CNEA** | Argentina | LEU | 900 | 46,800 | Converted |
| **MALLINCKRODT** | Netherlands | HEU | 3,500 | 182,000 | 2017 |
| **IRE** | Belgium | HEU | 3,500 | 182,000 | 2016 |
| **NTP** | South Africa | HEU/LEU | 3,500 | 182,000 | 2014 |

* Actual production is often less, as processing capacity is technically available 52 weeks while irradiated targets are not delivered 52 weeks of the year for all processors. This may have the effect of some processing capacity not being fully used if there is not sufficient irradiator capacity to supply the processor with irradiated product (Reproduced from (OECD –NEA Report 2014[14])

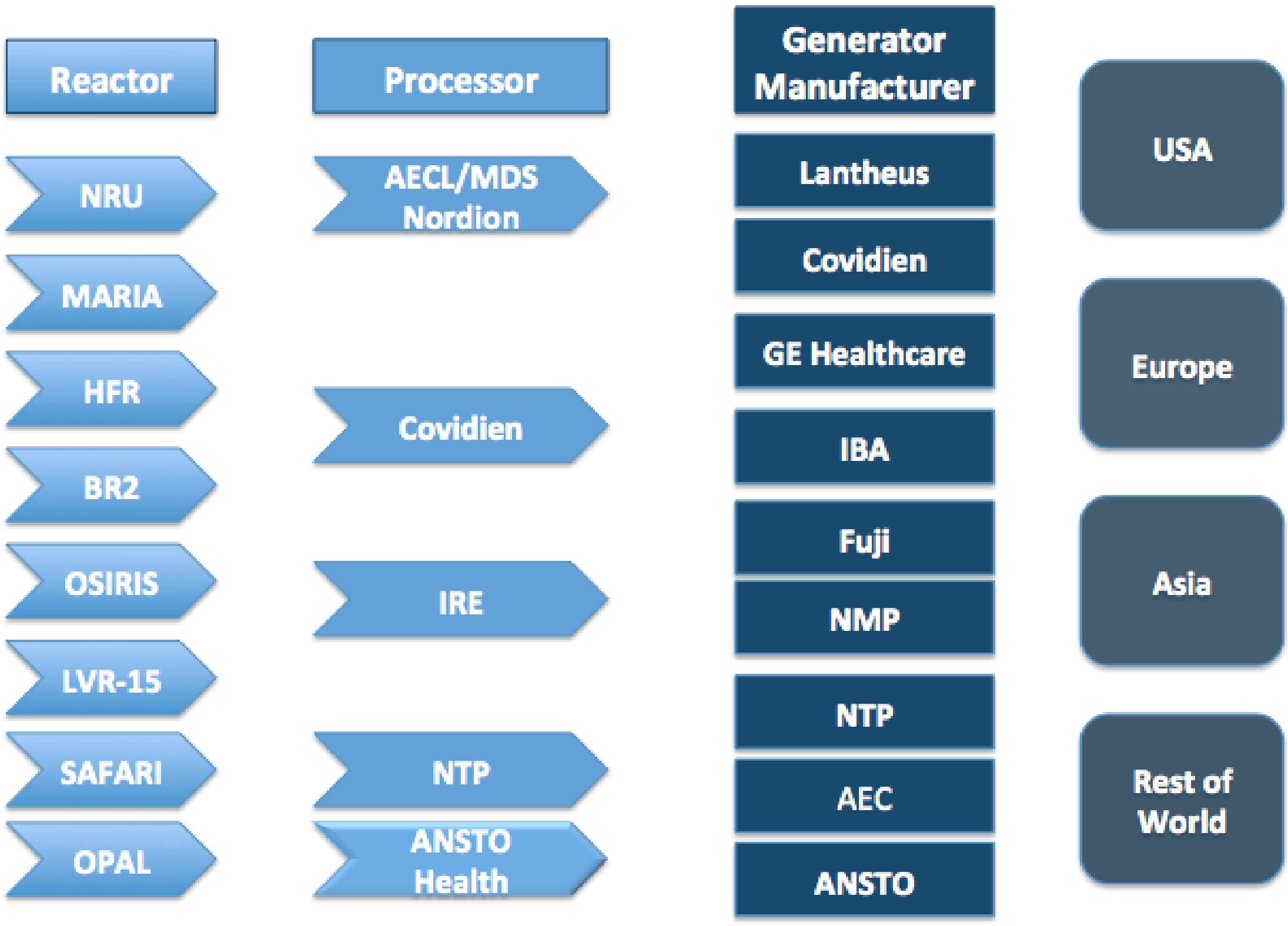

**Figure 1.5** Diagram showing the Molybdenum-99 distribution network from producer to end-user

## 1.4 Full-Cost Recovery and Outage Reserve Capacity

The origins of the radioisotope supply issue reside in the past use of research nuclear reactors that were developed exclusively using 100% funding from national governments, and primarily for nuclear materials research. $^{99}$Mo was initially a by-product of the main reactor activity, and when the importance of $^{99}$Mo increased there was no corresponding sustainable economic assessment or pricing to reflect the proper contribution to production costs and deterioration of equipment. The revenues to $^{99}$Mo producers remained small whilst demands increased, and tariffs were set by long-term contracts. This meant that increasingly the governments of $^{99}$Mo-producing countries effectively subsidized foreign healthcare systems and the companies supplying them. There was therefore no incentive for investment as those reactors aged and became increasingly vulnerable to outages and closures. Furthermore there was no incentive to participate in outage reserve schemes to mitigate the impact of unforeseen and unexpected closures of those same production facilities.

In recognition of the importance of production economics to the future secure supplies of radioisotopes, the OECD HLG-MR recommended that economic return must be improved compared to economic model in existence. In particular, it was recommended that reactors should operate their $^{99}$Mo production on a full-cost-recovery basis so that outage reserve capacity is provided, valued and recompensed to the producers. The following definitions were adopted:

### *1.4.1 Full-Cost Recovery (FCR)*

All costs associated with producing $^{99}$Mo are identified and covered through prices set for irradiation services. This should include both the operational and capital costs. FCR will increase $^{99}$Mo prices to healthcare providers significantly although the increase will be much smaller at the hospital end-user level. Higher prices at reactors must be passed on throughout the supply chain. FCR is important for economic sustainability and security of supply and to encourage new and ongoing investment in future $^{99}$Mo production capacity.

### *1.4.2 Outage Reserve Capacity (ORC)*

OCR is defined as the capacity within the system to account for occasional unplanned or extended shutdowns of reactors. Processors should hold ORC equal to at least their largest source in the supply chain (n-1 criterion). Market-based ORC prices allow for full cost recovery. ORC costs are expected to cover transaction and fixed capital and operational costs. Exercising an ORC option in the market will trigger additional payment for variable costs.

The scale of the problem associated with the economic model presently used in financing the global supply of $^{99}$Mo/$^{99m}$Tc is illustrated in Figure 1.6. It is acknowledged by the OECD NEA HLG-MR that this financial model is not sustainable. Inevitably the cost of $^{99m}$Tc per unit to the end-user will have to increase.

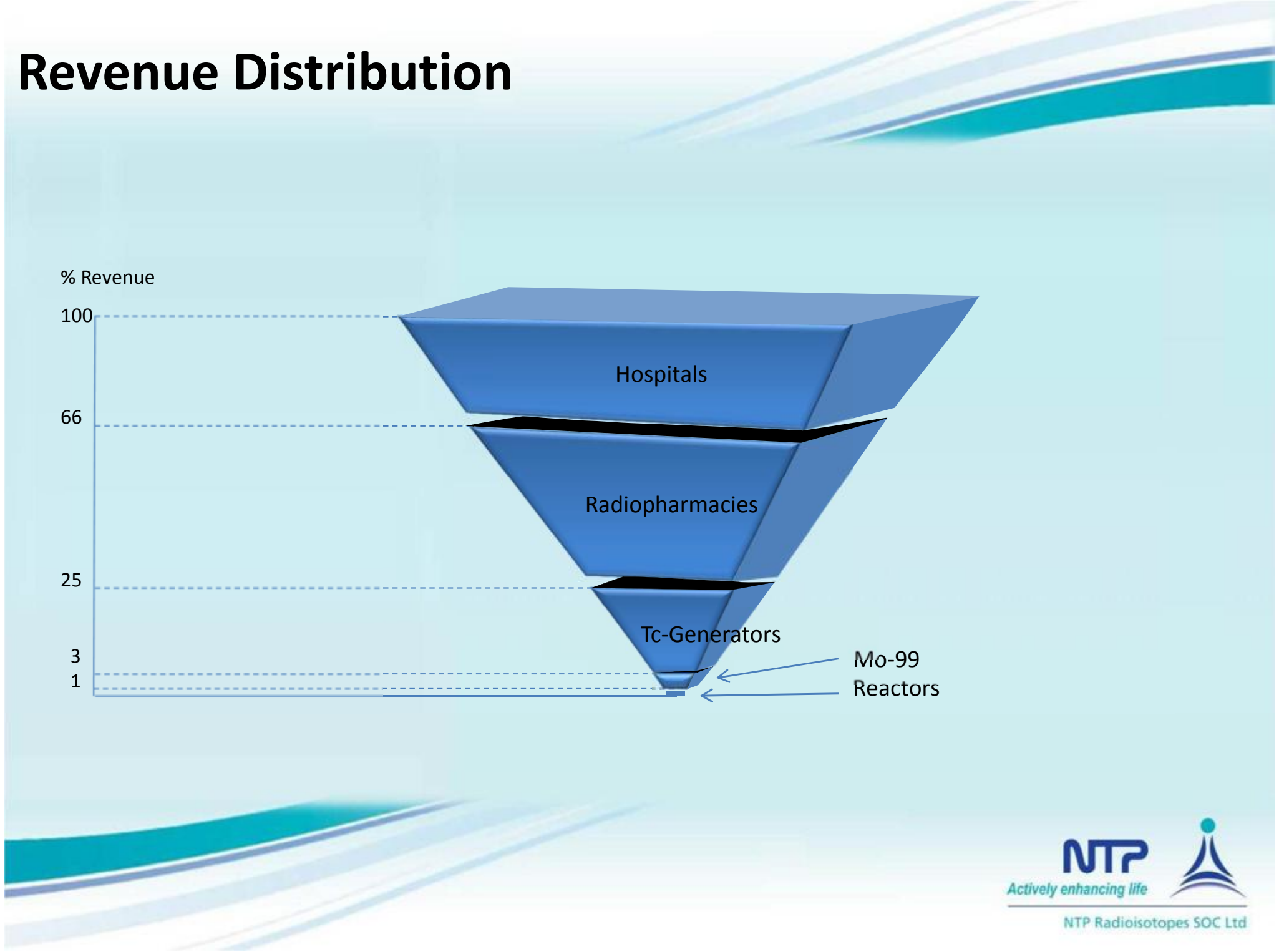

**Figure 1.6** Problem associated with the present financial model used to support global Molybdenum distribution (Courtesy of Piet Louw, NTP)

## 1.5 Future Demand and Supply

As outlined above, and based on returns from national surveys, the OECD-NEA estimates that demand for $^{99}Mo/^{99m}Tc$ will continue to rise within the forecast period out to 2030. The rate of rise will be different in the mature compared to the developing markets; within the mature markets, the need for $^{99}Mo/^{99m}Tc$ will increase at a rate of 0.5% per annum whereas the forecast for the developing markets is of the order of 5%. There are a number of factors contributing to this overall rise in demand including an increasingly older population and an increasing prevalence of dementia and malignancy.

The estimate of the true global demand for $^{99}Mo/^{99m}Tc$ also needs to incorporate a requirement for outage reserve capacity (ORC). It is not possible to measure ORC directly as outages are often unpredictable. The HLG proposal was that a processor should hold sufficient reserve capacity to replace the largest supplier of irradiated targets in the supply chain (the n-1 criterion). Experience has shown that the additional need to cope with supply outages lies somewhere between the n-1 and the n-2 criterion. This translates to an additional 35% to 62% capacity as shown in green and red lines in the accompanying graphs (Figures 1.7, 1.8, 1.10).

Since 2009, a number of initiatives have developed options, both reactor-based and non-reactor based, that hold out promise to maintain/increase the availability of $^{99}Mo/^{99m}Tc$. The following information derived from the 2014 OECD-NEA Report (14) provides an update on irradiator and processing capacity now and in the future.

### *1.5.1 Present Global Irradiator Capacity*

As shown (Figure 1.7) the assumption is that the global demand for $^{99}Mo/^{99m}Tc$ will continue to rise throughout the prediction period 2015 to 2020 (line labeled 'Demand with no ORC'). As shown global $^{99}Mo$ capacity (Line labeled 'Current irradiation Capacity') is predicted to be sufficient prior to the start of the BR-2 refurbishment project in 2015, after which it drops sharply as there is no immediate replacement. The sustained fall also reflects the loss of output from the OSIRIS reactor due to close in 2015.

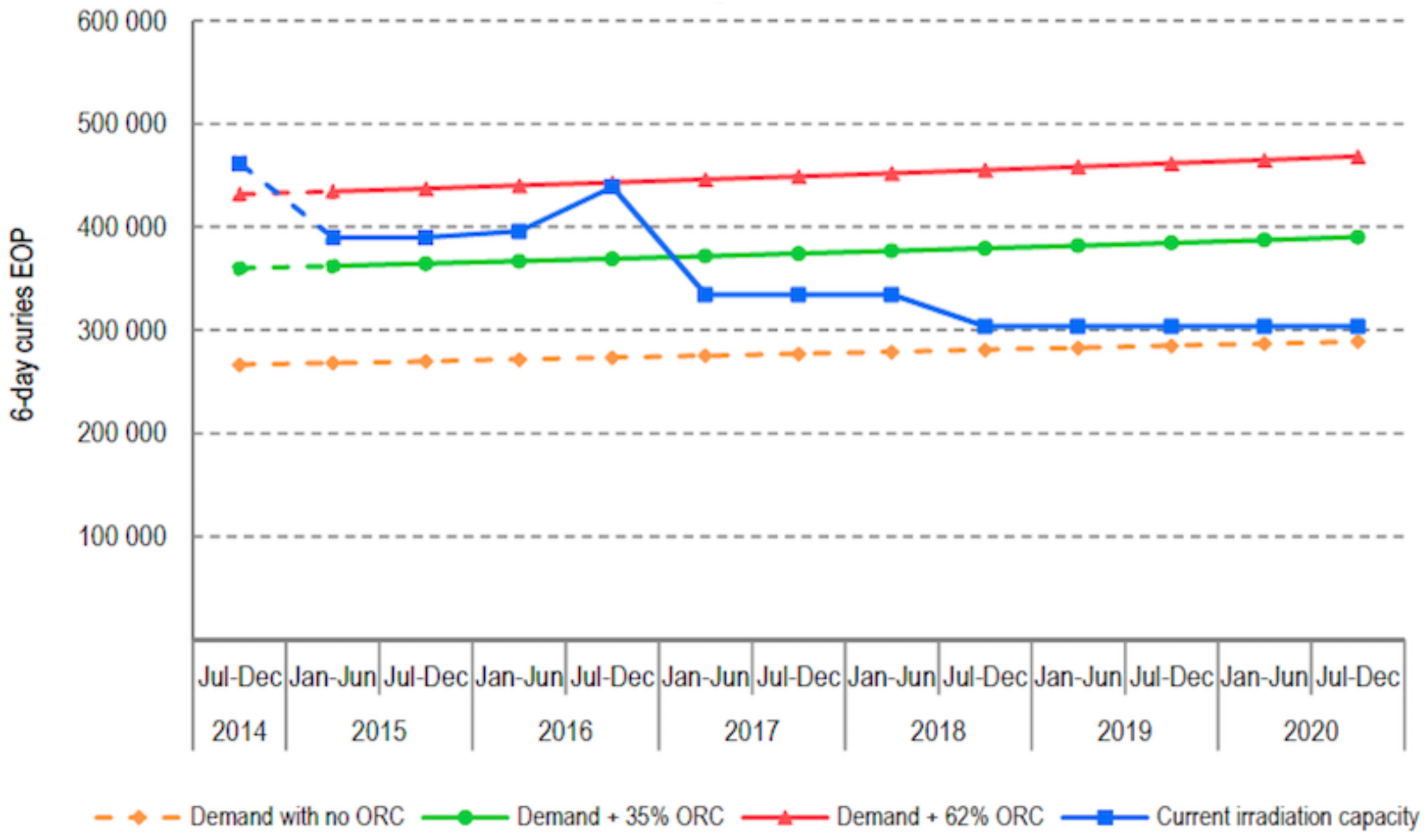

**Figure 1.7** Current irradiation capacity and demand for the forecast period 2015-2020 based on the current fleet of irradiators (Reproduced from (OECD – NEA Report 2014[14])

The increase in irradiation capacity in the second half of 2016 reflects the planned resumption of service from BR-2. The drop in capacity during the first part of 2017 reflects NRU's exit from the global supply chain in 2016. Assuming there is no additional input to the supply chain, the supply and demand lines converge introducing a significant risk of a shortfall of supply and substantially below the desired ORC. The ageing fleet of reactors, the impact of LEU conversion and the uncertainty as to whether operating licenses will be extended all increase the likelihood of shortfall in supply.

Based on the above considerations it is clear that it will be necessary to identify additional sources of $^{99}$Mo/$^{99m}$Tc supply.

### *1.5.2 Present Global Processing Capacity*

The rate-limiting factor in the supply chain for $^{99}$Mo is not confined to irradiator capacity but also is dependent on processing capacity. The reasons for this include the lack of on-site processing facilities at some reactor sites, and the loss of AECL/Nordion's processing facility with the closure of NRU in 2016.

As shown in Figure 1.8, processing capacity keeps pace with irradiation capacity until late 2016 when NRU closes and BR-2 re-enters the market. Unless there are new processing facilities it is predicted that there will be a gap between irradiation and processor capacity in the forecast period 2015-2020. It is also clear from Figure 1.9 that the years 2016 to 2017 will be a period when processing capacity falls below the ORC (Demand + ORC) line and therefore there could to be insufficient supply during this time. Thereafter, and assuming new irradiator and processing capacity comes on line, the total processing capacity appears sufficient to supply global demands.

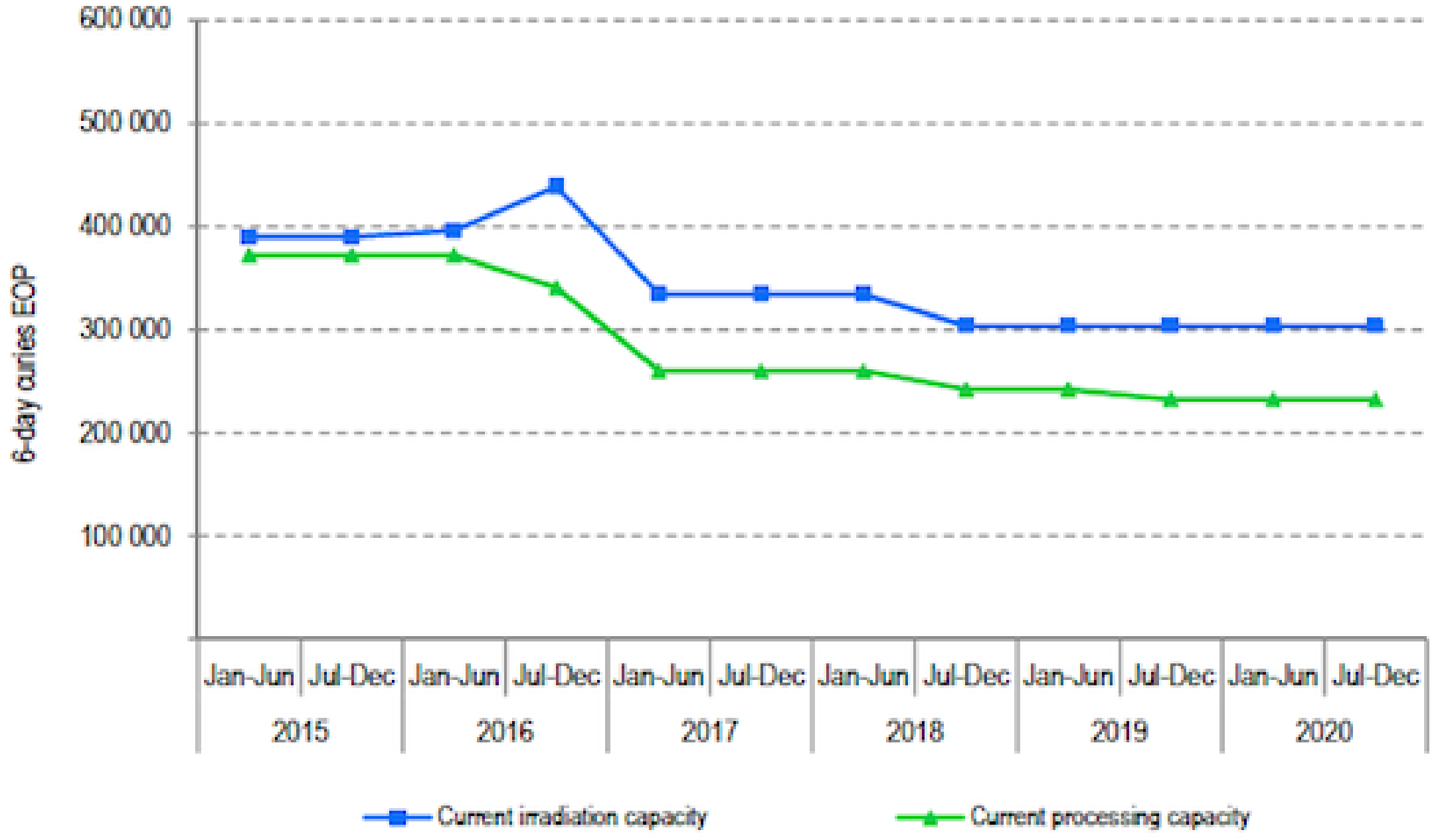

**Figure 1.8** Projected processing capacity compared to irradiation capacity and demand, 2015-2020 (OECD NEA 2014[14])

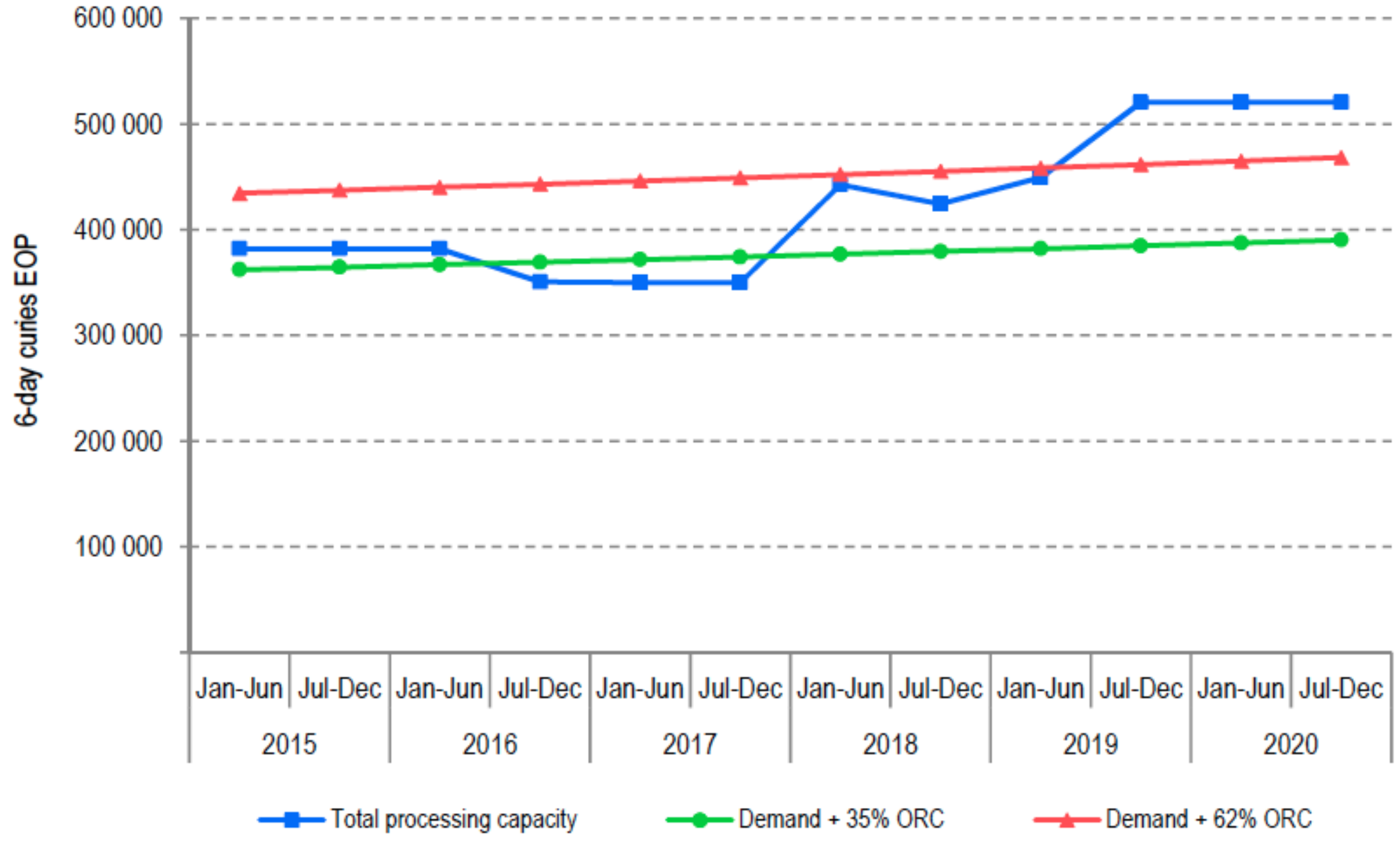

**Figure 1.9** Processing capacity compared to irradiation capacity and demand, 2015-2020 (OECD NEA 2014[14])

## 1.6 Global Initiatives To Address The Supply Problem

Possible paths to sustained availability of $^{99m}$Tc for medical imaging have been summarized in recent publications [20]. Individual nations have responded variously to the problems in the supply chain and as a result of the HLG-MR initiative, a number of different strategies are being pursued, some already underway, and some yet to commence. These are summarized in Table 1.3 and further detail is provided in Appendix 1.1 for each site.

**Table 1.3** National programmes for additional Molybdenum 99 capacity

| **Reactor** | **Targets** | **Weekly capacity (6-d Ci)** | **Annual production (6-d Ci)[1]** | **Estimated start date** | **Status** |
|---|---|---|---|---|---|
| RIAR (Russia) | HEU in CRR* | 1200 | 60000 | 2015 | Started |
| Karpov Institute | HEU in CRR* | 300 | 14800 | 2015 | Started |
| NORTHSTAR/MURR (USA) | Non-fissile/CRR* | 2750/3000 | 39100/156400 | 2015/17 | Phase 1 |
| MORGRIDGE/SHINE (US) | LEU solution with DTA* | 3000 | 144000 | 2017 | NYS* |
| NORTHSTAR (USA) | Non-fissile LINAC* | 3000 | 144000 | 2018 | NYS* |
| FRM-II (Germany) | LEU in CRR* | 1 600 | 54300 | 2017 | Started |
| OPAL | LEU in CRR* | 2600 | 111400 | 2017 | Started |
| KOREA | LEU in CRR* | 2000 | 85700 | 2018 | NYS* |
| CHINA Advanced RR | LEU in CRR* | 1000 | 50000 | 2019 | Modification |
| Brazil MR | LEU in CRR* | 1000 | 41400 | 2019 | Preliminary |
| RA-10 (Argentina) | LEU in CRR* | 2500 | 120000 | 2019 | Preliminary |
| Jules Horowitz RR (France) | LEU in CRR* | 3200 | 100600 | 2020 | Under Construction |

*CRR = conventional research reactor, DTA = deuterium-tritium accelerator, HEU = highly enriched uranium, LEU = low enriched uranium, LINAC = linear accelerator, NYS = not yet started, 6-day Ci = 6-day Curie

### *1.6.1 Potential impact of the new global initiatives*

The potential impact of these new reactor-based initiatives is illustrated in Figure 1.9, taken from the latest OECD Report [14]. It can be seen that, assuming entry into the market of new sources of $^{99}Mo/^{99m}Tc$, global irradiation capacity will remain above the lower ORC estimate of 35% but below 62% ORC during 2015 to 2016. During this period there is the potential for demand to outstrip supply.

Later in the forecast period it is anticipated that there will be an upturn in irradiation capacity due to commissioning of new reactor capacity particularly in South America and in Asia. The graphs separate conventional (reactor-based) from total irradiation capacity due to the fact that alternative methods of $^{99}Mo/^{99m}Tc$ production have not yet been used on a large-scale basis.

The model of production and supply of medical radioisotopes is vulnerable to disruption at many points along the supply chain. It is therefore vital that new strategies are developed to secure future supplies. Further details of options for alternative strategies towards sustainable radioisotope availability have been published in a recent article in the Journal of Nuclear Medicine [20] and further details are discussed in Chapter 4 of this report.

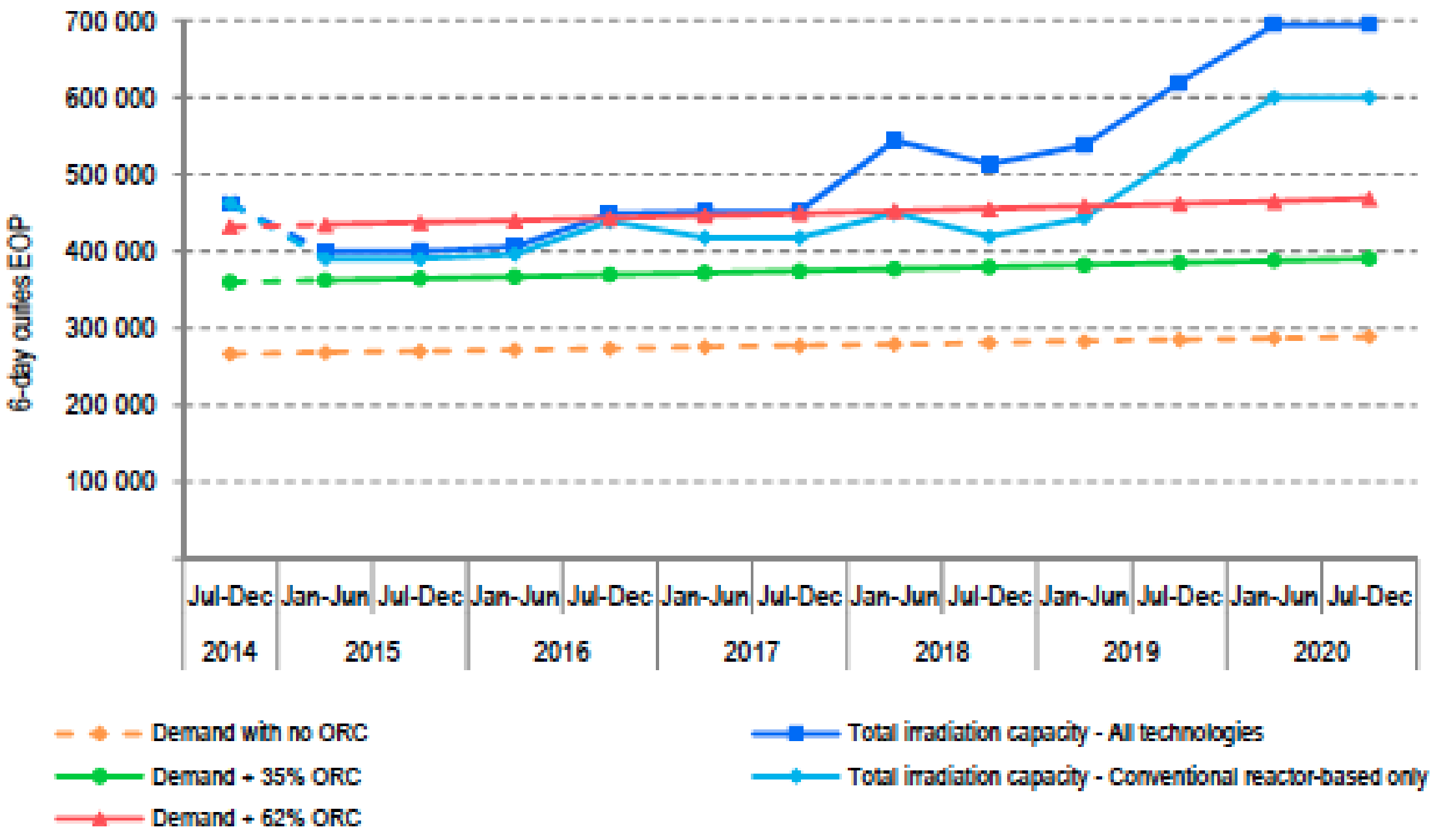

**Figure 1.10** Current and selected new irradiation capacity and demand, 2015-2020 (OECD NEA 2014)

## 1.7 Non-Reactor Initiatives

Some countries are exploring non-reactor based solutions for the supply of $^{99m}$Tc. Following the decision to shut the NRU reactor the Canadian government invested \$35 million between 2010 and 2012 from the Non-reactor-based Isotope Supply Contribution Program (NISP) and supported R&D of $^{99}$Mo/$^{99m}$Tc production using cyclotrons and other methods. The objectives were to diversify sources of isotope supply to reduce radioactive waste and to eliminate the use of HEU. In the period 2012-2016, the Canadian Government is investing a further \$25 million to further develop cyclotron and linear accelerator technologies for the production of $^{99m}$Tc on a commercial basis by 2016. Research and development activities are focused on the development of commercial, cost-efficient production of $^{99m}$Tc using high current cyclotrons (for example using the 19 and 24 MeV models manufactured in Canada by Advanced Cyclotron Systems Inc.). Because it is a direct production method, the cyclotrons must be based as a network in which the distribution distance by vehicle is no more than around 4 hours. An advantage of cyclotrons over some other non-reactor methods is that they can produce not only $^{99m}$Tc, but also a wide range of other PET and SPECT radioisotopes such as $^{18}$F, $^{64}$Cu, $^{123}$I, $^{124}$I, and $^{68}$Ga. This advantage is particularly relevant in the UK, and the national situation here is described in detail in Chapter 2.

## 1.8 Conclusion

From the foregoing discussion it can be seen that there is the potential for shortage of $^{99m}$Tc to the global market from 2016 through to 2020. The exit of NRU and OSIRIS from the market will be offset to an extent by the resumption of increased production at BR2 during 2016 and the fact that funds have been committed to keep HFR in production in the short to medium term. It is likely that other reactors will come on line but time scales are difficult to estimate and these are subject to frequent revision. The Maria and LVR reactors have been able to increase production in the short term. The situation is fluid and the estimates of the $^{99}$Mo supply are subject to change. For example the Australian Nuclear Science and Technology Organization (ANSTO) has recently announced that a \$168.8 million investment by the Australian Government will enable it to increase production to supply up to 25 to 30% of global demand [21]. This is likely to serve the needs of the southern hemisphere and given the uncertainties that persist in relation to demand and supply, it is recommended that the UK should explore alternative options for supply of its medical radioisotopes to maintain clinical nuclear medicine services. The following chapters discuss what measures are already in place in the UK and what options are available to mitigate the impact of the anticipated shortages that will arise from the reactor and processor changes that have been outlined above.

## Appendix 1.1 Initiatives of Individual Countries

**Russia** is developing additional irradiation and processing capacities although the contribution to the global market remains uncertain. Recently $^{99}Mo$ produced in Russia has been assessed as being of suitable quality to meet western regulatory standards.

**USA.** The US Department of Energy established a partnership with four US commercial companies to facilitate the establishment of independent technologies to produce $^{99}Mo$ without the use of HEU. These are as follows:

(i) **Babcock & Wilcox** in collaboration with Covidien has worked on the development of a Medical Isotope Production System (MIPS) to produce $^{99}Mo$ by irradiation of LEU fuel in solution in a low power 240 kW Aqueous Homogenous Reactors. Covidien withdrew from the project in October 2012 as the project was thought to be commercially non-viable.

(ii) **GE-Hitachi** completed a conceptual design to produce $^{99}Mo$ in BWR power plants by the neutron capture reaction $^{98}Mo(n,y)^{99}Mo$. However, GE-HITACHI decided in February 2012 to suspend progress on the project due to market conditions.

(iii) **NORTHSTAR** is working on two projects to produce $^{99}Mo$:

(a) Production of $^{99}Mo$ in MURR (Missouri University Research Reactor) by neutron capture ($^{98}Mo(n,y)^{99}Mo$) on natural and/or enriched $^{98}Mo$ targets. Production is expected to start in early 2014 and could be upgraded to 3000 Ci(6-d) per week by 2016.

(b) Production of $^{99}Mo$ by photonuclear production (100Mo(y,n)$^{99}Mo$) using a 24 MeV electron linac is also planned from 2016. Both technologies produce low specific activity $^{99}Mo$ that will require improvements to generator design.

(iv) **SHINE** Medical Technologies plans to produce $^{99}Mo$ in a subcritical LEU solution where fission is driven by an intense D-T neutron source. A low-energy deuterium beam is directed onto a tritium-loaded target to obtain fast (14 MeV) neutrons that are then moderated in a beryllium prior to entry into the LEU. SHINE expects to start commercial operations in mid 2016 with a licensed capacity of about 3000 Ci (6-d) per week from a single accelerator.

**FRM-II, Munich, Germany.** The FRM-II reactor in Munich has a project to install new equipment for the irradiation of LEU targets from 2015 depending on target availability. This project can be seen as a replacement for the $^{99}Mo$ production in Julich by the FRJ-2 reactor that was shut down permanently in 2006. The FRM-II research reactor is looking to irradiate LEU targets for $^{99}Mo$ production using the processing facility at IRE (Belgium). Transport arrangements are currently under consideration. This is unlikely to contribute to pharmaceutical grade production before 2017.

**OPAL, Australia.** The OPAL reactor (Lucas Heights) covers current national needs by the irradiation of LEU targets. Australia is building a new processing and waste facility with full production that is expected to start in 2016 providing 3500 6-d Ci per week. ANSTO's $^{99}$Mo produced from LEU targets is already FDA approved in the US.

**Korea.** Korea has announced $3B (US) government funding for a project to provide self-sufficiency in medical welfare. Presently there is a conceptual design for two parallel irradiation lines with five hot cells for production of 2000 Ci per week although there are no firm plans for construction.

**China**. The demand for $^{99}$Mo/$^{99m}$Tc is increasing in China. A 6MW reactor (CARR – China Advanced Research Reactor) has been operating since 2008 for medical radioisotope production. The range of materials produced includes $^{99}$Mo, $^{125}$I, $^{131}$I, $^{177}$Lu, $^{90}$Y, $^{14}$C and $^{188}$W.

**Brazil.** At present all $^{99}$Mo is imported. There is a plan to develop a 30MW multipurpose open pool reactor in cooperation with Argentina for production of 1000Ci $^{99}$Mo per week to satisfy domestic demand; the project is based at the Aramar Technological Centre in the State of Sao Paulo. Funding is to be provided from government and private partnership (around $500M US).

**France.** The license of the OSIRIS (France) reactor expires at the end of 2015. The new Jules Horowitz (JHR) reactor under construction in Cadarache will replace the OSIRIS reactor. Started in 2011, the reactor dome was put in place during 2013. It is planned to start operation in 2017 with production of $^{99}$Mo by 2020/21. JHR is expected to have the potential to provide up to 50% of European needs irradiating 32 to 48 targets per week giving up to 2400 6-day Ci per week.

**Belgium.** SCK.CEN plans to upgrade its BR2 reactor operating regime from five/six to up to eight operating cycles per year after a refurbishment, depending on the economics of production; BR2 currently operates 120 to 140 days per annum but from 2016 could operate from 180 to 200 days per annum. Refurbishment will allow operation until 2026 or until the start of the replacement facility MYRRHA scheduled to start operation in 2024.

**Argentina**. The Argentinian RA-10 30 MW reactor intends to satisfy the increasing national and regional demands for medical radioisotopes. The operator CNEA supports it through $200 million national funding ('Applications of Nuclear Technology in Health Industry and Agriculture'). RA-10 is planned to be operational by the end of 2018. The project plans a radioisotope processing and manufacturing facility with capacity for 2500 Ci $^{99}$Mo per week plus production of $^{131}$I. This will largely cater for the domestic and South American market.

# Chapter 2: Nuclear Medicine in the UK

**Brian Neilly and Alan Perkins**

## Summary

- Nuclear Medicine as currently practiced in the UK is a multidisciplinary specialty employing a workforce of more than 2,500 individuals
- There has been a significant increase over time in the NHS of the availability of gamma cameras with SPECT and SPECT-CT capability. These cameras are optimal for imaging using $^{99m}$Tc-based radiotracers.
- The demand for conventional (non-PET) imaging and non-imaging diagnostics in the UK is in the region of 600,000 per annum. Bone, cardiac and renal investigations make up more than half of the requests for these investigations.
- There is an increasing availability of PET-CT across the UK for a growing number of indications. In most of these centres there is limited capacity for further growth due to high use for FDG-PET-CT for oncological indications. Non-oncological and non-FDG PET indications presently make up only 10% of the total.
- The cyclotrons available in the UK for the production of $^{18}$FDG and other PET-tracers are not suitable for the large-scale production of $^{99m}$Tc. Higher-energy cyclotrons would have to be installed if commercial quantities of $^{99m}$Tc are to be produced.
- Nuclear medicine services in the UK receive generators from only three commercial companies. The majority of radiopharmacy services in the UK have been affected by generator shortages.
- Most nuclear medicine services in the UK have developed contingencies to deal with periods of shortage. These include re-scheduling the patient, employing alternative radiopharmaceuticals, or using lower administered activities, none of which are optimal. In a small proportion of cases this change has affected future patient referrals.
- There is evidence of unnecessary waste of radioactivity arising from the current generator delivery schedules to departments. Delivery and elution schedules for $^{99m}$Tc generators are highly complex and software options exist to facilitate the correct choice. This could be improved through partnerships with the commercial generator suppliers.

## 2.1 Introduction

The UK, like other countries, has been affected by the global shortage of $^{99}$Mo and $^{99m}$Tc as well as experiencing shortages of other reactor-derived radioisotopes such as $^{131}$I. Arguably the UK is more vulnerable to breaks in the supply chain since there is no domestic production either of $^{99}$Mo or of $^{99m}$Tc. Breaks in the supply chain can occur at reactor, processor or generator manufacturer stages with additional disruption possible during transportation to the UK. The use of medical radioisotopes in the UK is not restricted to $^{99m}$Tc but includes other radiopharmaceuticals including radioisotopes of iodine, indium and also PET radioisotopes other than FDG. The past 10 years has seen an increasing availability of PET imaging facilities. This chapter aims to review the present status of nuclear medicine services in the UK.

### *2.1.1 Nuclear Reactors*

There are no nuclear reactor target irradiation or molybdenum processing facilities for medical radioisotope production within the UK. As far as can be ascertained there are no plans to build a research nuclear reactor in the UK for or any other civilian nuclear purpose. Even in the event that a research reactor was proposed for these purposes the lag between proposal and the manufacture of radionuclides is considered to be up to 10 years. Given the timescales of shortages and the clinical need, the option of a research reactor has therefore been discounted in this report.

### *2.1.2 Generator Suppliers*

The UK relies exclusively on the importation of $^{99}$Mo/$^{99m}$Tc generators or processed bulk molybdenum (for generator manufacture at the GE plant at Amersham) from overseas. Three companies distribute $^{99m}$Tc generators to vendors throughout the UK: GE Healthcare; IBA/Alliance; and Mallinckrodt Pharmaceuticals. The share of the UK market for $^{99m}$Tc generator supply is in the region of 40%, 30% and 30% for GE, IBA/Alliance and Mallinckrodt Pharmaceuticals respectively. Generators are distributed to approximately 150 outlets in the UK; these outlets include central radiopharmacies and individual hospital departments. The generator suppliers source their raw material for generator manufacture from several suppliers and have in place contingency arrangements to attempt to secure supplies in times of shortage. Communication with the generator suppliers to the UK market to inform customers of anticipated times of shortage has improved in recent years allowing end users to invoke contingency measures to maintain clinical services.

The objectives in this section are to provide a snapshot of nuclear medicine in the UK and to provide an overview of how it is practiced. Data is drawn from several sources including the British Nuclear Medicine Society (BNMS) national survey and the UK Radiopharmacy Services Review in 2012.

The chapter also describes the current status of PET radiopharmaceuticals and cyclotrons for the production of PET tracers in the UK. It also sets out the initiatives that have been undertaken to mitigate the impact of shortages of $^{99m}$Tc and other reactor derived radioisotope.

## 2.2 BNMS Survey 2010

The BNMS survey is a web-based resource offering an ongoing annual snap shot of nuclear medicine within the UK. The last survey was undertaken in 2010 and was completed by 132 out of 210 (62%) known nuclear medicine departments. Data from a recent scoping exercise undertaken as part of the ongoing BNMS survey re-design is that the true estimate of NHS departments in the UK offering nuclear medicine services is in excess of 300. This fact, coupled with the participation of only 62% of departments in the 2010 BNMS survey means that the 2010 data likely represents an under-estimate of nuclear medicine practice in the UK. Nevertheless, the data is drawn from across a wide spectrum of regions providing figures to provide a reasonable snapshot of current nuclear medicine practice.

The survey was completed by participating departments from each of the English regions and from each of the devolved administrations including: England NE 6, England NW 20, Yorkshire 7, East Midlands 10, West Midlands 10, East of England 10, England South West 11, England South Central 9, England South East 9, London 23, Northern Ireland 4, Scotland 11 and Wales 3. The survey revealed information about the staffing (Table 2.1), equipment (Table 2.2), the scope and number of diagnostic studies undertaken (Table 2.3), and the number of procedures using therapy radionuclides (Table 2.4).

### *2.2.1 Nuclear Medicine Staffing*

The 2010 BNMS survey provided an estimate of the workforce employed within Nuclear Medicine in the UK and emphasizes its multi-disciplinary nature.

**Table 2.1** UK Workforce employed in Nuclear Medicine in 2010

| Staff Category | Number of Staff |
|---|---|
| Administrative & Clerical | 331 |
| Clinical Practitioners (Technicians/ Radiographers) | 1048 |
| Clinical Scientists | 281 |
| Health Care Assistants | 110 |
| Nuclear Medicine Clinicians | 89 |
| Nuclear Medicine Trainees (Science + Clinician Trainees) | 142 |
| Nuclear Medicine Radiologists | 389 |
| Nurses | 160 |
| Radiochemists* | 10 |
| Radiopharmacists* | 89 |

* Doubt has been expressed about the accuracy of the estimate of numbers of Radiochemists and Radiopharmacists in the survey.

### *2.2.2 Nuclear Medicine Equipment*

While the absolute numbers of gamma cameras and PET-CT scanners is likely to be underestimated (see above), it can be seen that the proportion of camera systems with SPECT, SPECT-CT and PET-CT has increased significantly over the four years since 2007 (Table 2.2). The number of planar gamma cameras has decreased in this time. The numbers of SPECT and SPECT-CT camera systems in place favour $^{99m}$Tc-based detection systems rather than PET systems at this time.

**Table 2.2** Nuclear Medicine Equipment in the UK in 2010

| NM Equipment | Equipment Type | 2007 | 2010 |
|---|---|---|---|
| Gamma Cameras | Planar | 78 | 27 |
| | SPECT | 185 | 240 |
| | SPECT-CT | 36 | 81 |
| PET Cameras | PET-only | Few | 0 |
| | PET-CT | 12 | 27 |
| Cyclotrons | NHS/University | Few | 8 |
| | Private Sector | Few | 5 |

### *2.2.3 Nuclear Medicine Procedures*

The BNMS survey provided a snapshot of the types of investigations undertaken using radionuclides in the UK in 2010 (Table 2.3). Also shown is the number of PET studies reported in the UK in 2010. The present estimate of numbers of PET-CT procedures is now estimated to be in the region of 40-50,000 per annum.

**Table 2.3** Types of Investigations using Radionuclides in the UK

| Nuclear Medicine Diagnostic Studies 2010 | NM Diagnostic Studies excluding PET in 2010 (Number) | PET Imaging Studies in 2010 (Number) |
|---|---|---|
| Brain | 11,718 | |
| Cardiac | 97,529 | |
| Endocrinology | 29,998 | |
| Renal | 74,403 | |
| Respiratory | 59,182 | |
| Bone | 218,408 | 81 |
| Oncology | 5,702 | 19950 |
| Infection/inflammation | 4,926 | |
| SLNB | 24,052 | |
| GI | 10,332 | |
| Miscellaneous | 62,402 | |
| Non-Imaging Diagnostics | 43,266 | |
| **Total Diagnostic procedures** | **673,156** | **20031** |

Corroboration of UK demand for $^{99m}$Tc and other non-PET radioisotopes has come from figures provided by industry. In 2012 industry figures estimated that the total number of diagnostic procedures using $^{99m}$Tc, $^{111}$In, $^{201}$Tl or isotopes of iodine ($^{123}$I, $^{131}$I), is in the region of 600,000 procedures per annum. Of these, the majority of procedures utilize $^{99m}$Tc. As shown in Figure 2.1, the main indications for $^{99m}$Tc-based radiopharmaceuticals are for bone, cardiac and renal scintigraphy.

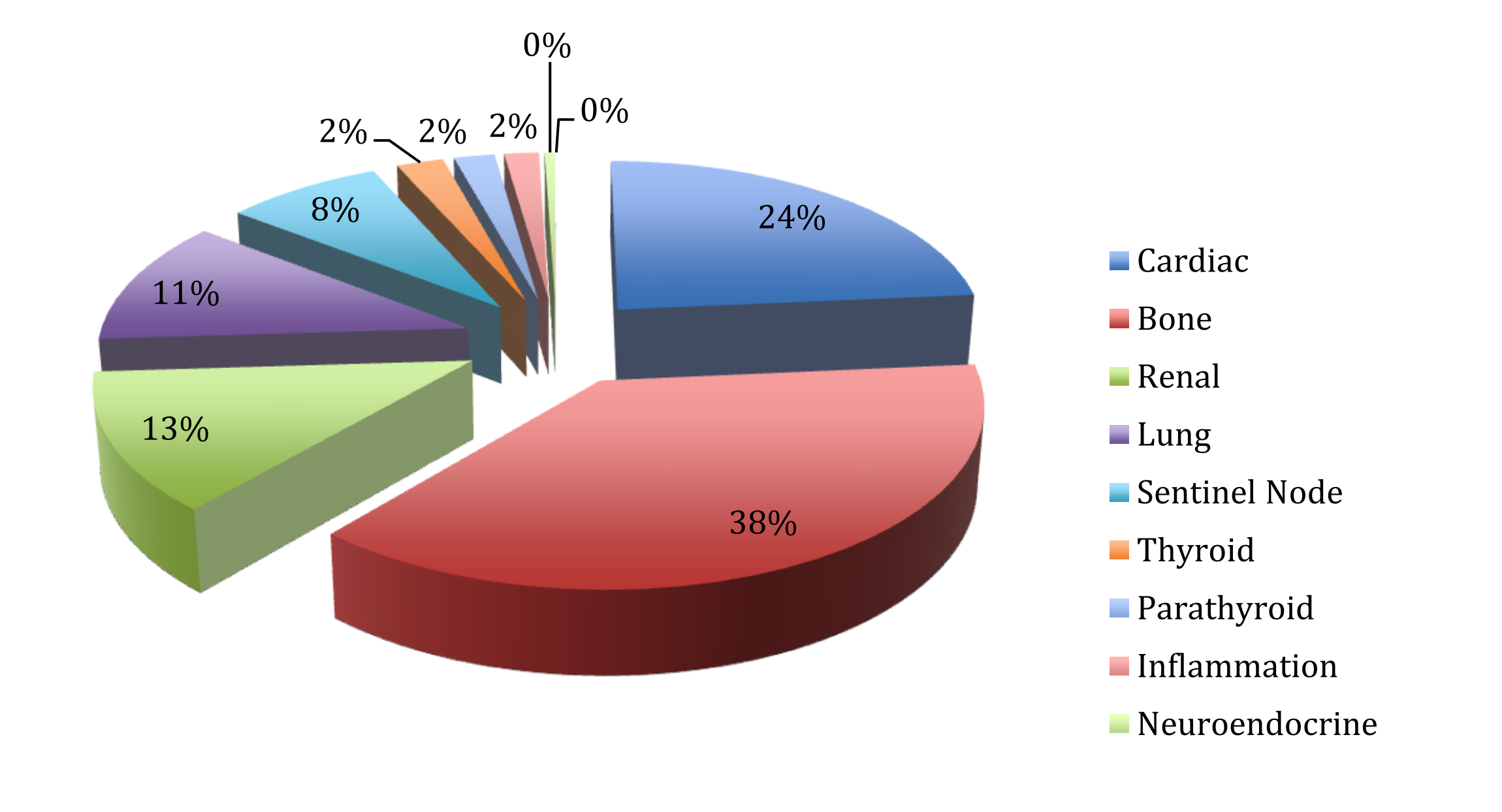

**Figure 2.1** Data sourced from Industry as to the breakdown of use of non-PET radiopharmaceuticals (mainly $^{99m}Tc$) used in the UK in 2012

### *2.2.4 Therapy Radionuclides Used in the UK*

The 2010 BNMS survey provided an estimate of the therapies undertaken in the UK during 2010 using radionuclides. The results are shown in Table 2.4.

**Table 2.4** Number of therapy procedures undertaken using radionuclides in the UK during 2010

| **Radionuclide Therapy UK 2010** | **Number in 2010** |
|---|---|
| In-patient Radionuclide Therapy | 2,842 |
| Outpatient Radionuclide Therapy | 9,044 |
| Total | 11,886 |

Treatment of benign thyroid disease (chiefly thyrotoxicosis) is the most common indication for radionuclide therapy using $^{131}I$. In addition to this 131I and other beta and more recently alpha radionuclides are used in the treatment of a wide range of oncological and non-oncological conditions as shown in Figure 2.2. Also shown are the numbers of centres in the UK involved in the administration of such therapies. This is expected to increase substantially with the introduction of $^{223}Ra$.

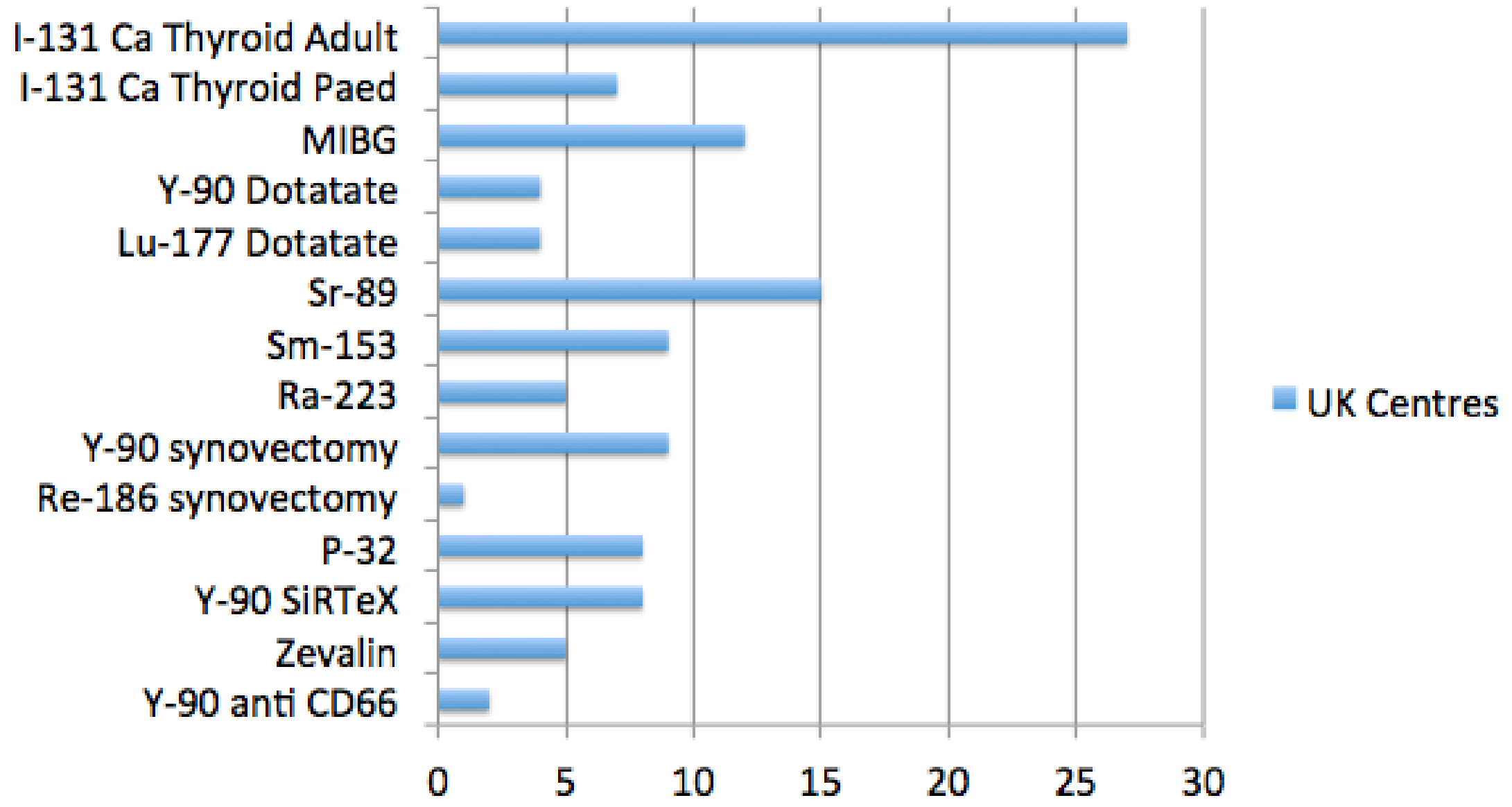

**Figure 2.2** Types of therapy using radionuclides (excluding $^{131}$I for benign thyroid disease) and number of participating UK centres 2012.

## 2.3 Radiopharmacy Services Review 2012

In the context of the recurrent shortages of medical radioisotopes, a survey of radiopharmacy services in the UK (2) was carried out by the BNMS by contacting all UK nuclear medicine departments held on a central contact list. Units were invited to respond to an online questionnaire managed through Survey Monkey™ between 3 and 31 October 2012.

The survey was completed by just over half of the 222 Nuclear Medicine departments known in the UK. More than two thirds of the responding departments (71.2%) held a manufacturing license with the MHRA. All departments received $^{99m}$Tc from a generator or a multi-dose vial. Of the departments responding, none stated that they received $^{99m}$Tc prepared as individual patient doses.

### *2.3.1 Generator Suppliers*

In the survey described above it was noted that $^{99m}$Tc generators were supplied by three commercial providers: Mallinckrodt; IBA and GE Healthcare. Twenty-eight different generator types and sizes were distributed from the different manufacturers. Generators are delivered on every day of the week except Sunday. The most common day for delivery was Wednesday (51%) followed by Saturday (42%) and Friday (24%) respectively. Delivery of the generator on a Friday and Saturday results in decay of valuable product given that most departments do not undertake clinical work on a Saturday and Sunday.

Analysis of the data returned in the survey showed that the use of customized software applications can provide a valuable means of optimizing weekly and daily activity profiles and can increase the amount of radioactivity available on a daily basis whilst reducing costs.

### *2.3.2 Generator Shortage Responses*

Nuclear medicine departments were asked if they had been affected by radioisotope shortages. The majority (93%) of nuclear medicine departments who responded had been affected by the shortages whereas 7% of departments claimed to have been unaffected.

The survey also revealed the actions taken by nuclear medicine services during periods of supply shortages due to reactor outages. Those departments that were affected were asked to state how they had maintained services during times of generator shortage (Table 2.5). Seventeen percent of departments felt that the shortages had affected subsequent patient referrals.

**Table 2.5** Responses to shortages of $^{99m}$Tc-generators

| Response to generator shortages. | % | Number of respondents |
|---|---|---|
| 1 Postpone/reschedule patient | 81.0% | 85 |
| 2 Use alternative radiopharmaceuticals | 18.1% | 19 |
| 3 Send patients to other diagnostic modalities | 25.7% | 27 |
| 4 Reduce activity administered | 48.6% | 51 |
| 5 Share activity with other units | 30.5% | 32 |
| 6 Reduce or stop supplies to other units | 9.5% | 10 |
| 7 Other (please specify) | 32.4% | 34 |

### *2.3.3 Radiopharmacy Services Review Summary*

- The nuclear medicine community receives its generators from three commercial companies.
- The majority of nuclear medicine services in the UK were affected by generator shortages.
- Most nuclear medicine services in the UK have developed contingencies to deal with periods of shortage. In a small percentage of cases this change had affected future patient referrals. These include re-scheduling the patient, employing alternative radiopharmaceuticals, using lower tracer activities thus having an adverse impact on patient services.
- There is evidence of loss of potentially useful radioactivity arising from the timing generator delivery schedules, which could be improved through better management.
- The type, size, delivery and elution schedules for $^{99m}$Tc generators are highly complex and software options exist to facilitate the correct choice and delivery of generators.

## 2.4 PET Radioisotopes UK

$^{18}$F-FDG PET-CT scanning is now accepted as a mainstream diagnostic test in most patients with lung cancer, lymphoma and colorectal carcinoma. The indications for PET-CT in the UK are varied and are set out in a joint document by the Royal College of Physicians and the Royal College of Radiologists document [4].

### *2.4.1 Demand for PET Services*

The current demand for PET-CT using $^{18}$F-FDG and other PET radioisotopes in the UK is uncertain. The 2010 BNMS Survey [1] estimated demand for PET-CT in the UK to be in the region of 20,000 investigations per year. However, for reasons previously stated, it is likely that this is an underestimate of the true demand; it is thought that the present number of PET-CT procedures currently undertaken in the UK is of the order of 40-50,000. Ninety percent of PET-CT procedures in the UK are performed using $^{18}$F-FDG for oncology indications; the remaining ten percent employ $^{18}$F-FDG for non-oncological indications or non-$^{18}$F-FDG PET radioisotopes for a variety of indications (oncological and non-oncological). In line with global trends it is anticipated that the demand for PET-CT in the UK will increase with time. The number of fixed site PET-CT Facilities in the UK has increased from 11 (6 in London) in 2005 to an estimated 34 (10 in London) in 2013 (Figure 2.3). In addition, under the existing national contract there are 40-50 PET-CT sites in England that are served by mobile PET-CT units.

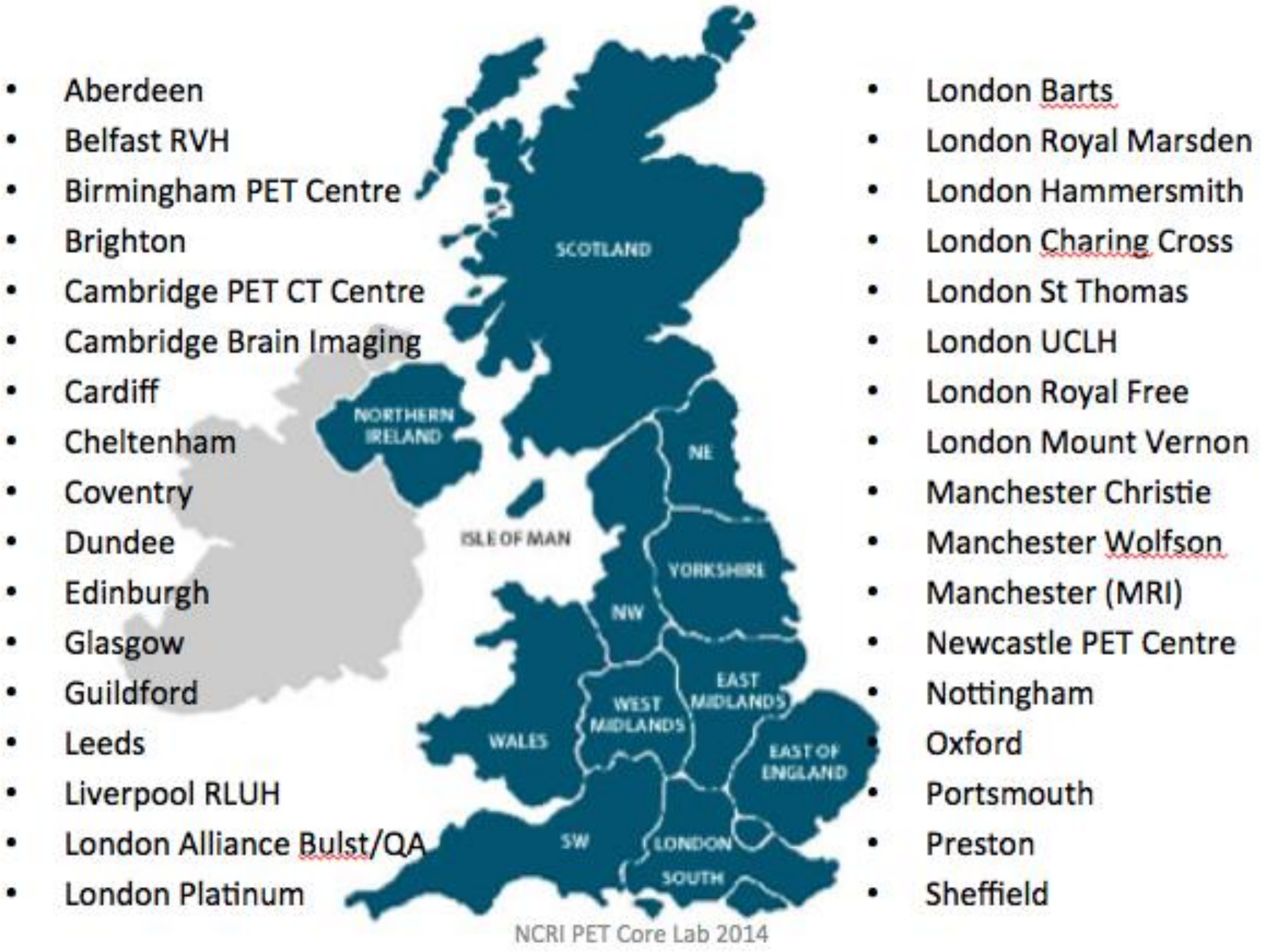

**Figure 2.3** Fixed site PET-CT Facilities in the UK 2013

### *2.4.2 Supply of PET Radiotracers*

$^{18}F$ and other PET radioisotopes are generally manufactured using cyclotrons and do not require a similar complex supply chain to that for fission $^{99}Mo$, involving nuclear reactors, uranium-bearing targets or remote processors. Because of the short half-life of $^{18}F$ (110 minutes), its derived radiopharmaceuticals such as FDG have to be manufactured close to where they are used; investigations using these ultra-short-lived PET tracers can only be carried out where there is co-location of the PET radiopharmacy facility and the hospital site. In contrast, a small number of radiopharmaceuticals such as $^{82}Rb$ and $^{68}Ga$ may be made available to hospital sites using generators (similar to those for $^{99}Mo/^{99m}Tc$).

At present FDG is produced in cyclotrons distributed throughout the UK, some by commercial manufacturers and some by the NHS/universities. The present locations of the cyclotrons that are located in radiopharmaceutical production units (RPUs) are shown in Figure 2.4.

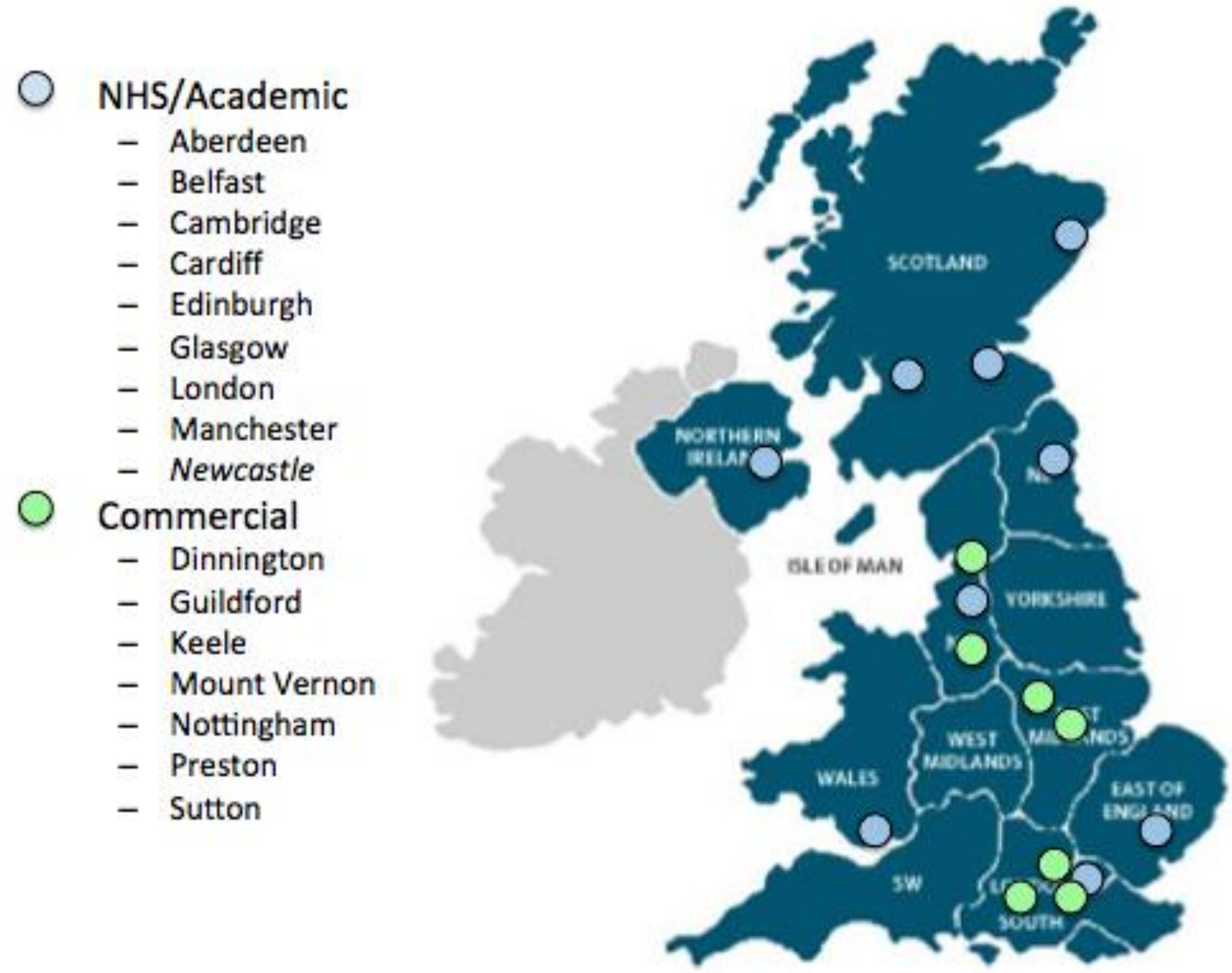

**Figure 2.4** UK PET cyclotrons (information derived from NCRI record of availability of cyclotrons in the UK [5])

The capacity of a PET radiopharmacy facility to manufacture FDG for PET scanning varies from one commercial manufacturer to another. This depends on a number of variables including cyclotron output capacity, number of clean rooms, number of synthesis boxes and dispensing units and the distance from PET centres served by the facility For a typical single-cyclotron PET radiopharmacy operating 2 target irradiations per day (Monday to Friday) the distribution radius is typically within a travel time of approximately 2 hours (about one half-life). Any greater distance than that will impact on customers who are likely to suffer a shortfall of activity due to the physical decay occurring during the delivery period.

Assuming a commercial site can deliver 20,000 patient doses of FDG per facility per annum, the potential number of doses of FDG that could be supplied from the 6 operational commercial sites in England is in the region of 120,000; as can be seen from the map (Figure 2.4), there are some areas of the country where PET radiopharmacy facilities are under-represented. It would seem that at present there is sufficient capacity to meet demand there is sufficient capacity although it must appreciated there are times when FDG production fails and also times when facilities are out of action for service and repair.

The Recommendations by ARSAC of the Impact of Recent Shortages of Molybdenum-99 and the Implications for Nuclear Medicine Services in the UK are set out in the Appendix 2.1 [6]. The response by the Department of Health to the ARSAC report is detailed in a 2012 report [7].

## Appendix 2.1: ARSAC Report 2010

In response to the ongoing problem of shortages in the supply chain for $^{99}Mo$ and $^{99m}Tc$, ARSAC undertook a review entitled 'A Review of the Supply of Molybdenum-99, the Impact of Recent Shortages and the Implications for Nuclear Medicine Services in the UK' [6]. The Report contained a number of recommendations as to how to mitigate the impact of periods of shortage of radiopharmaceuticals.

**Recommendation 1**
Nuclear medicine services are an integral part of healthcare and, despite the increased use of PET CT, for the near future, diagnostic nuclear medicine will continue to be based on $^{99m}Tc$-labelled products. Security therefore of $^{99}Mo$ supply is essential. This can be achieved in the long term at international level by investment to upgrade existing reactors and to build new research reactors, to ensure provision of $^{99}Mo$ at a sufficient level to meet current and expanding demand during all foreseeable events. We recommend that the Department of Health and other government departments should continue to support European Commission initiatives to ensure a stable supply of $^{99}Mo$ for Europe. In addition, the Commission and other international agencies should be supported in their efforts to amend European legislation to facilitate the transport of irradiated targets across national borders to ensure timely supply to generator manufacturers.

**Recommendation 2**
While the existing model for $^{99}Mo$ production has proved effective for many years, the recent shortages have demonstrated that over-reliance on a few suppliers can prove problematic. We recommend that new technological solutions be monitored, including the potential long-term solution of UK-based production of $^{99m}Tc$ in medical cyclotrons.

**Recommendation 3**
Currently the three European companies that manufacture $^{99}Mo$ generators do so during the working week rather than during weekends. This is not the case in the USA. We recommend that the Department of Health should explore with generator suppliers the feasibility of weekend production and alternative delivery schedules to ensure that $^{99m}Tc$ availability matches the traditional, efficient service delivery model for healthcare.

**Recommendation 4**
During periods of $^{99}Mo$ shortages many hospitals have adapted traditional working practices, e.g. changing the radiopharmaceutical production model to provide a second elution and production run per day or working at weekends to make more efficient use of the available $^{99}Mo$. We recommend that where practicable hospitals should consider adopting some of these strategies on a permanent basis, and that traditional employment contracts should be amended to introduce flexible working patterns, to include weekends and extended days within the current pay constraints.

**Recommendation 5**
Compared with Wales, Scotland and Northern Ireland, radiopharmaceutical production and supply within England are essentially local and are undertaken on a small scale. This is for historical rather than logistical reasons. While some larger central radiopharmacies exist, these are very much in the minority. The introduction of PET CT services in the UK using $^{18}$F-FDG has demonstrated that for routine diagnostic nuclear medicine services, a production and supply service based on fewer centralized facilities is viable. We recommend that the Department of Health undertake a review of radiopharmacy services to explore whether the number of large central radiopharmacies, which are more resilient during shortages, should be increased to provide the majority of $^{99m}$Tc-based radiopharmaceuticals for safe and patient-focused nuclear medicine services within England.

**Recommendation 6**
Nuclear medicine imaging demonstrates functional information. It relies upon the administration of radiopharmaceuticals that in general are produced outside the UK. It has already been highlighted that $^{99}$Mo production using traditional technology is not practicable at UK level. PET-CT services, which can provide physiological information as an alternative to conventional nuclear medicine in some circumstances, rely upon local production of radiopharmaceuticals in cyclotrons. This is a robust supply chain; however, a large percentage of PET-CT services in England are currently provided through time-limited independent sector contracts. We recommend that adequate PET-CT services continue to be provided across the UK. Optimal coverage of these services should be ensured following the completion of the independent sector provision in England.

**Recommendation 7**
The UK has well-established regulatory frameworks designed to ensure the safe production, use and disposal of radioactive materials and radiopharmaceuticals. As a consequence of recent shortages, the Environment Agency in England and Wales has demonstrated a flexible approach regarding its own regulatory and enforcement policies to enable the most efficient use of available $^{99}$Mo while maintaining overall safety strategies. We recommend that the Department of Health, with other government departments and agencies, address options regarding transfer of radioactivity, radiopharmaceuticals and generators between different hospitals, with a view to issuing clear guidance.

**Recommendation 8**
Technological developments in gamma camera hardware and software design have the potential for nuclear medicine investigations to be performed using less activity. This has benefits for the patient with regard to radiation dose but also for the nuclear medicine service in terms of costs and efficient use of available $^{99}$Mo. We recommend that a multi-centre, multivendor trial is undertaken in order that the viability of resolution recovery software with reduced activity is evaluated. If proved to be successful, we strongly recommend that nuclear medicine services purchase this software for new and existing gamma camera systems where applicable.

# Chapter 3: Alternative Strategies for Imaging Diagnostics in the UK

**John Buscombe and Andrew Scarsbrook**

## Summary

- Most alternatives to $^{99m}Tc$ using radionuclides will be more expensive.
- Some of the alternatives deliver much higher radiation doses.
- Non-radionuclide alternatives may be available but may not be as accurate, may not be usable in all patients and would displace work on other imaging machines such as CT and MR which themselves may be limited.
- Clinical studies would be required to evaluate objectively the relative benefits of alternative tracers for $^{99m}Tc$ as evidence is currently lacking.
- A business case would be needed to estimate the current costs and availability of $^{18}F$ Fluoride and $^{82}Rb$ and how costs will change if they became the de facto standard.
- Patients will be poorly served by not having a cheap, plentiful supply of $^{99m}Tc$.
- For both imaging and therapy there is incomplete knowledge as to the numbers and types of procedures/therapies carried out in the UK. For strategic planning of services it is vital that this information is readily available. Where information is lacking this will have to be determined.
- It will be necessary to investigate the potential of UK network development for the objective evaluation of new radiopharmaceuticals and cost/benefit analysis.

## 3.1 Introduction

Technetium-99m remains the main radionuclide used in nuclear medicine, the advantages being its ready availability via a generator and its gamma emission of 140 keV that present gamma cameras are optimized for. It also has good dosimetric properties in that reasonable counts can be obtained by using administered amounts of activity generally less than 800MBq $^{99m}Tc$. With the exception of myocardial perfusion scintigraphy, which uses two injections (stress and rest) the typical radiation doses of 2-4mSv similar to a single year's exposure to background radiation in the UK. The cost of $^{99m}Tc$ agents is often reasonable especially if many patients are imaged from the same vial of radiopharmaceutical.

The use of $^{99m}$Tc in nuclear medicine has evolved over the past 40 years and forms established diagnostic functional imaging techniques for a variety of medical complaints. $^{99m}$Tc is most commonly used to investigate bone, cardiac, renal and pulmonary conditions. In recent years worldwide shortages in the production of $^{99}$Mo (the parent isotope of $^{99m}$Tc) have intermittently affected the ability to perform these techniques. In November 2010 the Administration of Radioactive Substances Advisory Committee (ARSAC) report reviewed alternate imaging techniques using non-$^{99m}$Tc based nuclear medicine investigations (including PET-CT), CT, MRI and ultrasound [1]. At that point in time it was felt that due to a variety of factors including availability, manpower and additional cost that this would require substantial further investment in personnel and equipment to make it a robust and sustainable option. The eight recommendations of the ARSAC Report are set out in the appendix in Chapter 2 [1].

There are often non-nuclear medicine tests available in these fields such as CT coronary angiography and echocardiography that may provide an alternative to gated myocardial perfusion imaging but may not provide all the information the clinician needs or may not be acceptable or useable in a particular patient. This means that despite the growth of other forms of imaging there remains a substantial practice using $^{99m}$Tc-labeled products.

## 3.2 Alternative Modalities

In some circumstances it is possible to use a non-radionuclide procedure as an alternative, for example vital blue dye can be used for sentinel node localization but the failure rate can be much higher by about 10% compared to 5% with radiocolloids [1]. Emerging data also suggests that micro-bubble contrast-enhanced ultrasound may provide a viable alternative to sentinel node lymphoscintigraphy in some settings [2].

Likewise it is readily accepted that MRI is a good method by which bone metastases can be detected and may be more sensitive than bone scintigraphy as it can identify the metastases directly whilst bone scintigraphy depends on a peri-metastatic sclerotic reaction. However, for many cancer types including the most common with bone metastases such as prostate and breast carcinoma it is important to cover the areas that metastases occur most commonly which is from skull to knees.

To achieve this with an MR scanner this may take some considerable time - up to 45 minutes per investigation. Capacity for additional MR scanning may be limited due to high demand and if MR was to replace the 10 bone scintigrams commonly performed daily in many nuclear medicine departments it would take a substantial portion of the MR capacity for that day. While this could be used on some occasions, if it was to become a permanent arrangement an additional MR scanner would be needed in most centers if other cases were not to be delayed. Also there is no guarantee the clinician who reads the bone scintigraphy is also trained in bone MRI so additional reporting manpower may be needed.

## 3.3 Use of Other Radionuclides

It may be possible to replace $^{99m}$Tc with other radionuclides (Table 3.1); for example either $^{123}$I or $^{131}$I imaging could replace $^{99m}$Tc thyroid imaging. However, there are disadvantages with both of these radioisotopes. $^{131}$I is fairly cheap but has poor dosimetry. $^{123}$I has better dosimetry but is more expensive. Also the world supply of $^{123}$I may be limited with most production based in only two centres Eindhoven and Vancouver.

### *3.3.1 $^{18}$F-Sodium Fluoride ($^{18}$F-NaF).*

$^{18}$F-NaF provides an excellent alternative to bone scintigraphy [3] but has three disadvantages. As modern PET requires a co-acquired CT for purposes of attenuation correction the radiation dose of $^{18}$F-NaF is greater than bone scintigraphy. The cost of a study is presently almost three times greater than for bone scintigraphy. The third disadvantage is that it takes around 30 minutes to acquire the whole-body images required. Therefore to accommodate the required number of $^{18}$F-NaF bone scans the other work the PET scanner is doing would need to be done on a different machine. Free capacity on the available PET-CT scanners is likely to be limited in many parts of the UK.

### *3.3.2 Rubidium-82 ($^{82}$Rb) for Myocardial Perfusion Assessment*

A recent meta-analysis of $^{82}$Rb PET and SPECT myocardial perfusion imaging in the evaluation of obstructive coronary disease [4] revealed that $^{82}$Rb PET was more accurate (0.95 and 0.90 area under the receiver operating curve) than SPECT myocardial perfusion imaging. PET offers advantages over SPECT, including increased patient throughput because of rapid scanning protocols, reduced radiation exposure to patients and the ability to quantify tracer distribution accurately and myocardial perfusion reserve. Although $^{82}$Rb PET has advantages over SPECT, there are no large-scale prognostic or cost-effectiveness data to support its use as the primary MPS technique. Additionally, the availability of $^{82}$Rb PET is limited and at present restricted to only a few centres in the UK (principally at University College Hospital, London and at Manchester Royal Infirmary). A wider use of absolute measurements of perfusion has the potential to improve diagnostic accuracy and to add prognostic value over relative assessment of myocardial perfusion.

### *3.3.3 Other PET Radiotracers for Imaging*

There are additional PET radiotracers that at present have limited availability and depend on access to local Cyclotron facilities. Certain of these tracers are in increasing clinical use while others are restricted and only available at a limited number of centres particularly those with research facilities. In summary:

- A large number of potential agents using $^{18}$F are emerging [5], including $^{18}$F-fluorothymidine ($^{18}$FLT), $^{18}$F-fluoromisonidazole ($^{18}$F-MISO), $^{18}$F-fluoro-ethyl-tyrosine ($^{18}$F-FET), $^{18}$F-fluorocholine ($^{18}$F-Choline), 3,4-dihydroxy-6-$^{18}$F-fluoro-l-phenylalanine ($^{18}$F-DOPA) and $^{18}$F-fluoroazomycin arabinoside ($^{18}$F-FAZA). A limitation of $^{18}$F products, common to most PET radioisotopes is that it only has a 110-minute half-life and so cannot be sourced from outside the UK.

- There is a potential for the introduction of more $^{68}Ga$ generators allowing on-site access to patients under investigation for neuro-endocrine malignancy [6].
- Other potential PET isotopes will become of increasing relevance, including $^{124}I$ (for thyroid and mIBG biokinetics), $^{89}Zr$ (EGF, PSMA etc.), $^{64}Cu$ (Hypoxia), $^{86}Y$ (proven to predict therapy response from $^{90}Y$ therapies and potentially useful to identify unnecessary therapy procedures).

It should be noted that the UK lacks a large cyclotron such as the Arronax facility at Nantes in France and that could have implications for the development of novel Cyclotron-produced medical radioisotopes.

**Table 3.1** Alternative tests for imaging studies

| **Nuclear Medicine Study** | **Cost £** | **Radiation mSv** | **Alternative Test** | **Cost £** | **Radiation mSv** |
|---|---|---|---|---|---|
| $^{99m}$Tc Bone Scan | 280 | 3 | Na$^{18}$F<br>MRI [7] | 800<br>150 | 9<br>0 |
| $^{99m}$Tc Renography | 150 | 1 | MRI [1] | N/A | |
| $^{99m}$Tc-DMSA (Renal Scan) | 150 | 2 | | 300 | |
| $^{99m}$Tc Thyroid | 100 | 1 | $^{123}$I/$^{131}$I | 300/80 | 4/8 |
| $^{99m}$Tc HIDA (Biliary Scan) | 150 | 3 | MRI with Primovist [8] | 300 | 0 |
| $^{99m}$Tc Myocardial Perfusion Scan | 500 | 8 | $^{82}$Rb PET-CT[9]<br>Cardiac Stress Echo [10]<br>Cardiac MRI [10,11) | 500<br>400<br>900 | 2<br>0<br>0 |
| $^{99m}$Tc White Cell Scan | 400 | 4 | $^{18}$F-FDG<br>$^{68}$Ga Citrate [12] | 900<br>300 | 4<br>9 |
| $^{99m}$Tc MAA Lung Perfusion Scan | 150 | 2 | CTPA[13]<br>MRA<br>$^{68}$Ga-MAA [14] | 90<br>300<br>700 | 4<br>8<br>? |
| $^{99m}$Tc Parathyroid | 250 | 4 | $^{18}$F -DOPA<br>Ultrasound [15]<br>CT [16]<br>$^{11}$C-Methionine [17] | 4000<br>40<br>80<br>1500 | ?<br>0<br>?<br>2 |
| $^{99m}$Tc Gut Transit | 150 | 0.5 | MRI capsules [18]. | N/A | 0 |
| $^{99m}$Tc Sentinel node localisation | 150 | 0.1 | Fluorescent dyes<br>Microbubble Ultrasound | N/A | N/A<br>N/A |

# Chapter 4: Non-Reactor Production of Technetium-99m

**Hywel Owen and Rob Clarke**

## Summary

This chapter summarises the technology options available for the non-reactor production of technetium-99m either directly or via its molybdenum-99 precursor. Based on an assessment of the relative maturity of the different options and the possible co-use for purposes such as manufacture of other radioisotopes, it is concluded that the most promising technology for provision of $^{99m}$Tc in the UK is its direct production using proton cyclotron bombardment at moderate energies between 18 and 24 MeV.

## 4.1 Reactor Fission vs. Other Methods

Whilst a variety of nuclear reactions are possible for the production of either $^{99m}$Tc, or more usually its precursor $^{99}$Mo, reactor-based fission of highly enriched uranium (HEU) targets remains the principle route by which $^{99m}$Tc is obtained around the world today [1-7]. Nuclear reactors of moderate to high power (i.e. >1 MWth) provide a copious density of thermal neutrons; these initiate fissions in suitable HEU targets that are placed in the reactor, as they do in the rest of the $^{235}$U that makes up the fuel of the reactor itself; the withdrawal and rapid chemical processing (within days of irradiation) of the HEU targets means that the fission products with mass A=99 are predominantly molybdenum – i.e. $^{99}$Mo - and this chemical extraction allows the separation of the original target material (HEU) from the molybdenum product. Other (stable) molybdenum isotopes are also present in the separated product, but the end-of-processing (EOP) specific activity is typically very high, around 3000 Ci/gm.

The intrinsically-high specific activity, and the historical fact that these high bombarding fluxes were first obtained with reactors, has led to there being a mature development of the reactor fission approach at the expense of leaving alternative (non-reactor) approaches relatively under-developed. That said, there are intrinsic advantages associated with reactor fission $^{99}$Mo production compared to other methods; the advantages and disadvantages in general terms may be summarised as follows:

*Reactor Fission $^{99}$Mo Advantages:*
Large flux of neutrons typically available
Resulting fission $^{99}$Mo available in large quantities; single reactor provides large capacity

*Reactor Fission $^{99}$Mo Disadvantages:*
Larger capital cost cf. accelerator installations
Licensing more difficult than most accelerator-based methods
Significant waste stream compared to other methods
Longer lead time for construction compared to other methods

An important issue when choosing methods is whether the radioisotope produced is $^{99}$Mo – in which case a generator is utilised from which $^{99m}$Tc is later eluted – or whether $^{99m}$Tc is produced directly; whilst nuclear reactors have generally been used to produce $^{99}$Mo, accelerator-based methods may be used for either $^{99}$Mo or $^{99m}$Tc. Direct production of $^{99m}$Tc is chosen if the bombardment route produces $^{99}$Mo with a low specific activity, which is typical of transmutation reactions where the precursor is another isotope of molybdenum. The consequence of choosing either $^{99}$Mo or direct $^{99m}$Tc production is that the latter method restricts the distance between production and usage due to the intrinsic difference in half-life between $^{99m}$Tc (6 hours) and $^{99}$Mo precursor (66 hours). The choice between these approaches thereby has a knock-on effect on the supply chain and infrastructure choices that are possible for the supply of a country or region. Loosely speaking, $^{99}$Mo may be produced in a different country from where it is used, whilst $^{99m}$Tc most likely must be used within a few hours' transport time from its site of production.

## 4.2 Particles and Reactions

Any chosen isotope may be manufactured by the bombardment of a suitable nuclide by an incident set of particles of appropriate energy; in this regard there is no particular difference between a nuclear reactor, a particle accelerator (whether conventional or laser-based) or any other method, other than what technology is able to produce the particles required. As such, a variety of reactions are possible both in reactors and in particle accelerators that may transmute a (typically) stable parent nuclide into either $^{99}$Mo or into $^{99m}$Tc. These reactions broadly fall into one of two groups: fission-based methods, and transmutation-based methods; the most prominent reactions are discussed below, followed by a discussion of how the accelerators providing for the bombardment may be provided.

An important distinction between fission and transmutation routes is that whilst fission-based approaches will naturally produce a large specific activity product they must inevitably involve the use of uranium-containing targets. Neutron-based fission approaches in practice require the use of $^{235}$U as the target nuclei due to the much lower fission cross-section of $^{238}$U, and so highly enriched uranium (HEU) targets may be required. However, HEU targets (defined as >20% enrichment, but usually much higher) minimise the waste stream volume for a given amount of molybdenum product, but they raise proliferation concerns and must therefore be managed carefully; the long-term aim of the USA RERTR programme [8], augmented by the American Medical Radioisotopes Production Act 2012, is to eliminate the civilian use of HEU either in reactor fuel assemblies or in targets, and to replace them with the use of LEU targets with enrichments of 20% or lower [6,7,9]. Whilst the proliferation concerns of LEU targets are reduced, the intrinsic presence of $^{238}$U atoms in the target plates gives rise both to an increased waste stream (around four times as much as when HEU targets are used, for a given amount of molybdenum product), and also gives rise to the generation of plutonium. Whilst these two issues are tractable, they are less suitable compared to other methods to be adopted for local production, for example at a hospital or industrial setting such as is done with cyclotron-based $^{18}$F production. These processing and waste issues are of relevance to any non-reactor methods that would utilise uranium targets, the most notable being photofission. Overall, whilst photofission may utilize LEU or depleted uranium targets, the waste stream and processing issues may give rise to licensing issues and processing complications similar to using a nuclear reactor, hence reducing the capital advantage of using an accelerator.

### *4.2.1 Neutron-Induced Fission*

Neutron-based fission of $^{235}$U atoms has already been described above, wherein reactor neutrons which typically have thermal energies (~1/40 eV) are able to induce fission in $^{235}$U atoms but not in the $^{238}$U atoms which have higher natural abundance in uranium ore; this gives rise to the need for enriched targets. In principle $^{233}$U- or $^{239}$Pu-containing targets could also be used for $^{99}$Mo production, but these are impractical due to their (non)availability, their having similar or greater proliferation concerns cf. $^{235}$U, and far more difficult handling and processing. For example, it has been proposed to utilize a $^{233}$U-based molten-salt reactor (i.e. with liquid rather than solid fuel) and to conduct online fission-product extraction to obtain $^{99}$Mo [10,11]. However, this technology is immature compared to the established irradiation of solid HEU targets in light-water-cooled research reactors.

Whilst fast neutrons (~1 MeV) may be used to induce fission in $^{238}$U (or even in transuranic nuclides), this is impractical in reactor systems due to the much greater complexity involved in the required fast reactor technology. However, fast neutrons are available in reasonably large fluxes using either low-energy nuclear reactions (for example c. mA currents of protons or deuterons incident on Li, Be, or C targets, as used for boron neutron capture therapy [12] or for neutron irradiation of materials for fusion reactor studies [13,14,15]), or high-energy spallation-based reactions such as protons on Pb targets [16,17]. There is a trade-off: neutron production using low-energy protons gives low yield but a high current is possible; spallation gives more neutrons per proton but requires a larger accelerator that can obtain a relatively lower current. Lower-energy reactions used for neutron production vary from the 2.8 MeV threshold energy for the Li(p,n) reaction on solid or liquid targets [12,16,18], through to the c.40 MeV deuteron energies proposed for hard neutron production on the proposed IMFIF and FAFNIR irradiation facilities [13,14]. Spallation reactions such as p on Pb benefit from the much higher neutron multiplicity available at proton bombardment energies above around 600 MeV, and a broad optimum for accelerator cost and current exists for proton energies around 1000 MeV, at which energy around 20-30 neutrons are produced per proton dependent on the target material and geometry; solid Pb, W and U and liquid Hg targets have all been utilised [13,14,15,16,17].

Both the low-energy and spallation methods have been proposed as methods for $^{99}$Mo production. Three examples are notable. The first is the proposed use of low-energy D-T neutron generators surrounded by either a depleted uranium ($^{238}$U) assembly or a subcritical assembly or subcritical uranium-containing solution. The c.100 kV D-T 'neutron generator' is now a widespread source that may be bought in a variety of geometrical configurations from a number of suppliers: whilst originally developed as neutron triggers for nuclear weapons, these generators are now most often used for pulsed neutron generation whose use for example is in oil well diagnostic imaging [19]. The accelerated species is ionised deuterium gas, whilst the tritium is present as a sputtered scandium hydride target on a copper target backing. Apart from the naturally low capture yield in the subcritical assembly (unless the neutron reproduction k value is brought close to k=1, leading to a system rather similar to a proper reactor from the point of view of licensing), two deficiencies of D-T approach are the low typical lifetime for the neutron generators, and the care required to avoid tritium leakage. The second approach is the proposal to use spallation neutrons from either a high-power neutron source - such as ISIS (Oxfordshire) [20] or the European Spallation Source (to be constructed in Sweden) [21] - or from a high-power linac for particle physics such as the CERN Linac 4 [22] or the Fermilab Project X proposal [23], where a similar spallation target would also be used. The final scheme is to utilize a medium-energy proton cyclotron (100 MeV to 350 MeV), such that each proton (impinging on a robust high-Z target such as tungsten or tantalum) produces a few neutrons per proton which then pass into a subcritical assembly; the subcritical assembly will contain $^{235}$U either in solid or solution form, the latter most often proposed as an aqueous solution. For example, the IBA ADONIS scheme proposes a 150 MeV, 2 mA proton cyclotron and a subcritical (k<0.9) assembly incorporating HEU targets and a combination of Be and heavy water moderator material to tailor the spallated neutron energy to values optimal for fission [24].

A different way to generate neutrons for fission is to utilize a two-stage combination wherein (e.g.) Bremsstrahlung in a tungsten target is used to generate c.15 MeV gamma rays from a high-current (c.20 to 100 mA) beam of electrons of energy around 50 MeV [25,26]. The Bremsstrahlung gamma rays can then be converted in the same target - or in another one - to produce photoneutrons using a (γ,n) photonuclear reaction [27]. Those outgoing neutrons then impinge onto a uranium target or subcritical assembly containing uranium. The low photonuclear cross-section is offset in part by the fact that it is much easier to generate very high currents of electrons than it is to generate high currents of protons; several technologies - including superconducting linacs [28,29] and rhodotrons [30,31] – are available that can in principle allow very large electron currents, and the initial gamma-producing target is typically where the power limitations will occur in practical designs.

A closely-related method to the photonuclear fission scheme is to utilize the Bremsstrahlung photons directly to initiate photofission in a uranium target, where the photofission cross section also peaks at gamma-ray energies around 15 MeV [32]; however, unlike neutron-initiated fission, photofission may be produced in both $^{235}U$ and in $^{238}U$, although the $^{238}U$ photofission cross section at such neutron energies (170 mbarn) is rather lower than for $^{235}U$ (400 mbarn) [27]. Both cross sections are much lower than the thermal neutron fission cross-section (600 barn), and so again much greater fluences of photons are required than neutrons for the same fission rate; but again, the lower cross-section is compensated for by the fact that the intermediate photonuclear conversion stage is not needed. This latter scheme has been proposed and developed by SHINE Medical Technologies in partnership with LANL, and uses a 100 mA deuteron current at 300 keV to generate around $10^{14}$ neutrons per second from a gas target, the neutrons are moderated with a Be multiplier before impinging on an aqueous solution of either LEU uranium nitrate of sulphate; estimated production is up to 500 6-day Ci per week of $^{99}Mo$ as well as other isotopes [32]. IBA have proposed a separate scheme using their rhodotron electron accelerator technology, which proposes a subcritical homogenous liquid target of LEU salts in heavy water and which utilises a HDPE reflector [33,34]; limiting the electron beam power to 450 kW – which is already very high - implies that a k-value very near to 1 (i.e. k=0.99) is needed, which has implications with regard to nuclear licensing. If a lower production rate is acceptable then a depleted $^{238}U$ target may be utilised directly (and in which $k<<1$); this scheme is similar to that utilised in several radioactive ion beam (RIB) nuclear physics facilities, but differs in whether the fission products escape from the target. The TRIUMF photofission proposal was originally recommended by a Canadian review of non-accelerator production as the best route for $^{99}Mo$ production [28], and which was considered as a beam current upgrade (to 100 mA) of the proposed 10 mA ARIEL RIB facility.

It may be seen that a number of neutron/fission schemes are possible. However, they are limited by one of three factors depending on the particular scheme:

- The need for very high proton currents at low energies, and the power limitation on the low-energy target;
- The need for moderate currents and a large capital cost/size for a spallation-based approach, which restricts this sort of scheme to regional-based production;
- The need for both very high electron currents and subcritical assemblies in the photonuclear/photofission approaches, the latter of which makes nuclear licensing more difficult.

The principal advantage of all the non-reactor fission-based methods is that the resulting product will be sufficiently similar that it may be substituted for reactor fission $^{99}Mo$.

### *4.2.2 Neutron Capture*

Neutrons have also been used to generate $^{99}Mo$ using neutron capture in $^{98}Mo$ targets, which targets are typically enriched to prevent the co-production of other molybdenum isotopes [4,6,7,28,35]. Reactor-based methods such as implemented recently at MURR are limited both by the available flux and that in most reactors the neutron spectrum is rather thermalised, which limits the resultant saturation specific activity to several Ci/gm; nevertheless, generator technologies such as that used by NorthStar are able to produce suitable $^{99m}Tc$ product by this route, and potentially from other transmutation reactions giving low specific activity. The harder spectrum available in some reactors may be used to achieve an enhancement of the capture rate through capture of neutrons in the epithermal resonances of the $^{98}Mo$ nucleus, and this has been proposed by several groups, notably in patents by Carlo Rubbia that outline the use of a liquid target [36,37,38]. However, the broad, mostly thermalised, spectrum means that the saturation activity is at best only increased a few times [39].

The spectrum of neutrons available from an accelerator target - whether spallation or otherwise - is typically much harder than that available from a reactor, and with appropriate shaping (using a tailored moderator) can be used to enhance the rate of neutron capture in a nuclide's resonance region. This resonance neutron capture technique was originally proposed as a method of enhancing the rate of neutron capture in $^{99}Tc$ and other long-lived fission product nuclear waste, by a number of researchers including Carlo Rubbia [40,41]. A number of accelerator and target (both solid and liquid) configurations have been proposed for the production not only of $^{99}Mo$, and several have been implemented [42,43]; for example, the company Advanced Accelerator Applications s.a. (AAA) in France currently utilize a 30 MeV proton linac and Be neutron-producing target coupled to a lead moderator, which is used for the generation of niche isotopes such as $^{154}Ho$ using resonant capture [37]. However, it is inherently difficult to shape a neutron spectrum using scattering-based moderators, so that the practical enhancement in capture rate even in accelerator-based methods is probably limited to less than a factor of 100.

As well as the developed AAA facility in France, several other proposals are of note. CERN have proposed the utilisation of a high-power proton spallation target located on Linac 4 to carry out neutron capture for $^{99}Mo$ production [44]. Whilst the possible capacity of such a production facility would be quite large, the large capital, infrastructure and expertise requirements to construct and operate such a facility means that it could only be done nationally (and perhaps only internationally) and not locally. That said, such a facility could in principle be located either at a national accelerator facility such as one of the STFC sites (Rutherford Appleton or Daresbury), or at a nuclear-based site. Another option would be to locate such a target at a spallation source such as ISIS or ESS, similar to the possibilities for a fission-based approach; some discussions have been had in the past about the use of both facilities for isotope production, but at present there are no firm plans.

A locally-based facility would require a compact, high-current proton source: one model for this is use of compact neutron sources developed for boron-neutron capture therapy (BNCT), such as the 2.8 MeV DC proton accelerator used at Birmingham to generate neutrons from a solid Li target [18]. Similar to other neutron facilities utilising low-energy accelerators, Li, Be or C targets may be used; beam powers greater than around 3 kW require the use of flowing liquid Li rather than solid Li targets [12]. Designs have been developed for applications such as BNCT and materials irradiation; one such example is the proposed 40 MeV FAFNIR deuteron linac to be used for materials studies for the ITER fusion project [13,14]. Co-location of a neutron capture target on such an irradiation facility, or duplication of the linac and target design on a separate facility, are both in principle possible.

Neutron capture methods have the advantage that there is no use of uranium targets or fission product processing. The disadvantages are the inherently low specific activity; there is no practical method for separating $^{99}Mo$ from the original $^{98}Mo$ target material, and so either a low-activity generator or direct dissolution of the target would be needed.

#### 4.2.3 Photonuclear Transmutation

As well as photofission, gamma-ray production from a high-intensity electron accelerator may be used to conduct photonuclear transmutation of a molybdenum target; in this case, the reaction of interest is $^{100}Mo(\gamma,n)^{99}Mo$ [6,7,35]. Similar to neutron capture, either solid or liquid Mo-containing targets may be used, but the conventional geometry utilises a set of thin Bremsstrahlung converter targets (typically tantalum or tungsten) followed by one or more solid (enriched) $^{100}Mo$ targets. The energy required from the high-current (10 mA or greater) electron beam must be enough to deliver a significant number of gamma rays around the c.15 MeV giant dipole resonance energy, which for practical purposes means an electron energy of 35 MeV or more. Hence the source and target technology is essentially the same as that for photofission production.

Whilst electron bombardment has an intrinsically lower cross-section, this is countered by the fact that electron accelerators of high current capability are generally simpler, more reliable, and able to deliver higher currents than proton accelerators for similar purposes. A side-benefit is that photons are also more penetrating, allowing for the use of windows and thicker targets. Provided a reaction channel exists for the desired isotope, the generally fewer reaction channels available result in an inherently reduced production of undesired isotopes and therefore waste compared to fission.

Several proof-of-concept photonuclear production facilities have been constructed. One such is the Canadian NRC-INMS prototype at Ottawa, which has used a 35 MeV, 2 kW electron linac to irradiate 2 cm diameter enriched $^{100}Mo$ targets using a water-cooled tungsten Bremsstrahlung converter target; automated direct extraction of technetium from the dissoluted irradiated targets ($Na_2MoO_4$ solution) has been demonstrated in a partnership with NorthStar technologies. The Canadian Light Source have since constructed a 20 MeV, 20 kW industrial electron linac to carry out the same irradiation using a stacked target with a likely 48-hour bombardment time 45]. Dissolved targets are then shipped as $Na_2MoO_4$ solution from which $^{99m}Tc$ is extracted in an automated separator such as that form NorthStar. Production rates are said to be able to scale to around 2 TBq/day, and the PIPE/NorthStar consortium is proposing a network of up to 16 linacs to supply 50% of US requirements [46].

#### *4.2.4 Proton-Induced Transmutation*

Transmutation caused by proton bombardment of a metal target was one of the earliest methods of using an accelerator to create radioactive species; indeed, Emilio Segre first identified the artificial element technetium in 1937 following the parasitic deuteron bombardment of a molybdenum deflector plate in the Lawrence 37-inch cyclotron. Subsequent to this a considerable number of measurements were made following Beaver and Hupf's 1971 proposal to manufacture significant quantities of $^{99m}Tc$ using proton cyclotrons [47,48]. Latterly, Tom Ruth's group at TRIUMF revived this idea [49] which was then developed by both that lab and at the University of Alberta. Work by Katie Gagnon and others have shown both theoretically and experimentally that an enriched $^{100}Mo$ target is required to avoid unwanted side reactions (see later chapters of this report) [50,51,52,53,54,55,56,57,58,59,60]. In this regard, proton irradiation of an enriched $^{100}Mo$ solid target is much the same as for photonuclear irradiation, except that the target plate used for proton irradiation is much thinner (see below). The processing of these targets will be rather similar to each other and to the processing of $^{98}Mo$ targets used for the neutron activation method.

The TRIUMF and University of Alberta approaches are similar, but differ in certain respects. In both cases it is recognized that an incident proton energy of at least 19 MeV is required to obtain a good trade-off between radionuclidic purity and yield. For example, at TRIUMF a $^{100}Mo$ target thickness of around 300 μm (on a tungsten backing plate) is used to tailor the exit proton energy to remain above 10 MeV. 19 MeV has been adopted in experimental bombardments to understand how existing 19 MeV PET-producing cyclotrons might be adapted for technetium production. To that end, test targets have been produced both for GE PET-TRACE cyclotrons (e.g. the 880) and ACSI TR-19 cyclotrons. In either case, thermal power is managed by tilting the angle of incidence of the beam on target; this is more easily done in the TR-19 target arrangement, which has subsequently been adapted by University of Alberta for use on TR-24 cyclotrons, and where they have adopted an aluminium backing plate. A higher-energy cyclotron will produce greater yield of product and also some commercial cyclotrons are available at these energies with currents approaching or even exceeding 1 mA. At higher currents dual extraction ports are likely to be typical, in which a carbon foil is adjusted at each port to determine the current extracted to each target. One model might be to run extraction ports alternately in different shielded areas, so that activation issues can be managed by moving irradiations to the other port if required. Some cyclotron designs allow then for dual operation (via extraction line switching dipoles) so that more than one target station (and therefore type) may receive protons from an extraction port. This would allow for bombardment to manufacture – for example - $^{18}F$ on one target, and then $^{99m}Tc$ on another target.

Much progress has been made in the last several years in developing suitable targetry and chemistry for proton bombardment [55,60,61]. Enriched $^{100}Mo$ feedstock material is available at around USD 500 per gram at present, and it is thought this cost may decrease if it is more widely used. Historical sources derived from stable isotope centrifuge enrichment have mostly been used to date – principally of Russian origin - but these are now being augmented by freshly enriched $^{100}Mo$ where it is possible tailor the relative composition of the other stable isotopes. This will assist in reducing the co-production of unwanted Tc isotopes as detailed later in this report, and thereby will assist in controlling the extra patient dose from these isotopes.

Target plates developed at the two Canadian sites mentioned typically utilise around 80 mg or so of $^{100}Mo$ which is then sintered and pressed into the backing plate. Initially it was thought that efficient recycling of the feedstock material was needed due to the high cost, and hence attention was given to achieving good recycling during $^{99m}Tc$ extraction; rates greater than 90% have now been demonstrated at TRIUMF. However, there remain issues about ensuring isotopic consistency of target batches if recycled material is used and how this might affect both actual patient doses and how such targets might be licensed. One model might be to have a national or even international single site where target recycling may be done such that isotopic consistency can be assured. However, if $^{100}Mo$ feedstock price falls significantly then it may not even be necessary to recycle it.

Experimental irradiation work on cyclotron-produced technetium to date is promising. Experimental yields now approach prior theoretical estimates, for example that a 200 uA, 19 MeV cyclotron can produce around 0.4 TBq of $^{99m}Tc$ in a single 6-hour bombardment; discussion remains about whether a single 6-hour or two 3-hour bombardments are best. Also, some testing will be needed of target irradiations at high cyclotron currents to examine the durability, but it is likely that currents exceeding 500 uA will be feasible.

Scaling the present production estimates, it is estimated that two 19 MeV, c.300-500 uA cyclotrons could produce around 1 TBq of $^{99m}Tc$ per day in a single bombardment each; this would be sufficient for the UK population demand of around 600,000 doses per year, but does not account for any downtime of the cyclotrons themselves [35].

### 4.2.5 Summary

We assess that cyclotron-produced technetium is at present the most mature technology available relevant for UK needs, based on several considerations. Firstly, the accelerator and targetry are effectively developed, and production is now being achieved at scales relevant to volume production. Secondly, understanding of how to control radionuclidic purity is better demonstrated with this method than with others. Thirdly, the possible co-use of such facilities to produce other isotopes such as $^{18}F$ is attractive when considering overall capital cost. Finally, such cyclotrons and associated processing lend themselves more easily to licensing and local hospital use. We next consider the various accelerator technologies available.

## 4.3 Particle Source Technologies

### *4.3.1 Electron Accelerators*

Electron linacs of up to several tens of MeV and several tens of kW of output power are available industrially, and are already used for a variety of irradiation applications such as cable sheath processing and sterilization. Most linac designs in industrial use are based on conventional 3 GHz radio-frequency structures, and the power and current limitation arises principally from limitations in structure heating and water cooling. Very high power electron linac designs over 10 mA, particularly for higher-energy applications above 50 MeV output, have looked to superconducting accelerating structures to limit the resistive heating in the cavity walls. Significant progress has been made in this area over the last twenty years, with both US (1.5 GHz) and European (1.3 GHz, so-called 'TESLA' type) designs capable in principle of delivering 100 mA of continuous beam current. For example, TRIUMF have previously proposed an upgrade to their Ariel photofission linac to obtain currents of 100 mA at 50 MeV for photofission production [28].

An alternative to linear accelerators is to use the commercial-only Rhodotron method, in which electrons are accelerated across the diameter of a cylindrical accelerating cavity [30,31]. After each crossing a magnet 'petal' returns the electron beam back to the same common cavity and with the correct phase for further acceleration, a process limited by the magnetic fields available in the petals; electron beam powers in excess of 200 kW are now available, with the final electron energy from a single Rhodotron being around 10 MeV. Whilst the number of crossings per Rhodotron is limited by geometry and power availability, it has been proposed to have a sequence of such accelerators where the output from one feeds to the next. Both IBA and AMIC have proposed such a scheme in which Rhodotrons supply electrons to a subcritical assembly for fission molybdenum production, but in principle such accelerators could also be used for photonuclear production.

### *4.3.2 Low-Energy Proton Accelerators*

Low-energy protons may be characterized as those which may be supplied by commercial accelerators. The vast majority of these accelerator sources are conventional proton or deuteron cyclotrons (i.e. that involve the use of a single magnet with static dipole field) operating with energies up to around 30/15 MeV (p/d) [62,63,64]. The primary market for these cyclotrons is for the production of short-lived isotopes such as $^{18}$F for PET, but they may also be used for other isotope production by changing the target assembly. Typical proton currents obtained over the 15-30 MeV (relevant for $^{99m}$Tc production) range from 30 to 1500 uA, and at higher currents power handling becomes an issue both in the beam extraction (if used) and in the isotope production target. A number of manufacturers offer cyclotrons in this energy and current range, nearly all of which utilize normal-conducting electromagnetic dipoles. A typical example of a compact cyclotron is the IBA Cyclone 18/9 (p/d), which utilises a 1.35 T magnet and which weighs 25 tons (c.2 m diameter and 2.2 m height), delivering around 150 uA of proton current to up to 8 target ports. Very high current cyclotrons will typically accelerator negative hydrogen ions so that beam extraction can be efficiently obtained to enable the highest currents. The primary current limit is in the target extraction foil which is typically arranged as a carousel wherein one of a number of carbon foils are presented into the beam to strip the electrons from the circulating $H^-$ ions and thereby extract them. One such design is the ACSI TR-24 used at the University of Alberta for $^{99m}$Tc production where two extraction ports are provided, and in which the relative currents extracted are determined by how far each foil is presented into the circulating beam. Stated lifetimes of foils are said to be several weeks. The very highest currents are achieved using extraction of $H^-$ ions; hence such cyclotrons cannot be of the 'self-shielded' variety and must utilise a conventional shielded bunker within which the cyclotron and targets are situated.

In the last few years attention has been given to the design of more compact cyclotrons which operate at lower energy. The reasoning is that the increased current and decreased size available at lower extraction energy offset the decreased isotope yield of for example $^{18}$F, with the argument that the smaller size of cyclotron enables more compact hospital-based radiopharmacies. As discussed above, the requirement for proton energies up to 24 MeV means that compact cyclotrons are less straightforward for direct $^{99m}$Tc production.

A method of reducing the cyclotron size is to utilize superconducting magnet technology, and several groups have developed this. The AMIT project, which includes the Spanish CIEMAT institute, is developing a 4 T, 8 MeV self-shielded proton cyclotron for $^{18}$F and $^{11}$C production using an internal target [65]. At 2 tons the weight is much reduced compared to higher-energy cyclotrons, and a little lower than comparable low-energy cyclotrons that utilize normal-conducting magnets such as designs from ABT Molecular (3 tons) and OSCAR (3.5 tons). Ionetix have recently demonstrated a cryo-cooled 12.5 MeV proton cyclotron using 6 T NbTi superconductor to allow the entire cyclotron to fit within a cryostat, and which operates using dewar LHe and 35 kW of wallplug power to manufacture $^{13}$N [66,67].

Although the use of either superconducting technology or lower energy may reduce the size and weight of the cyclotron magnets, it does not eliminate their use entirely. One approach to eliminate the need for magnets is to utilize DC acceleration, i.e. with a Van de Graaf or equivalent DC accelerator. Historically these have needed to be rather large (many meters in length) to hold off any significant voltage (1 MV or greater), and so have been restricted to being used for heavy ion acceleration. Tandem acceleration may be used to double the effective voltage, in which a negative ion accelerates across a DC voltage, is stripped of electrons by passage through a foil, and then re-accelerated through the same potential difference. Tandem pelletrons are now available commercially that can accelerate a variety of ion species to reasonable energies; for example, the now-operating Dalton Cumbrian Facility will eventually obtain up to 100 uA of protons up to 10 MeV [68,69]. The practical limit for voltage has been around 10 MV (20 MV tandem), for example at the now-decommissioned

Daresbury Nuclear Structure Facility. Recent innovations in semiconductor technology have allowed the development of a novel geometry called the Oniac [70]. Siemens and the STFC Rutherford Appleton Laboratory (RAL) have collaborated to demonstrate this accelerator, in which the DC stack is in the form of a set of concentric shells with the high voltage and stripper foil at the centre; no magnetic fields are thought to be required, reducing the weight and cost of the accelerator, but with a similar limit on available voltage. The overall size of a 10 MV accelerator is estimated to be around 1 m in diameter, and tests are proceeding at RAL at which so far a few shells have been tested up to perhaps 1 MeV.

Another approach to obtain medium-energy protons is to use conventional proton linacs, and whilst a number of configurations are possible a notable example is the Front-End Test Stand (FETS) at RAL [71]. Although designed as a front-end for a future high-power (> 5 MW) spallation source, the FETS injector has so far demonstrated 6 mA of average ($H^-$) current at 65 keV; at present this is delivered with a pulsed time structure (around 10% duty factor) that may not be compatible with isotope target irradiation, but this may perhaps be changed. FETS is due to complete the addition of a 324 MHz radio-frequency quadrupole (RFQ) to deliver 3 MeV protons [72]. A later addition of linac structures could provide a very intense beam of $H^-$ ions that could in principle be utilised for isotope production. However, it must be pointed out that whilst the proposed current from FETS is somewhat higher than from present cyclotrons, it has not yet been demonstrated nor a cost comparison been carried out. Given the complexity of the design it is not clear if either the capital cost or operation of a linac would be favourable, and indeed the cyclotron was originally invented as a method of alleviating the complexities of linacs when obtaining moderate-energy protons.

A final approach to obtaining moderate-energy protons is to adapt the design of compact cyclotrons to obtain a low-energy FFAG (fixed-field, alternating-gradient) accelerator. These may be thought of as cyclotrons with stronger focusing, and one such design led by University of Huddersfield proposes four separate magnets and an internal target for $^{99m}$Tc production [73]. At present this is a theoretical study, and a comparison of advantages and disadvantages compared to operating cyclotrons has not yet been presented in the literature.

### *4.3.3 High-Energy Protons*

High-energy proton production has mostly aimed at provision of either rather low currents at very high energies (many GeV) for uses in particle colliders for particle physics, or at lower energies (around 1 GeV) at higher currents for use in applications such as spallation neutron production for scientific use, neutron generation for accelerator-driven subcritical reactors, or a variety of neutrino- or muon-based experiments that require a high intensity beam of protons as a generating source. At the lower c.1 GeV energies the present limitation in current is the continuous beam power of c.1 MW that is delivered by the Paul Scherrer Institute 590 MeV cyclotron, which delivers an effectively continuous beam current up to 2.2 mA [74,75]. This current has been achieved after some years of study and development, and such machines are generally reserved for use by accelerator laboratories with the manpower to develop and support them.

In contrast to cyclotrons, spallation sources used for neutron scattering typically deliver pulsed bunches of protons, and thereby neutrons for users; the pulsed neutron production is required to enable various experimental techniques valued in condensed matter physics, such as neutron diffraction. The main two existing spallation sources - ISIS in the UK (160 kW) [21] and the SNS in the USA (1.1 MW) [76] - are based around a linac pre-injector followed by a synchrotron, but the next-generation 5 MW European Spallation Source utilises a linac-only design as the user-required pulse structure is different [77]. With appropriate beamline components it is in principle possible to extract a small fraction of the beam current for other purposes, and this is already done on ISIS for muon production using a thin carbon target upstream of the main tungsten spallation target. Spallation sources are very large installations costing ~£1 billion, and so are only trans-national projects. Their use for isotope production must take into consideration the complexities of extracting a portion of the beam, the reliability of the facility, and the lack of facility redundancy in the case of equipment failure.

It should be noted that large cyclotrons delivering powers in excess of 1 MW are also in development. For example, the DAEDALUS neutrino project proposes an 800 MeV superconducting $H_2^+$ cyclotron to deliver around 10 MW of continuous beam power [75,78], but whilst ion source and cyclotron magnet development are becoming mature the demonstration of an integrated design remains some years away.

Whilst both linacs and cyclotrons are in principle capable of extremely high beam power, both these large-scale machines are complex and therefore difficult to operate, and are probably not appropriate to consider in the context of radioisotope production. That said, isotope production has been proposed at both the CERN Linac 4 project [44] and for the Argonne National Laboratory Project X proton accelerator.

At lower energies more conventional accelerator designs become accessible. A reference point are the cyclotrons developed for proton therapy; these are limited in practice to proton energies up to around 250 MeV, and have demonstrated currents up to around 0.5 to 1 uA that are sufficient for delivery of therapeutic beams [79,80]. Several of the manufacturers of these cyclotrons have proposed higher-current, lower-energy cyclotrons for isotope production. For example, IBA have proposed a 150 MeV, 2 mA cyclotron for spallation neutron production either for subcritical reactor studies or for $^{99m}$Tc production (see below) [6,81]. As yet no demonstration of these cyclotron parameters has been made.

### *4.3.4 Neutron Production*

Neutron production may be performed using a variety of methods. As indicated earlier, production methods are generally either low-energy nuclear reactions such as $^{7}$Li(p,n)$^{7}$Be [12], or spallation reactions in high-Z targets carried out most efficiently using c.1 GeV protons. At low energy the neutron multiplicity (neutrons per proton) is rather poor: a typical example is the existing 2.8 MeV Birmingham dynamitron which presently delivers around 1 mA (3 kW) of protons onto a solid lithium target [18]. Power is limited by the available cooling of the target, and significantly higher powers almost certainly require the use of liquid targets, for example as at the Israeli SARAF facility [12,82]. Most of the energy given to the protons during their accelerator is simply lost again through ionization slowing, and this is a feature of all low-energy production methods. At somewhat higher energies the neutron production cross section in beryllium targets becomes favourable, and is utilised for example by AAA for neutron capture isotope production (see above). A novel recirculation method through a thin target becomes possible at higher proton energies; this method – called ERIT (energy recovery internal target) [83] - utilises an FFAG accelerator with large beam acceptance to cope with the transverse scattering induced by each proton passage through a Be foil target with a thickness of a few microns [73]. The use of a sufficiently thin target, plus re-acceleration and recirculation after each pass, allows the re-use of the vast majority of protons that did not produce neutrons on prior passes, whilst at the same time optimizing the energy of the proton as it passes through the target to maximize the neutron production cross section. This method is already in practical use at KURRI in Japan to produce neutrons for boron neutron capture therapy (BNCT) [83], and has been considered for neutron production for both isotope production and security scanning [73]. A similar target concept might be used directly for $^{99m}$Tc production in which the Be foil were replaced by a $^{100}$Mo foil. However, given the heat load suffered by more robust carbon foils in conventional cyclotrons, it is likely that using this method for $^{99m}$Tc production will present formidable difficulties.

At higher energies, and with higher-Z targets, spallation starts to dominate neutron production, and the neutron multiplicity rises above 1 neutron per proton for incident proton energies above c. 150 MeV [16,17]. Although the multiplicity in large targets continues to rise with proton energy, consideration of the accelerator and energy cost for accelerating each proton results in an optimum incident proton energy around 1 GeV, although this peak in incident energy is quite broad; a 1 GeV proton incident on a lead target will produce 20-30 neutrons depending on the target size and shape; the size of target is determined by the requirement to make best use of the inter-nuclear cascade of produced protons and neutrons generated from the initially-struck target nucleus. As evidenced in the design considerations for the European Spallation Source (ESS), a variety of solid and liquid targets have been considered, including solid uranium (previously used at ISIS), solid tungsten/tantalum (used at ISIS), liquid mercury (used at SNS), liquid lead (used at Paul Scherrer Institute) and liquid lead-bismuth eutectic (studied for the MYRRHA reactor). As well as the demonstrated solid and liquid targets used at ISIS and SNS for neutron scattering, PSI have already demonstrated around 1.2 MW beam power on a liquid lead target in the context of accelerator-driven subcritical reactor development. Liquid targets help both to overcome the beam power (melting) limitations of solid targets and to cope with the shock and ablation caused by the incident proton beam pulses. Other methods which have been considered include the use of powder jet targets, and rotating solid targets to assist in power dissipation; the ESS project has adopted the latter of these in its baseline design that is currently under construction in Sweden [84].

### *4.3.5 Laser Acceleration*

As discussed above, the direct production of $^{99m}Tc$ from $^{100}Mo$ targets bombarded by moderate-energy (18-24 MeV) protons is the most promising as an alternate technology to reactor fission. However, conventional accelerators are not the only choice; laser-based accelerators may utilise the same $^{100}Mo$ target, extraction mechanisms and recovery cycle as that proposed for cyclotrons and other conventional accelerators. Since the early 2000s there has been a great deal of research on the acceleration of high-energy protons using one of several laser-plasma based acceleration mechanisms [85,86,87,88]. This work has in the majority been directed towards providing suitable beam properties for hadron therapy where proton energies up to around 250 MeV with narrow energy spread (c. 1 MeV) are desired [79,80]. The work in this field has seen progressive increases in achieved energy and fluence is now approaching the above specifications [88,89,90,91,92]; it is predicted that, with sufficient power, protons up to several GeV energy might be obtained.

Several groups have also performed experiments to investigate the production of radioisotopes, concentrating initially on the production of PET isotopes such as $^{18}F$ and $^{11}C$ [93,94,95]. This research has utilised the existing broadband proton output which has been achieved to generate radioisotopes through (p,n) reactions typically using the giant resonance peaks: $^{11}B(p,n)^{11}C$ and $^{18}O(p,n)^{18}F$. The most notable technique at present is target normal sheath acceleration (TNSA), which phenomenon dominates ion production at laser intensities of $10^{19}$-$10^{21}$W/cm$^2$, and which uses thin (few um) foil targets. Single-shot production of 10s MBq of $^{11}C$ and 100s kBq $^{18}F$ were demonstrated in 2004 [94]. More recently, investigations into direct $^{99m}Tc$ production using the same mechanism has yielded 8 kBq [96] of $^{99m}Tc$ from natural molybdenum targets in each laser shot, which would to ~90 kBq if an enriched (99.54%) $^{100}Mo$ target were used. Each laser shot comprises typically a ~1 ps laser pulse of between 50-200 J focused onto a single target which (though multiple mechanisms) accelerates the protons; TNSA gives a broad output proton energy spread from which the giant resonance peak results in around one $^{99m}Tc$ nucleus for 1000 protons (18-24MeV). To generate clinically-relevant yields the laser repetition rate must be increased above present sub-hertz values; it is estimated that pulse energies of c.50 J at repetition rates over 10 Hz would yield a 1 GBq patient dose in a 20-minute irradiation using current laser and targetry designs.

Presently, given the overall efficiency, size and cost of such a laser system, there is no particular advantage over a cyclotron, as the latter is able to produce around a factor of 100 greater yields over the same exposure time. To be competitive the laser method would have to be around 1/100$^{th}$ of the cyclotron cost. Over a 10-15 year timescale this might be possible; the introduction of diode-pumped pulsed lasers has increased laser efficiency dramatically and significantly reduced thermal loads. This has enabled laser systems to increase repetition rates from a few shots per hour towards 10 Hz in only a few years; given the role of such laser systems in research and industry significantly, the achievement of significantly higher repetition rates seems possible. So whilst not competitive at present, laser acceleration of protons may become a practical alternative in around 10 years.

# Chapter 5: Radiochemical, Pharmaceutical, Dosimetric, and Operational Considerations for Cyclotron Produced $^{99m}$Tc:

**Jim Ballinger and Louise Fraser**

## Summary

- International efforts have demonstrated that production of $^{99m}$Tc by cyclotron is a feasible technology, though it requires infrastructure and incurs ongoing costs of daily production and transport.
- It has been demonstrated on a small scale that standard $^{99m}$Tc based radiopharmaceuticals can be prepared using cyclotron produced $^{99m}$Tc and that their chemical and biological properties appear unchanged.
- There is an increased radiation dose associated with the co-produced longer-lived radionuclides of Tc. There may be other regulatory issues related to the licensing of kits and acceptable levels of radiochemical impurities.
- Provision of $^{99m}$Tc based radiopharmaceuticals across the UK using $^{99m}$Tc produced in a small number of cyclotrons will likely involve a combination of central and local radiopharmacies in order to optimize service provision. There should be public sector oversight of this process.
- Workforce issues and training shortages both within radiopharmacy and to the wider nuclear medicine community in the UK cannot be underestimated – a full impact assessment must be undertaken as part of any move away from the traditional supply chain for $^{99m}$Tc radiopharmaceuticals.

## 5.1 Good Manufacturing Practice (GMP) Compliant Processing of Target Material

This chapter starts at the point when a $^{100}$Mo target is removed from a cyclotron vault. As this will almost certainly be a solid target system it involves a physical transfer process. The subsequent steps are:

- Dissolution of the target
- Recovery of $^{100}$Mo
- Purification of $^{99m}$Tc

Dissolution of the metal target under oxidizing conditions, such as using $H_2O_2$, yields molybdate and pertechnetate in solution. It is not practical to separate the two using standard generator technology (i.e. aluminium oxide column) due to the large quantity of Mo present, which would require an excessively large column in order to achieve adequate separation [1, 2]. Solvent extraction with butanone (methyl ethyl ketone) has been used in some parts of the world for many years, but is difficult to perform safely and efficiently on the scale required [3]. Sublimation has also been evaluated and offers the advantage of recovery of $^{99m}$Tc in a small volume of saline [4]. However, anion exchange column chromatography would appear to be the most promising approach, as it is relatively straightforward to automate [1, 5].

In one such process [1], the target is dissolved in $H_2O_2$ and passed through an anion exchange cartridge. The $^{100}$Mo is washed through using 3 M ammonium carbonate and collected. The cartridge is washed with 1 M sodium carbonate and the eluate taken to waste. Finally $^{99m}$Tc is eluted with distilled water through a cation exchange cartridge into the collection vial.

An automated device has been designed at TRIUMF in Vancouver [5]. The irradiated $^{100}$Mo target (80 mg) is dissolved in 10 mL $H_2O_2$ at 60$^{o}$C and filtered through a polyethylene frit before being mixed with 20 mL 5 N KOH. The solution is passed through a 500 mg ABEC-2000 (aqueous biphasic extraction chromatography) column and washed with 3 mL 3 N KOH, with the waste going to the $^{100}$Mo collection bottle. Flow rates up to 5 mL/min do not affect trapping or recovery. $^{99m}$Tc is then eluted with 5 mL distilled water and the solution is neutralized by passage through a strong cation exchange cartridge. The eluted $^{99m}$Tc is trapped on an alumina cartridge and washed off with saline. The isolation efficiency was 84% in less than 30 minutes. The extent of recovery and recycling of $^{100}$Mo is critical to the economics of the project. The consistency of purity of recycled $^{100}$Mo is also critical as it affects the profile of radionuclidic impurities produced, discussed below in the dosimetry section.

## 5.2 Specifications for Cyclotron-Produced $^{99m}$Tc

The current pharmacopoeial specifications are as follows. The European Pharmacopoeia (Ph. Eur), harmonised with the British Pharmacopoeia (BP), has different specifications for radionuclidic purity depending on the method of production of $^{99}$Mo. The Ph. Eur has begun deliberations on specifications for cyclotron produced $^{99m}$Tc and intends to have a draft by early 2015. The current United States Pharmacopoeia (USP) specifications are presented for comparison (Table 5.1).

**Table 5.1:** Specifications for Cyclotron-Produced $^{99m}Tc$

| **Parameter** | **BP/Ph. Eur** | **USP** |
|---|---|---|
| Appearance | Clear, colourless solution | Not stated |
| Tonicity | In normal saline | In normal saline |
| pH | 4.0 to 8.0 | 4.5 to 7.5 |
| Aluminium | <5 ppm | <10 ppm |
| Sterility | Sterile | Sterile |
| Endotoxins | <175/V EU/mL | <175/V EU/mL |
| Radionuclidic purity – fission | $^{131}I$ <0.005%<br>$^{99}Mo$ <0.1%<br>$^{103}Ru$ <0.005%<br>$^{89}Sr$ <0.00006%<br>$^{90}Sr$ <0.000006%<br>Other gammas <0.01%<br>Alphas <$10^{-7}$% | $^{131}I$ <0.005%<br>$^{99}Mo$ <0.015%<br>$^{103}Ru$ <0.005%<br>$^{89}Sr$ <0.00006%<br>$^{90}Sr$ <0.000006%<br>Other gammas <0.01%<br>Alphas <0.001 Bq/MBq |
| Radionuclidic purity – non fission | $^{99}Mo$ <0.1% | $^{99}Mo$ <0.15 kBq/MBq (0.015%)<br>Other gammas <0.5 kBq/MBq (0.05%)<br>Max 92 kBq/dose |
| Radiochemical purity | >95% $TcO_4$ | >95% $TcO_4$ |

A number of papers (Table 5.2) have included evaluation of their results against the USP specifications:

**Table 5.2:** Comparison of published results against USP specifications

| Parameter | USP Spec | Gagnon [1], new target (n=4) | Gagnon [1], recycled target (n=3) | Morley [5] | Galea [6] |
|---|---|---|---|---|---|
| pH | 4.5 to 7.5 | 5.0 to 7.0 | 6.0 to 6.5 | 5.5 | 6-7 |
| Radiochemical purity | >95% $TcO_4$ | >99% | >99% | - | 97-99% |
| Aluminium content | <10 ppm | <2.5 | <2.5 | <5 | <10 |
| Radiochemical identity | Rf 0.9 | - | - | 0.9 | - |
| Molybdenum content | Not specified | - | - | <5 ppm | - |
| $^{99}$Mo content | <0.015% | - | - | <0.001% | <0.015% |

Results for radionuclidic purity will be discussed below in the dosimetry section. In summary, it appears that all of the current pharmacopoeial specifications can be met except for radionuclidic purity. A detailed theoretical and experimental assessment by Lebeda et al [7] predicts a best case of 0.06% contribution from Tc radioisotopes other than $^{99m}$Tc, principally $^{95}$Tc (t½ 20 h) and $^{96}$Tc (t½ 4 d).

It must also be remembered that the radionuclidic purity decreases with time after production, since longer-lived radiocontaminants contribute more as the primary radionuclide decays. The limits are set for the time of use (or expiry) of the $^{99m}$Tc solution. This could be a problem if the irradiation ends at, say, 04:00 but the last injection is planned for 16:00, by which time the radiocontaminants could have increased 4-fold. This is considered by Hou et al [8].

## 5.3 Preparation of $^{99m}$Tc Labelled Products

Once the $^{99m}$Tc pertechnetate has been purified, a range of $^{99m}$Tc products would be prepared from kits. Although there are some concerns that problems may arise, the available information suggests that current kits perform satisfactorily with cyclotron-produced $^{99m}$Tc, at least on the scale studied so far.

The main concern is with specific activity due to the presence of larger amounts of $^{99}$Tc than is seen with generator $^{99m}$Tc and also other radioisotopes of Tc. This increases the total chemical quantity of Tc present in the labelling solution, which may affect:

- The ability of the stannous ion in the kit to reduce $^{99m}$Tc to the correct oxidation state
- The stability of the resultant $^{99m}$Tc complex as there will be less excess ligand present; this could mean a shorter shelf life, which could be a problem for supply from central radiopharmacies

As discussed recently by Qaim et al [9] the specific activity of cyclotron-produced $^{99m}$Tc starts out roughly equivalent to "Monday pertechnetate", with which there are recognized problems of the nature mentioned above. With decay of $^{99m}$Tc during the day the situation becomes even more unfavourable for preparation of kits at later times. There is also a theoretical concern that this lower specific activity product could have subtly different biodistribution properties. The following table (Table 5.3) summarises the published results on products prepared using cyclotron-produced $^{99m}$Tc:

**Table 5.3** Published results of products prepared using Cyclotron-produced $^{99m}Tc$

| Reference | Product | Cyclotron $^{99m}Tc$ results | Comparison |
|---|---|---|---|
| **Radiochemical studies** | | | |
| Gagnon [1] | MDP | 97% | 99% generator |
| Nagai [4] | MDP | >99% | Meets USP spec |
| Richards [10] | MDP | >99% | Meets spec |
| Galea [6] | MDP | 93-97% | 94% |
| | Tetrofosmin | 96-97% | 97% |
| Jalilian [11] | Sestamibi | 94±3% | Meets spec |
| | DTPA | 98±1% | Meets spec |
| | DMSA | 97±2% | Meets spec |
| | Phytate | 96±2% | Meets spec |
| Benard [12] | Sestamibi | 99±1% | Meets spec |
| | MDP | 98±3% | Meets spec |
| | HMPAO | 91% | Meets spec |
| **Pre-clinical studies** | | | |
| Gagnon [1] | MDP | Normal rabbit bone scan | Similar to generator product |
| Galea [6] | MDP | Bone uptake = 114<br>Bone/liver = 2.95 | Bone uptake = 109<br>Bone/liver = 2.75 |
| | Tetrofosmin | Similar myocardial accumulation and retention; Heart/liver = 2.5 | Heart/liver = 2.5 |
| Richards [10] | MDP | Normal mouse bone scan | Expected biodistribution |
| Jalilian [11] | Sestamibi | Heart, kidney, and GI tract highest | Expected biodistribution |
| | DTPA | Kidney and bladder highest | Expected biodistribution |
| | DMSA | Kidney and bladder highest | Expected biodistribution |
| | Phytate | Liver and spleen highest | Expected biodistribution |
| **Clinical studies** | | | |
| McEwan [13] | Pertechnetate | Normal biodistribution | No difference between cyclotron and generator |
| Selivanova [14] | Pertechnetate | Normal biodistribution | No difference between cyclotron and generator |

As pointed out by Qaim et al [9], in the validations performed so far the kits may not have been "stressed" with maximum activities nor worst-case specifications of cyclotron produced $^{99m}Tc$ (e.g. long irradiations, higher level of co-produced Tc isotopes). In recent years, the regulatory authorities have imposed restrictions on the use of eluates of generators which have not been eluted for more than 24 h (i.e. "Monday pertechnetate"), which could have implications for the acceptability of cyclotron produced $^{99m}$Tc.

It is also of concern that there is still no published clinical evaluation of cyclotron-produced $^{99m}$Tc, only an abstract in 2012 from the Edmonton group [13] and another in 2014 from the Sherbrooke group [14], both using pertechnetate.

## 5.4 Dosimetry Considerations When Using $^{99m}$Tc Produced by a Cyclotron

There are three potential routes that can be used to produce $^{99m}$Tc using a Mo(p,x)X reaction:

1) $^{98}$Mo(p,γ)$^{99m}$Tc
2) $^{100}$Mo(p,2n)$^{99m}$Tc
3) $^{100}$Mo(p,pn)$^{99}$Mo -> $^{99m}$Tc

Impurities in cyclotron-produced $^{99m}$Tc have implications for radiopharmaceutical chemistry and patient dosimetry. These impurities include a large number of various radioactive and stable isotopes of technetium, molybdenum, niobium and zirconium.

In contrast, $^{99m}$Tc produced from a reactor-produced $^{99}$Mo generator is 'pure' in that it only contains $^{99m}$Tc and its decay product $^{99g}$Tc, though it can contain up to 0.1% $^{99}$Mo breakthrough and much lower levels of other radionuclides.

There is a need to optimise the cyclotron operational parameters to maximise the production of $^{99m}$Tc and to minimise contaminants. These parameters include:

- Beam energy
- Irradiation time
- Target enrichment
- Cooling time (time after end of beam)

Non-technetium impurities (including Mo, Nb, Zr) can be removed via chemical separation provided this is performed correctly and efficiently. These impurities will constitute radioactive waste and become potential sources of contamination from the cyclotron production lab. Chemical separation cannot isolate $^{99m}$Tc from the remaining technetium isotopes, so the radiopharmaceutical preparations will contain these impurities and will increase the patient radiation dose. The radioactive technetium isotopes are shown in Table 5.4 (those isotopes produced with T½ >$10^3$ years are considered stable):

**Table 5.4** Half-lives of Important Technetium Isotopes.

| Isotope | Half life |
|---|---|
| $^{99m}Tc$ | 6.01 h |
| $^{97m}Tc$ | 91.4 d |
| $^{96m}Tc$ | 51.5 min |
| $^{96g}Tc$ | 4.28 d |
| $^{95m}Tc$ | 61 d |
| $^{95g}Tc$ | 20 h |
| $^{94m}Tc$ | 52 min |
| $^{94g}Tc$ | 293 min |
| $^{93m}Tc$ | 43.5 min |
| $^{93g}Tc$ | 2.75 h |

Hou et al [8] have produced theoretical dosimetry data for three commonly used radiopharmaceuticals: sestamibi, phosphonates and pertechnetate. They also investigated the effect of target composition, beam energy, irradiation and cooling time on patient dosimetry estimates. These theoretical dose estimates were compared to those from $^{99m}Tc$ obtained from a $^{99}Mo$ generator. The calculations from Hou et al demonstrate that the target enrichment must be carefully considered. The percentage of $^{100}Mo$ enrichment must be high but the relative amounts of other Mo isotopes, particularly $^{95}Mo$, $^{96}Mo$ and $^{97}Mo$, should be minimised to reduce the production of $^{95g}Tc$ and $^{96g}Tc$ which contribute significantly to patient dose.

Hou et al [8] suggest that the most favourable proton energy for cyclotron-produced $^{99m}Tc$ would be in the region of 16-19 MeV. At higher proton energies, the production of other technetium isotopes increases and these impurities contribute considerably to patient dose.

In addition, Hou et al [8] suggest short irradiation times and cooling times less than 12 hours. Using a 6 h irradiation time, a target enriched to 97.39% $^{100}Mo$, proton energy through target from 19-10 MeV, and a cooling time of 2 h, the percentage increase in patient effective dose is 0.69, 0.54 and 0.63% for sestamibi, phosphonates and pertechnetate respectively. Selivanova et al recently estimated similarly low increases in effective dose with optimal irradiation times, target enrichment values, and beam energies [14].

These theoretical calculations need to be verified by direct experimental measurement. Lebeda et al [15] have recently reported on radiochemical purity quality control measurements on a range of radiopharmaceutical kits manufactured using cyclotron-produced $^{99m}Tc$. Their results showed no impurities in seven out of the nine kits tested. These results are promising, but direct comparison to $^{99}Mo$-produced $^{99m}Tc$ was not performed and no measurements were made at a later time, where a contribution from longer-lived technetium isotopes may be demonstrated. Further work is still required in this area.

## 5.5 Feasibility and Commercial Analysis

The OECD Nuclear Energy Agency report of 2010 predicts that a 6 h irradiation in a suitable high-power cyclotron could yield >1 TBq of $^{99m}$Tc, enough to supply a population of about ~10 million people [16]. Lebeda claims to have produced 1.8 TBq (unpublished). The TRIUMF group recently demonstrated that 350 GBq $^{99m}$Tc can be obtained following 7 h irradiation at 18 MeV and 220 μA [12].

An economic analysis of cyclotron $^{99m}$Tc costs would have to include the following:

- Capital costs: cyclotron, laboratories, targetry, purification equipment
- Consumables: targets, reagents, testing
- Personnel: numbers and grades; unsocial hours payment
- Transport: from central production sites

A big question is whether $^{99m}$Tc production could be piggy-backed onto existing current $^{18}$F (PET) production in a single facility. In terms of timing, this should be possible, as a $^{99m}$Tc production run might be 22:00 to 04:00 followed by processing, during which time the $^{18}$F bombardment could take place. In terms of capital costs this might make sense, provided that there is access for transfer of the solid target and sufficient space for the $^{99m}$Tc purification process to take place. However, there is increased risk of impact on the PET production if there are problems with the cyclotron or facilities. There is also concern about "flogging" the cyclotron, with additional wear and tear caused by continuous running. Dual operation with other isotopes also limits $^{99m}$Tc production to one (overnight) run per day, and makes a second bombardment for $^{99m}$Tc problematic.

It is emerging that it might be better to plan two runs per day. As discussed in section 5.3, one of the factors that result in increased levels of radionuclidic impurities is prolonged irradiation times. Prolonged irradiation periods may be required in order to achieve an adequate yield of $^{99m}$Tc, particularly if the cyclotron beam current is low [17]. This would likely require a dedicated cyclotron, which would increase the capital cost. In any case, there would need to be back-up supply from a nearby unit for the inevitable downtime due to equipment failures.

An economic analysis has been obtained from the Edmonton Radiopharmaceutical Centre, where a 24 MeV cyclotron has been installed solely for $^{99m}$Tc production. The parameters in the model include:

- Capital cost of purchase and installation, amortized over 20 years;
- Operating costs;
- Staffing;
- Consumables;

- Regulatory;
- Maintenance.

The bottom line is that operation at a level of 200,000 doses per year would result in an average cost of Canadian $20 per dose (equivalent to £10.80) for the pertechnetate component. A recent survey of UK radiopharmacies indicated an average crude cost for pertechnetate of £8.54 per dose, with unit dose at £7.66 and multidose at £10.32. Thus, the cost of cyclotron-produced $^{99m}$Tc is quite comparable to current generator prices, and would become even more competitive as generator prices increase due to anticipated full economic costing of reactor-derived $^{99m}$Tc.

There are two cyclotrons explicitly marketed for $^{99m}$Tc production, and another that has been proposed for adaptation for $^{99m}$Tc. These are summarised in Table 5.5. Other cyclotron manufacturers are likely to market the potential of their products to manufacture $^{99m}$Tc, and/or offer integrated solutions.

**Table 5.5** Notable Cyclotron Manufacturers relevant to $^{99m}$Tc production

| Manufacturer | Model | Energy (MeV) | Beam current (µA) |
|---|---|---|---|
| Advanced Cyclotron Systems Inc. | TR24 | 18-24 | 300-1000 |
| Best Cyclotron Systems Inc. | B15p<br>B25p | 15<br>20 or 25 | 400<br>400 |
| GE Medical Systems | PET trace 880<br>Upgrade<br>Future | 16.5 | 130<br>250<br>400 |

## 5.6 Radiopharmaceutical Preparation Using Cyclotron-Produced $^{99m}$Tc

At first glance, it would seem to be most efficient for the bulk $^{99m}$Tc to be used for preparation of kit-based products at a single location adjacent to the production site, rather than being sent in batches to a number of smaller radiopharmacies. Here the model of large North American radiopharmacies could be studied in order to make best use of the available $^{99m}$Tc.

However, centralized radiopharmaceutical preparation might not be possible if the stability of the products, and hence the shelf life after labelling, were insufficient for efficient clinical use after transport. It could be that a combination of central and local preparation is required for optimal quality of products, efficiency of use, and flexibility of provision of clinical service.

This was discussed more thoroughly in the 2010 ARSAC report [18]. Among the potential problems to be considered are the following:

- *Greater impact of equipment failure at central radiopharmacy*. If there is no local back up, a failure in the air-handling unit would affect a large number of hospitals. Microbiological contamination or a radiation spill within the radiopharmacy could have a similar effect. Recovery times could be of the order of 1-2 weeks.
- *Reliance on a daily supply of active starting material*. Daily supply of bulk $^{99m}$Tc rather than generators that can be used over a period of up to three weeks means that there is no back-up supply should the cyclotron fail. A large number of hospitals would be affected. The recovery time would be dependent on the nature of the problem with the cyclotron.
- *Fewer alternatives for back-up supply*. This is less catastrophic than the unit failure mentioned above, but could include shortages of staff, a particular product or occasional failure of a $^{99}$Mo generator.
- *Access to technical procedures*. Services that require direct radiopharmacy involvement as part of the procedure (e.g. gastric emptying, blood labelling or denatured red cells) may be limited at remote sites due to the unavailability of trained staff and a lack of specialised facilities. However, arrangements can be made for provision of some of these services, such as sending trained staff to remote sites or transporting blood to the radiopharmacy for radiolabelling.
- *Access to professional services*. Services that require senior staff involvement for patient facing roles (e.g. patient counselling, therapeutic administration, medicines history taking and medicines interaction reviews) may not be provided as the employment of senior staff solely for this purpose would not be financially justifiable.
- *Critical dependence upon transport*. Transport introduces a source of uncertainty due to traffic conditions and reliance upon the availability of trained drivers and equipped vehicles. The logistics of supplying ~200 nuclear medicine departments daily from a very small number of central locations are seen as challenging.
- *Delays in obtaining extra doses*. The transport time will have to be taken into account when extra or urgent doses are required. This may not be possible for all sites.
- *Limitations on weekend supply*. Individual hospitals would have to negotiate with the central radiopharmacy for weekend or extended hours' supply.
- *High extremity radiation doses to radiopharmacy staff*. This could be reduced by introduction of automation that may be economically feasible in a smaller number of specialist centres.
- *Reduced quality of life for central radiopharmacy staff*. Due to early start – this could lead to problems in recruitment and retention of staff.
- *Loss of scientific role of radiopharmaceutical scientist*. As provision becomes primarily a technical service, although this would probably be more of a problem in the commercial sector. Within the NHS the scientific role would remain, but only if large radiopharmacies were located within clinical and research environments [19].

- *Workforce issues* Multi-professional training in radiopharmacy practice of allied professional groups in nuclear medicine (e.g. physicists, clinicians, radiologists, radiographers, clinical practitioners and technologists) could not be accommodated in larger units where rationalisation has occurred because of the higher throughput pressures and the increased demand as a result of fewer numbers of units.
- *Learning and development within radiopharmacy* The Modernising Pharmacy Careers Board has proposed in a report to Health Education England that there should be national workforce planning for the specialist pharmaceutical technical areas [20]. As a result, the Clinical Pharmaceutical Scientist Training programme was established. Radiopharmacy training is an integral part of this training programme, and on completion, graduates are able to work in all technical areas of pharmacy. A reduction in capacity for education and training would affect this training programme as well as other training schemes which require radiopharmacy input.
- *Lack of professional leadership* Reduction in the headcount of senior professional staff in radiopharmaceutical services will diminish both radiopharmacy and nuclear medicine services for the reasons besides affecting a technical supply:
    - Lack of professional leadership
    - Loss of corporate memory
    - Lack of innovation
    - Loss of service development
    - Reduced cross-sector audit
- *Loss of radiopharmaceutical expertise in the wider community* Although consultant radiopharmaceutical scientists would be available in the central radiopharmacy there would be less access to local expertise for advice on clinical issues and research. This can also be a limitation for training programmes in nuclear medicine technology, medical physics, and specialist medical registrars in nuclear medicine and radiology.

If the independent sector takes the initiative to develop cyclotron based production of $^{99m}Tc$, there should be an impact assessment held by the public sector on the current model of radiopharmacy services and the knock on effect to nuclear medicine (both diagnostic and therapeutic) prior to any development taking place. BNMS and the UKRG are best placed to do this.

# Chapter 6: Radionuclides for Therapy and Imaging (Theragnostics)

**Glenn Flux and Adrian Hall**

## Summary

- There is currently no record in the UK of the numbers of treatments delivered with radiotherapeutics, the number of centres offering treatment, or of the details or outcome of the treatments themselves. Only two surveys have been held to obtain a snapshot of current activity.
- Radiotherapeutics are particularly prone to shortages, as these cannot be stockpiled due to the natural decay of the radioisotopes and the limited time for stability of radiolabelling. Many radiotherapeutics are produced at single sites that may be outside Europe. Supplies can be halted at short notice, preventing completion of a course of treatment. There is no evidence or protocol governing continuation of treatment.
- In many cases, two or more radiotherapeutics are available for treatments. In the absence of head-to-head clinical trials, the treatment of choice at any given centre is dependent on local logistics and clinical judgement rather than on evidence
- The costs of radiotherapeutic drugs for cancer treatment is estimated to have increased from £2.5m - £8m from 2007 – 2012 and is projected to increase to ~ £50m in the coming years, dependent on NICE approval of commercial products.
- MRT is a highly multi-disciplinary field. Resource costs and the workforce required to deliver radiotherapeutic treatments effectively and safely have not to date been estimated. These will rise in line with the drug costs.
- The EU directive 2013/59, mandating dosimetry-based treatment planning for MRT as for external beam radiotherapy and brachytherapy, will have a significant impact on treatment procedures and on resource issues.
- Treatment protocols are currently standardised only where these have been formulated by the company providing the radiopharmaceuticals, following commercially funded clinical trials. A limited number of academically driven trials have been conducted, although these are difficult to fund and perform. Support for academic development would ensure UK competitiveness, a national supply and rapid translation of the drug into clinical practice.

## 6.1 Introduction:

Many radionuclides (radiotherapeutics) used for molecular radiotherapy (MRT) of cancer and benign disease also act as 'companion diagnostics' that permit functional medical imaging. This allows staging and treatment planning prior to therapeutic procedures and response monitoring following therapy. The potential to directly image the agent responsible for treatment is unique to MRT and is a leading example of the emerging field of 'theragnostics'. Radiotherapeutics are also vulnerable to reactor, processor or distribution failures that can disrupt not just diagnostic information from scanning but the treatments themselves.

At present, there is no national registry for the treatment of cancer with radiotherapeutics and consequently no record of the treatments administered, the centres offering treatment, or of outcome data. The figures given below of UK usage are taken from a BIR report [1] that followed a survey in 2007, and a recent survey conducted by the UK Internal Dosimetry Users Group [2].

## 6.2 Current Practices

There are numerous options for treatments using beta-emitting radioisotopes, although currently there are 5 main treatments:

### *6.2.1 Thyroid Cancer and Benign Thyroid Disease*

Iodine-131 [$^{131}$I]) has been used extensively in the treatment of benign thyroid disease, for thyroid remnant ablation following surgery for thyroid cancer and for the treatment of thyroid metastases in differentiated thyroid cancer for over 60 years. This remains the treatment of choice for the majority of diagnosed cases. A recent survey from the UK Internal Dosimetry Users Group [2] found that ablation procedures following surgery increased by 33% over the years 2007 – 2011.

### *6.2.2 Adult Neuroendocrine Tumours*

Refractory adult neuroectodermal tumours, such as phaeochromocytoma and paraganglioma, are routinely treated with Yttrium-90 ($^{90}$Y) or Lutetium-177 ($^{177}$Lu) radiolabelled peptides, referred to as Peptide Receptor Radionuclide Therapy (PRRT), or with $^{131}$I-*m*IBG in at least 11 centres in the UK (an increase from only four centres in 2007). For the period 2007 – 2011 the number of treatments with radiopeptides increased by 173%.

### *6.2.3 Paediatric Neuroblastma*

$^{131}$I-*m*IBG is also used to treat primary refractory and relapsed neuroblastoma in children in three centres in the UK. Two of these centres are in London.

### *6.2.4 Palliation of Bone Metastases*

Samarium-153 ($^{153}$Sm) EDTMP, Strontium-89 ($^{89}$Sr) and Rhenium-186 ($^{186}$Re) HEDP are used to palliate pain due to skeletal metastases in patients with prostate cancer. In 2013 the alpha-emitter Radium-223 ($^{223}$Ra) Dichloride was approved by the FDA for the treatment of patients with castration resistant prostate cancer (CRPC), symptomatic bone metastases and no known visceral metastatic disease, following clinical trials that demonstrated improved survival and symptom control in comparison with a placebo [3].

### *6.2.5 Microsphere Treatments of Liver Cancer and Liver Metastases*

In recent years, two commercial products have emerged for the treatment of liver tumours with $^{90}Y$ microspheres. Both products use a similar administration technique whereby the radioactive spheres are directly injected into the hepatic artery. The microspheres differ only in terms of the size, composition and activity concentration of the spheres themselves. This new treatment modality has expanded rapidly (from 23 treatments in the UK in 2007 to 155 treatments in 2011). These products are currently undergoing commissioning through evaluation.

### *6.2.6 Additional Therapies*

Further treatments are performed on an *ad hoc* basis in various centres in the UK. These include $^{90}Y$ Zevalin for non-Hodgkin's lymphoma, P-32 for the treatment of craniopharyngioma in children and $^{90}Y$ labelled anti-CD66 for haematopoietic stem cell transplantation.

## 6.3 UK Stakeholders

In recent years a number of groups have emerged to deal with one or more aspects of this highly multi-disciplinary field. These include:

IDUG (Internal Dosimetry Users Group). Primarily concerned with internal dosimetry for molecular radiotherapy.

CERT (CRUK/ECMC UK Radiopharmacy Task Force) – Focussed on the translation of new academic radiotherapeutics into clinical practice.

CTRad (Clinical and Translational Radiotherapy Research Working Group) – providing support for clinical trials.

MetroMRT (Metrology Institutes of Europe, led by the UK National Physical Laboratory) - concerned with standardisation of quantitative imaging and dosimetry.

The British Nuclear Medicine Society have fostered ongoing communication and collaboration between these groups via meetings and teleconferencing and a newly formed sub group of the Radiotherapy CRG will focus on MRT.

## 6.4 Supply of Radiotherapeutics

The supply of commercial radiotherapeutics is particularly prone to shortages, due to the very short shelf life in comparison with chemotherapeutics. For example, in the case of the treatment of bone metastases, the chemotherapeutic agents Docetaxel and Abiraterone have shelf lives of 36 months and 12 months respectively, while the half-lives of $^{223}Ra$ and $^{89}Sr$ are 11 days and 55 days respectively. Many radiotherapeutic agents may have a shelf life of only several days. It is therefore not possible to stockpile radiotherapeutic agents, even in the event of an expected shortage.

With few exceptions, the supply of radiotherapeutics is governed by pharmaceutical companies, as the introduction of commercial radiotherapeutics into clinical practice increases. This has led to a number of issues that have interrupted or prevented patient treatment and associated medical imaging. Two examples are given:

#### *6.4.1 $^{177}$Lu DOTATATE Treatment of Neuroendocrine Cancer.*

The number of PRRT treatments administered has increased rapidly in recent years and growth is forecast to continue significantly as more centres adopt this treatment. A potential shortage occurred in 2013 when, following in-house production of the $^{177}$Lu DOTATATE by hospital radiopharmacies, a patent was granted to Mallinckrodt to manufacture and distribute DOTATATE. Exclusive rights to develop and commercialise $^{177}$Lu DOTATATE were subsequently acquired by Advanced Accelerator Applications (AAA). This resulted in a sudden threatened shortage of the product and a significant rise in costs. Only a limited number of batches were produced each week by the company from a single site in Italy. The patent, although potentially an item of dispute, remains in force at present, although it is recognized that within Europe, only the UK abides by this.

#### *6.4.2 $^{223}$Ra for the Treatment of Bone Metastases (Xofigo)*

This product is currently being evaluated by NICE as a potential alternative to abiraterone following docetaxel in CRPC. Xofigo is supplied by Bayer Healthcare Pharmaceuticals. All production is based in Norway. There has been a worldwide shortage of $^{223}$Ra (from early October 2014), imposed at short notice, due to contamination of the drug in production, which caused an interruption of treatment for a large number of patients, although the lack of a UK registry prevents the actual numbers from being known. Some treatments have been continued, although the radiobiological consequences of pausing treatment part way through a course have not been considered.

The supply of other radiotherapeutics is often not easy to discern although in the majority of cases, European supplies are produced in a single manufacturing plant. No commercial radiotherapeutic is produced in the UK and in some cases there is no production in Europe. For example, the European supply of $^{90}$Y resin microspheres (Sirtex Medical Ltd), used to treat liver cancer and metastases, comes from a site in Singapore, whereas $^{90}$Y glass microspheres (BTG plc) are exclusively supplied from Canada.

## 6.5 Alternative Products

There are potential alternative products for many radiotherapeutic radioisotopes in common use, although it should be considered that while there is often little or no evidence for the potential benefit of one radioisotope over another, there is much conflicting opinion based on clinical judgment and experience.

To consider the main therapy procedures outlined above:
$^{131}$I is obtained from irradiation of tellurium-130 ($^{130}$Te) in a nuclear reactor. While there are at least 2 suppliers of $^{131}$I, it is unclear at this stage whether the impending closure of reactors will impact on these treatments.

$^{177}$Lu DOTATATE is obtained from a single supplier (Advanced Accelerator Applications) and manufactured in Italy. Alternative radiotherapeutics in routine use include $^{90}$Y DOTATATE, a product that may be conjugated in a radiopharmacy, and variants of ligand, including DOTATOC. $^{131}$I *m*IBG is also used to treat neuroendocrine tumours.

Similarly, the treatment of primary refractory and relapsed paediatric neuroblastoma with $^{177}$Lu DOTATATE, often treated with $^{131}$I *m*IBG, is currently the subject of a clinical trial at UCLH. Unfortunately the vagaries of the NHS reimbursement scheme have resulted in at least one further centre not able to join the trial.

In addition to $^{223}$Ra, metastatic bone deposits have been treated with a range of radionuclides, including $^{89}$Sr, $^{153}$Sm, $^{186}$Re, $^{188}$Re and $^{32}$P. There have been no clinical trials to directly compare these products, although large multicentre studies to demonstrate increased survival without concomitant chemotherapy have only been performed for $^{223}$Ra.

$^{90}$Y microspheres are available from 2 companies. Previously, similar treatments have been administered with $^{131}$I lipiodol. At least one further product is in development, using Holmium-166 ($^{166}$Ho) in place of $^{90}$Y.

## 6.6 Projected use of Radiotherapeutics

### *6.6.1 Novel Therapeutic Radiopharmaceuticals*

A significant number of radiotherapeutics are known to be either in development or in early phase clinical trials [4]. Therapeutic radionuclides with potential therapeutic benefit but currently unavailable clinically include Astatine 211 ($^{211}$At), Copper 67 ($^{67}$Cu) and Bismuth-212 ($^{212}$Bi). The costs of new agents to the NHS, and the associated diagnostic scanning required, will be significant and will have a large impact on resources. In addition, it can be expected that the indication for some radiotherapeutics will be extended. For example, it is feasible that bone metastases from other primary cancers, including breast cancer, could be treated with $^{223}$Ra.

### *6.6.2 Costs*

While at this stage projected costs are difficult to estimate, it is certain that an exponential rise in the costs of radiotherapeutics is underway as commercial products emerge (Figure 6.1). This marks a significant change from traditional low cost radiotherapeutics either supplied directly or manufactured in house.

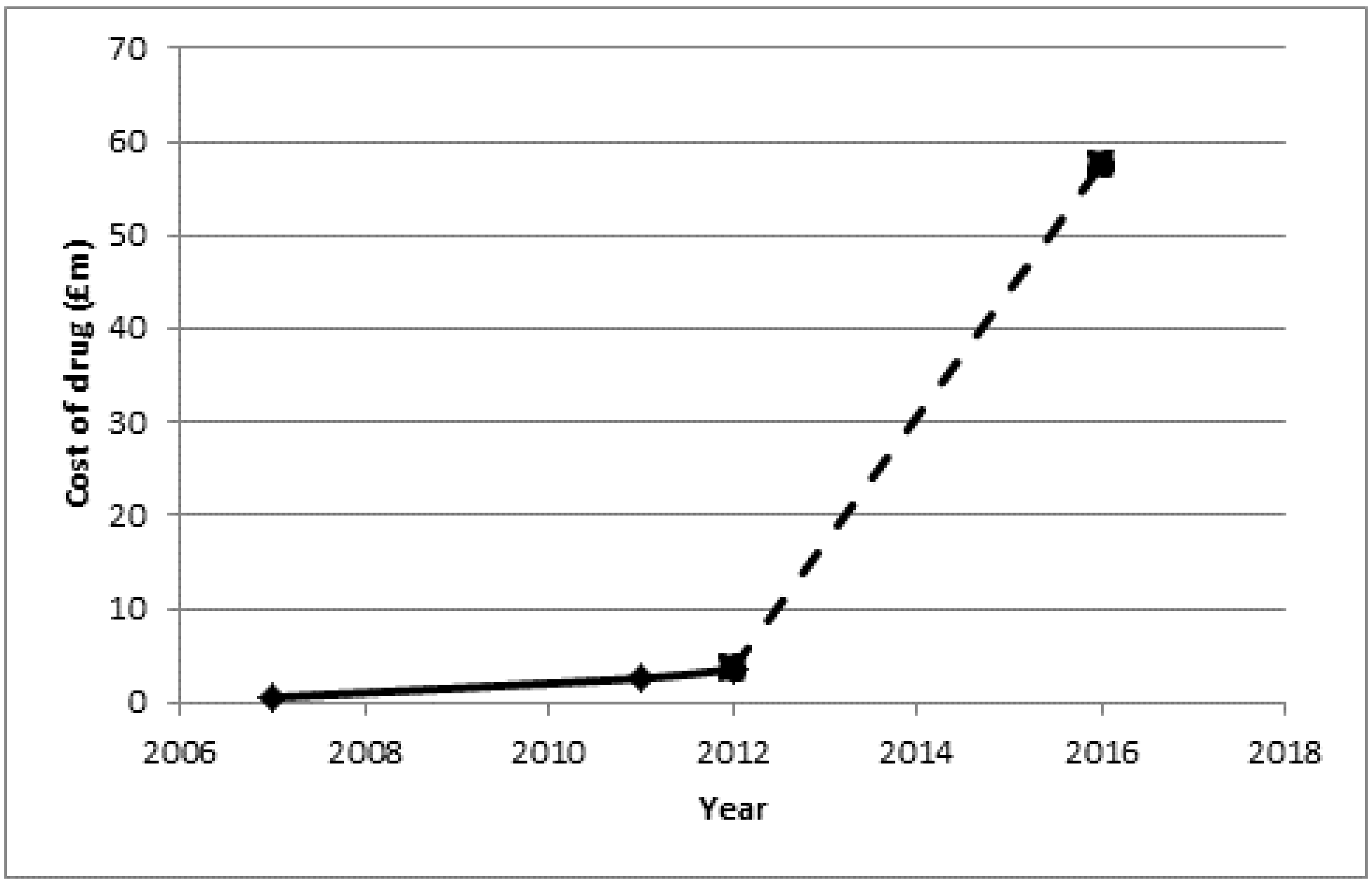

**Figure 6.1**:Previous and predicted costs of radioactive drugs (estimated based on previous survey figures and current costs, and projected usage)

As an example of increasing costs, the recent iDUG survey found that a total of 111 radiotherapeutic treatments were administered for bone metastases from prostate cancer [2], at an average cost of ~ £1000 per treatment. It has been estimated that with NICE approval of $^{223}$Ra, 2000 patients *p.a.* in the UK will be eligible for treatment, which according to the current administration protocol will result in 12,000 administrations each year. $^{223}$Ra is currently funded through the Cancer Drugs Fund at a cost of ~£20,000 for course of six administrations.

### *6.6.3 Euratom Directive*

The Euratom directive 2013/59 states that
'For all medical exposure of patients for radiotherapeutic purposes <including nuclear medicine for therapeutic purposes>, exposures of target volumes shall be individually planned and their delivery appropriately verified taking into account that doses to non-target volumes and tissues shall be as low as reasonably achievable and consistent with the intended radiotherapeutic purpose of the exposure.'

It is further stated that 'Member States shall bring into force the laws, regulations and administrative provisions necessary to comply with this Directive by 6 February 2018.'

Compliance with this directive will entail a dramatic change in the treatment of patients with radiopharmaceuticals, as this is now recognised as a form of radiotherapy whereby patient-specific dosimetry is mandatory, rather than a form of chemotherapy, for which imaging and dosimetry are not possible.

An increasing body of evidence that demonstrates correlations between the absorbed doses delivered and response and toxicity underpin the need for dosimetry-based treatment planning, in the absence of such evidence based on the administered activities alone [5].

### *6.6.4 Clinical Trials and Research Funding*

As demonstrated by a lack of national or international guidance concerning either the levels of activity to administer or the frequency of administration for any therapy procedure, there is a paucity of evidence-based clinical trials to evaluate the optimal use of any radiotherapeutics currently administered. Standardised protocols have been produced only by pharmaceutical companies, rather than as a result of academic driven investigations. As examples, Metastron ($^{89}$Sr) is commonly administered with a fixed activity of 150 MBq, whereas both Quadramet ($^{153}$Sm) and Xofigo ($^{223}$Ra) are administered according to a simple weight-based formula (a single administration of 37 MBq/kg and 6 administrations of 50 kBq/kg respectively). However there are an increasing number of clinical trials of therapy radiopharmaceuticals. In addition to the ALSYMPCA ($^{223}$Ra) trial [3] mentioned, commercial clinical trials have been conducted for $^{177}$Lu DOTATATE [6] and $^{90}$Y microspheres [7]. A small number of academic trials have been conducted although funding and performing these trials has proven difficult and often prohibitive.

### *6.6.5 Workforce Issues*

The workforce and resources required to deliver a radiotherapeutic treatment safely and effectively are highly specialised. In many cases procedures are more resource intensive than is the case for chemotherapeutic drugs due to the need for radiation protection, radiopharmaceutical labelling and dispensing, imaging and dosimetry. The projected increase in the use of radiotherapeutics for cancer treatment, in common with the need for increased capacity to meet the EU directive 2013/59 will require a dramatic increase in resources for nuclear medicine (Figure 6.2).

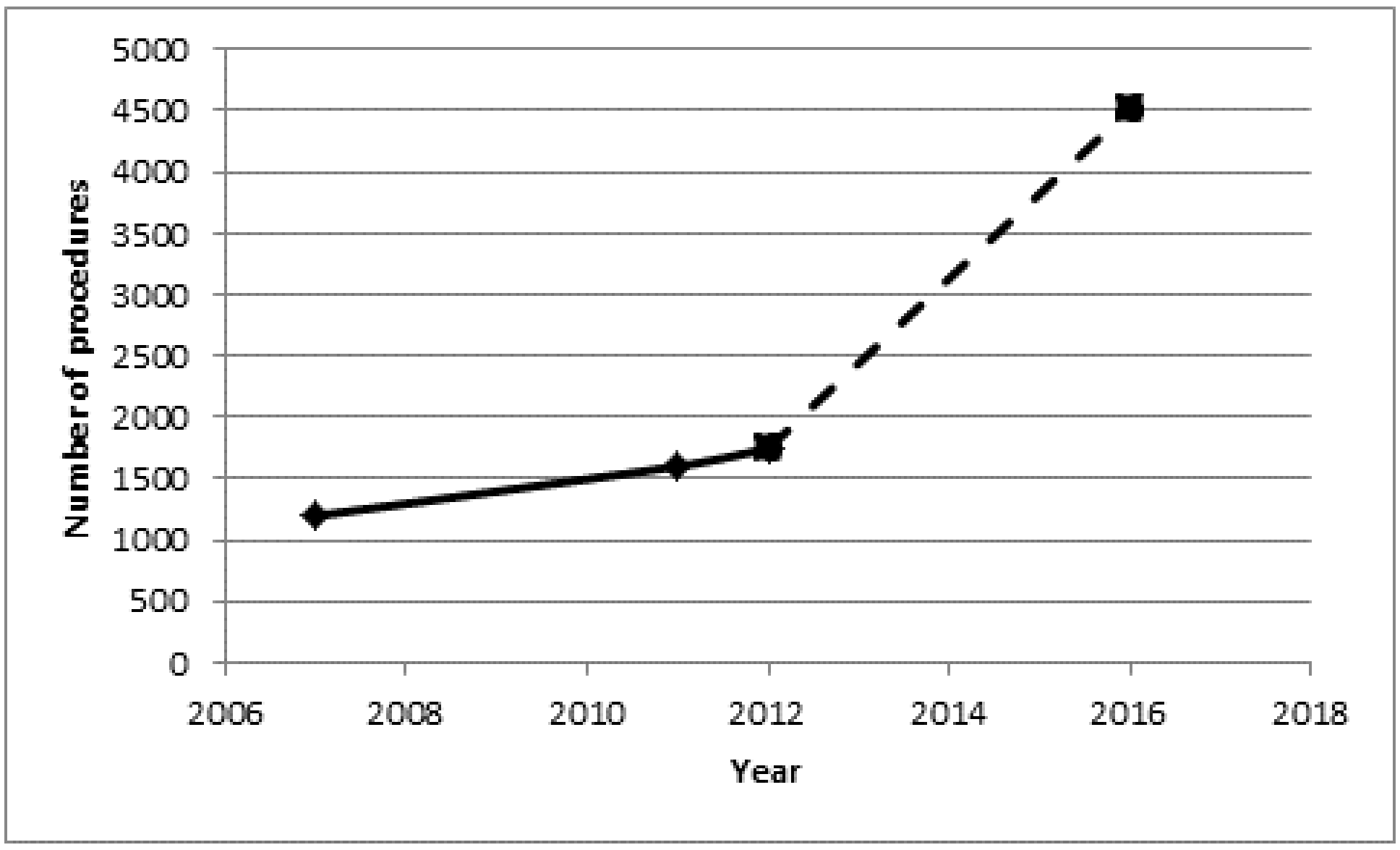

**Figure 6.2** Previous and predicted numbers of radiotherapeutic procedures requiring nuclear medicine support

## Recommendations

- A UK registry for molecular radiotherapy practice is required urgently to monitor the use of radiotherapeutics for cancer treatment, particularly as new radiotherapeutics are introduced to the clinic.
- As MRT is a highly multi-disciplinary area, the various stakeholder groups should be fully consulted and supported to ensure collaboration
- The reimbursement scheme for radiotherapeutics should be reviewed to enable a clear funding pathway to be identified that will ensure support for safe and effective delivery.
- A full investigation is needed of the supply of radiotherapeutics, as it is not possible to stockpile such drugs. Consideration should be given to the development of a national radiopharmacy / radiochemistry resource that may collaborate with commercial pharmaceutical companies and with academic environments as well as with NHS centres to ensure that shortages are minimised
- To ensure optimal and cost effective use of radiotherapeutics, clinical trials should be encouraged to determine treatment regimens and to compare products that offer a similar treatment.
- A full cost analysis of the current and projected use of radiotherapeutics should be conducted, to include the resource implications for Nuclear Medicine and associated disciplines as well as drug costs themselves.
- An investigation of the impact of the EU Directive 2013/59 should be conducted so that this can be met by 2018.
- A national strategy for the use of radiotherapeutics for cancer treatment should be developed to address the supply of radiotherapeutics, projected costs of drugs and resources, the clinical introduction of new radioactive drugs, national equality of access to treatments and resource planning. Consideration should be given to centres of excellence supporting satellites.

# Chapter 7: Nuclear Medicine Workforce

**Sarah Allen, Beverley Ellis, Audrey Paterson and Brian Neilly**

## Summary

- Accurate information on the size of the nuclear medicine workforce in the UK is lacking. This is due to a number of factors including the nomenclature used and to diverse ways of recording such data by the professional bodies. An initiative is required to bring collect and collate this data for the entire UK (not just for NHS England) for future workforce planning.
- Despite the lack of accurate data, there is good evidence of a workforce shortage at all levels of the multi-disciplinary team. This is evident from surveys done by professional bodies, difficulties with recruitment, limited access to training pathways and from projected retirement of the present workforce.
- The nuclear medicine workforce managed the challenge of past shortages of radioisotope but such a response is not sustainable in the long-term. The same consideration applies to the wider imaging workforce outside of nuclear medicine, especially where alternative imaging studies are involved.
- Technological innovations in the field of nuclear medicine have driven the need for further workforce training but to date initiatives to update and make available training programmes are at best patchy.
- Further initiatives such as 7-day working if applied to nuclear medicine will put further pressure on an already stretched workforce and to be effective would have to be fully costed and funded.

## 7.1 Introduction

The bulk of this report discusses in detail the technical options and developments for achieving sustainability of supply of medical radioisotopes within and for the UK. However, the technical solutions that can provide sustainability of supply in the short, medium and longer term cannot be considered in isolation from the workforce who carry out the in-vivo and in-vitro radionuclide imaging studies, and treatments using radionuclides. This chapter sets out the nuclear medicine workforce considerations and identifies the need to ensure that these are addressed.

The term 'nuclear medicine workforce' is used throughout this chapter to mean the essential core disciplines in the multi-disciplinary team. These include clinical practitioners (comprising nuclear medicine technologists and radiographers), clinical scientists, nuclear medicine physicians, nurses, radionuclide radiologists, radiochemists, radiopharmacists, and radiopharmacy technicians. Other members of the multi-disciplinary team, particularly administrative and clerical, and portering staff are also very important.

Additionally, the nuclear medicine specific workforce is drawn from a large number of professional groups, including physicians, healthcare scientists, healthcare science practitioners, nurses, physicians, radiochemists, radiographers, radiologists, and radiopharmacists. Data on sub-specialisation such as radionuclide radiology is difficult to access. The difficulty is compounded by the use of various professional titles, for example, the terms nuclear medicine radiographer, nuclear medicine practitioner, and nuclear medicine technologist are all in current use and describe essentially the same job role. It would be helpful if the nuclear medicine community would agree and use titles consistently, and work with NHS ESR (Electronic Staff Record) [1], to ensure that it captures accurately the whole nuclear medicine workforce.

There are particular difficulties in determining the reality of the current situation relative to the nuclear medicine workforce. It is small overall carrying out approximately 600,000 nuclear medicine, PET or SPECT examinations in the UK. This contrasts with 34.5 million imaging tests in England over the period 1 November 2012 to 31 October 2013 (full year data were not available at the time of writing this chapter). However the workforce carry out important and complex investigations and therapy using radionuclides.

It is the case that the current nuclear medicine workforce was instrumental in maximising the number of studies using $^{99m}$Tc carried out during the 2008/09 shortage, and minimising disruption and delay to patients. This was achieved by: a combination of extending and altering working hours to fit in with the availability of radionuclide activity; modifying (reducing) the administered amount of radioactivity in doses used to nevertheless achieve an appropriate diagnostic result; and transferring patients for relevant alternative imaging modalities that did not rely on $^{99m}$Tc.

These strategies were sustainable for the short term but, for a longer or long-term reduction or failure in the supply of $^{99m}$Tc, it would not be possible for the current, already short-handed, nuclear medicine workforce to work extended hours, days and weeks indefinitely. Similarly, it might not be easy for non-$^{99m}$Tc nuclear medicine, CT and MRI services to absorb the volume of patients referred for alternative imaging.

There may also be considerations for the wider imaging workforce outside of nuclear medicine, especially where alternative imaging studies such as computed tomography (CT) or magnetic resonance imaging (MRI) are used to overcome temporary or permanent shortfalls in supplies of medical radioisotopes. Implications for the wider workforce are outlined towards the end of this chapter.

A further complicating factor is the rate of technological innovation; for example new radiotracers and nuclear-medicine-based radiotherapeutics is growing fast and drawing on an already overstretched workforce. Additionally, if some of the novel technologies identified in this report are to become part of the solution to the reduced availability of $^{99m}$Tc, then the current workforce will require considerable updating and training in those technologies.

Finally, the recommendations of the future hospital commission include seven day care wherever the patients need it, that acutely unwell patients should have the same access to medical care on a weekend as on a weekday and that care for patients should focus on their recovery enabling them to leave hospital as soon as their clinical need allows them to do so [2]. This will have implications for how the service including access to nuclear medicine is delivered and will have implications for the nuclear medicine workforce.

## 7.2 Nuclear Medicine Workforce: The Present Situation

Information and data on the nuclear medicine workforce as a whole is difficult to identify as such information has traditionally been gathered by the professional or workforce groups, and not all of those who are involved in delivering nuclear medicine services will have been part of any relevant workforce survey.

### *7.2.1 Clinical Practitioner Workforce*

As discussed above, Clinical Practitioner is the new umbrella term for radiographers and technologists working in nuclear medicine. This group forms the largest workforce group within nuclear medicine (see Table 2.1 Chapter 2). Despite being the largest group, there is acknowledged to be a shortage of clinical practitioners in nuclear medicine. A 2011 Society of Radiographers survey with responses from 40 Departments in the UK [3] concluded the following:

- Nearly half of the workforce delivering nuclear medicine/radionuclide imaging in the responding departments are radiographers.
- There has been a net increase of 2.1 WTE posts across the 40 responding departments in the last two years (less than 1% of the total establishment).
- The current vacancy rate across all 40 responding departments is 5.0% and the 3-month plus vacancy rate is 2.1%.
- One-fifth of the workforce delivering nuclear medicine/radionuclide imaging in the responding departments is due to retire in the next ten years.
- Around half of the responding departments report an increase in referrals in the last two years with around half reporting a decrease, indicating that workload appears to have remained steady overall across the responding departments.
- The most prevalent qualifications held by staff delivering nuclear medicine/radionuclide imaging are postgraduate diplomas and diplomas of radionuclide imaging.

There remains a concern about the training of clinical practitioners. The issues are summarized in a 2013 editorial in Nuclear Medicine Communications [4]. The Modernising Scientific Careers (MSC) initiative to train practitioners by commissioning a Practitioner Training Programme for nuclear medicine led to the funding of six trainees by the London Local Education and Training Board (LETB) but it remains unclear as to the roll out across the UK and the final numbers or commitment in the long term.

### *7.2.2 Clinical Scientist Workforce*

Accurate data about the numbers Clinical Scientists working in Nuclear Medicine is lacking. The Modernising Scientific Careers (MSC) Scientist Training Programme (STP) is now established. The first cohort to be trained completed in 2014 but the impact on the workforce from this programme is unclear. Important factors are an increase in the demand and complexity in external beam radiotherapy treatments leading to a growing need for clinical scientists to work within this area of Medical Physics. Longer-term this could end in shortages in the Nuclear Medicine clinical scientists workforce unless training numbers are increased. Higher Specialty Scientist Training (HSTT) has started but the pathway is not clear for Medical Physics at present. Recruitment remains difficult at high grades as experienced staff members are unlikely to move especially to high costs areas. There is limited data on the risk to this workforce from budget reductions.

### *7.2.3 Medical Workforce*

**Consultants:** Estimates of the medical workforce in nuclear medicine are derived from a number of sources but principally from the Royal College of Physicians (RCP) annual census of consultant physicians [5]. Current estimates are that there are 75 consultant physicians working in nuclear medicine in the UK [5]. However, this underestimates the true medical workforce since reporting of nuclear medicine imaging procedures is carried out by clinical radiologists with an interest in nuclear medicine imaging. Since radionuclide radiology is a subspecialty there is no formal record of the numbers involved with this group. Additionally a number of specialists from other disciplines such as cardiology, clinical oncology, endocrinology, haematology and neurology hold ARSAC certification for a limited number of imaging, non-imaging or therapy serials. The true size of the nuclear medicine workforce is therefore difficult to assess making workforce planning difficult.

The specialty of nuclear medicine is a consultant-led service. The RCP estimated in 2010 that for nuclear medicine the ratio of consultants to population should be one whole time equivalent (WTE) per 300,000 population [6]. This figure greatly exceeds projections based on current supply and future need based on population growth. The 2012 RCP survey revealed that the this figure was achieved only in London and the South East and that (subject to the caveat mentioned above) the number of WTE Consultant posts per population was well below the recommended level and in certain areas of the UK there is no dedicated clinical input and the region has relied heavily on the input from clinical scientists to maintain the nuclear medicine service in the region [5]. Up to one quarter of the identified consultant workforce is expected to reach the age of 65 years in the next 10 years.

**Trainees:** Recruitment into the specialty has been a problem in recent years with unfilled training posts at specialty training 3 (ST3) level. The origins of this are manifold but include uncertainty about future employment and preparedness to lead a modern nuclear medicine service given the changes in technology (e.g. SPECT-CT, PET-CT, PET-MR) and changes in working practice (e.g. integration of the imaging dataset and leadership of the multidisciplinary team meetings [MDTs]). In response to this a change to the nuclear medicine training programme was proposed and accepted by the General Medical Council (GMC) in August 2014 to commence in August 2015. This new nuclear medicine training programme will extend the period of training from four to six years, provide the trainee with access to the FRCR examination and the Postgraduate Diploma in nuclear medicine and allow application for inclusion on to the GMC Specialist Register in both nuclear medicine and in clinical radiology. The impact of this training is that the new trainees will not complete their training until 2021. Those presently in training will be offered the opportunity to convert, which will extend the period of training. The impact of this is that it will reduce the number of available applicants for substantive nuclear medicine consultant posts in the short-term.

### *7.2.4 Radiopharmacy Workforce*

Radiopharmacy services may be located in pharmacy or nuclear medicine departments or may be in stand-alone units. The workforce delivering the services is multidisciplinary but broadly divided into scientific and technical groupings. Scientific staff may be radiopharmacists, radiochemists or radiopharmaceutical scientists and may perform similar roles. Technical staff may be pharmacy technicians, science graduates, radiographers or nuclear medicine technologists. All these technical groups may be full-time or rotational through radiopharmacy and may perform the same technical roles.

There is a critical shortage of senior-level specialist radiopharmaceutical scientists/radiopharmacists to provide professional leadership and to co-ordinate education and training. There are also shortages of radiopharmaceutical technologists and no defined career pathway for this staff group. (Centre for Workforce Intelligence–Presentation from WRT's 2007 multi-professional imaging review [7]).

A new scientist training scheme (STP) in clinical pharmaceutical science was approved and set-up in 2013 under the MSC framework which was a joint initiative with modernising pharmacy careers (MPC). This training scheme is for pharmacists and scientists and aimed at the pharmaceutical technical specialties (which includes radiopharmacy). Although this has been a positive step for long-term succession planning for scientists/pharmacists it does not address the short-term succession planning for senior posts. The current staff undertaking this scheme will not be registered until 2017.

There is currently no structured training scheme for radiopharmaceutical technologists. Consideration could be given to utilising and training registered pharmacy technicians who are trained in pharmaceutical aseptic services to work in radiopharmacy services.

The MPC Board has proposed in a report to Health Education England (HEE) that there should be national work force planning for the specialist pharmaceutical technical areas. [8].

There may be challenges to collecting the workforce data from the NHS electronic staff record (ESR) owing to the different disciplines (pharmacists, pharmacy technicians, healthcare scientists and practitioners, radiographers etc.) involved in delivering radiopharmacy services and this may not provide information on the sub-specialisation of these groups. At present it is not known how many of the pharmacy workforce are trained in radiopharmacy.

Consideration needs to be given on how accurate workforce information can be obtained for workforce planning, development and commissioning to meet future nuclear medicine services

#### *7.2.5 Workforce Development and Commissioning*

Workforce development solutions are not achieved quickly, with education and training taking anything from four to twelve years depending on profession and specialism. Traditionally, there has been minimal consideration of the workforce as a whole although, in England, the establishment in 2013 of Health Education England may help to change this. However, bringing together and coordinating postgraduate medical, healthcare science, and clinical practitioner education commissioning is likely to prove difficult, and is unlikely to produce tangible, positive outcomes in time for the next predicted shortage of $^{99m}Tc$, predicted for 2016.

# Chapter 8: UK Delegation to assess the viability of the Canadian schemes for cyclotron production of $Na^{99m}TcO_4$

**Alan Perkins and Beverley Ellis**

## Summary

- Two production methodologies with the potential to supplement UK $^{99m}$Tc supplies on a commercial basis have been identified.
- The production of $^{99m}$Tc on high-powered cyclotrons has been accomplished with the potential for routine production of TBq amounts of radioactivity.
- $^{99m}$Tc cyclotron production requires high-energy cyclotrons and cannot be achieved using the existing medical cyclotron facilities in the UK currently used for the production of positron emitting radiopharmaceuticals.
- The key innovation and intellectual property concerning cyclotron production of $^{99m}$Tc rests with the beam targetry and target plate technology.
- Efficient methods for the radiochemical separation and recovery of $Na^{99m}TcO_4$ from $^{100}$Mo targets have been developed.
- The radionuclidic purity of cyclotron-produced $^{99m}$Tc is well understood.
- Small amounts of $^{93}$Tc, $^{94m}$Tc, $^{94}$Tc, $^{95}$Tc, $^{\text{and 96Tc}}$ impurities may be present in variable amounts and with optimisation of cyclotron energy these radionuclidic contaminants may contribute an additional 10% in effective radiation dose to the patient.
- Initial work on radiolabelling standard kit formulations has been carried out and products have met the current QC specifications.
- The position relating to the regulatory approval of cyclotron produced $^{99m}$Tc radiopharmaceuticals is unclear and further information should be sought from both the Canadian regulatory authorities and the MHRA.
- Economic assessment have shown that cyclotron production of $^{99m}$Tc could be achieved at a cost of <$1Ca/mCi.
- The radioactive waste implications of routine cyclotron production of $^{99m}$Tc have not been quantified.
- The reliability of routine large-scale production of $Na^{99m}TcO_4$ is still to be demonstrated.

## 8.1 Introduction

Given that the UK is entirely dependent on the importation of $^{99}$Mo for the production of $^{99m}$Tc it was considered necessary to investigate further into the most highly developed methodologies with potential for domestic production. Of the various technologies outlined previously in this report, those at a most advanced stage with potential for clinical production in the near future have been developed at two sites in Canada. A delegation funded by the Foreign and Commonwealth Office and the STFC and led by the BNMS visited the Medical Isotope Cyclotron Facility (MICF) at the University of Edmonton and TRIUMF in Vancouver during October 2014. This fact-finding mission provided first hand information on the status of these Canadian $^{99m}$Tc production technologies.

## 8.2 The UK Working Party

| | |
|---|---|
| Project lead: | Brian Neilly |
| Delegation: | Alan Perkins |
| | Beverley Ellis |
| | Hywel Owen |
| | Chris Bee, STFC |
| | Anthony Gleeson, STFC |

## 8.3 Cyclotron Technology

The alternative technologies for medical radionuclide production have been described in Chapter 4. Canada's interest in developing alternatives to the traditional reactor based technology resulted in two funding initiatives led by Natural Resources Canada. In 2010 the Non-reactor-based Isotope Supply Contribution Program (NISP) granted $35M for two-year projects. An additional $25M was committed for four years from 2012, for the Isotope Technology Acceleration Program (ITAP). These funding initiatives have led to two promising multi-institutional collaborations. The first is a collaboration between the Universities of Edmonton and Sherbrooke and the second is a consortium led by TRIUMF. The Alberta and TRIUMF approaches are both based on the $^{100}$Mo(p,2n)$^{99m}$Tc reaction with proton bombardment of an enriched $^{100}$Mo target (see Figure 8.1).

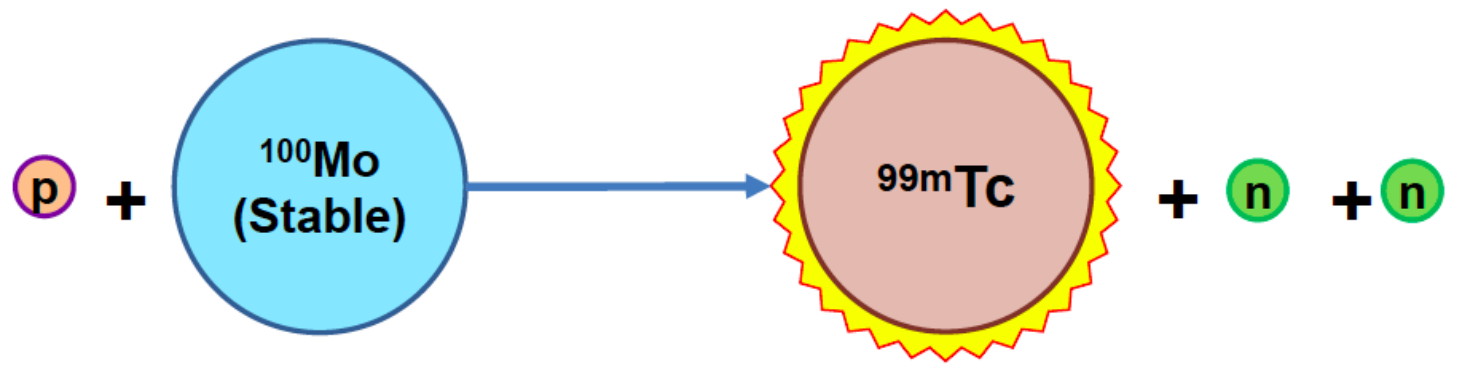

**Figure 8.1** The p,2n reaction for the production of $^{99m}$Tc from enriched $^{100}$Mo

Initial experiments at both Edmonton and TRIUMF were performed with ACSI TR19 cyclotrons, however both centres are going forward with $^{99m}$Tc production based on the TR24 cyclotron. In both cases it is recognized that an incident proton energy of at least 19 MeV is required to obtain a good balance between radionuclidic purity and product yield. Experimental work carried out to date has demonstrated that there is a dynamic balance between production of $^{99m}Tc$ with lower energies offering maximum radionuclidic purity (see Figure 8.2).

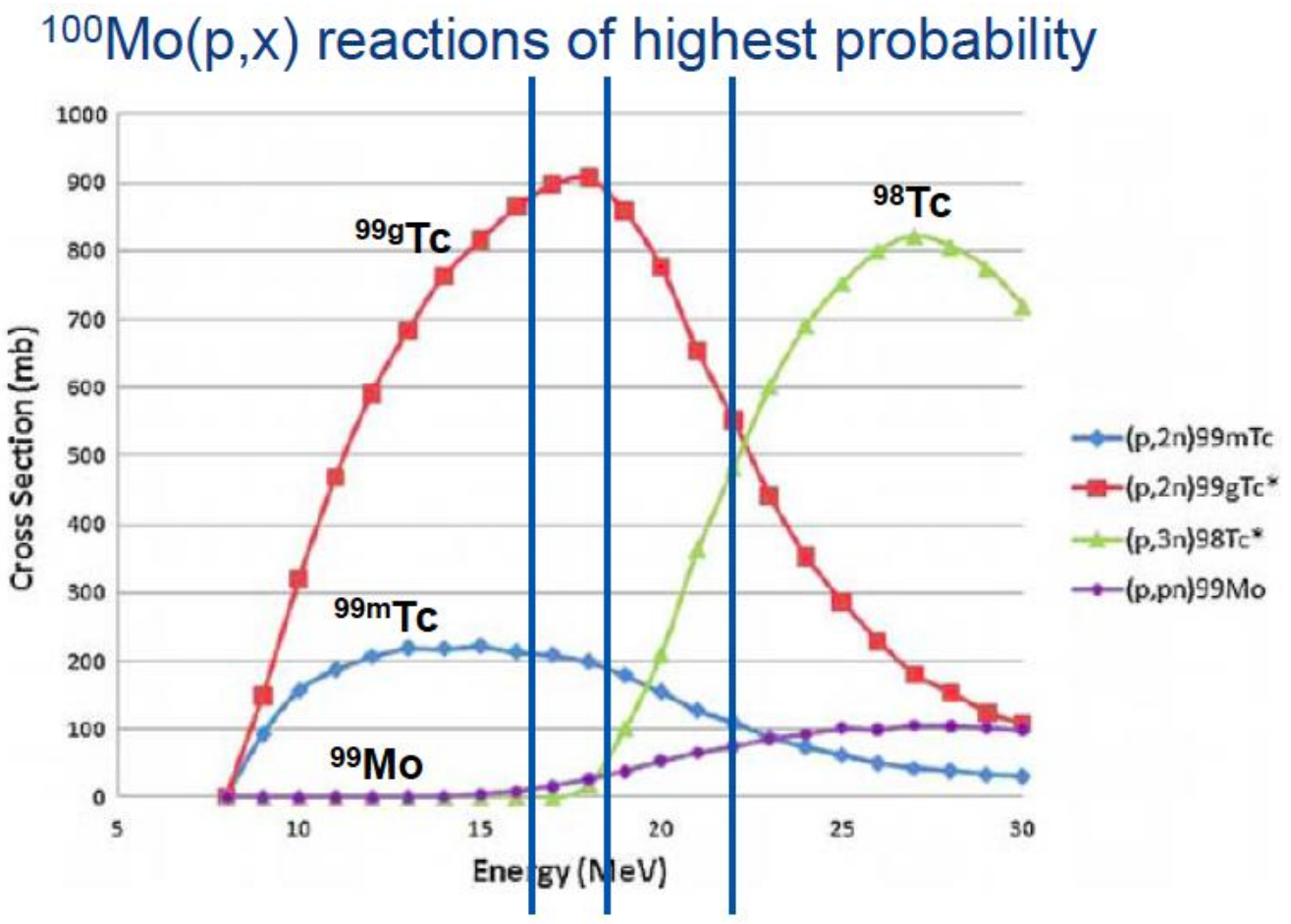

**Figure 8.2** Predicted cyclotron product yields. Taken from A. Celler et. al. [3]

Experiments have shown that cyclotron energies of between 20 and 22MeV appear optimum for the production of $^{99m}Tc$ for patient use. Any molybdenum cyclotron products (including $^{100}Mo$) can be separated out chemically, however the other impurities including $^{93}Tc$, $^{94m}Tc$, $^{94}Tc$, $^{95}Tc$, and $^{96}Tc$ will remain in the final radiopharmaceutical product. These radionuclides will contribute to the final patient effective dose following administration. Both Edmonton and TRIUMF are working on a worst case scenario/benchmark dose threshold of increase in patient dose, which is considered to be of the order of an additional 10%, this being considered acceptable from a patient protection point of view.

## 8.4 Cyclotron Targetry

Given that the feasibility of production by the $^{100}Mo(p,2n)^{99m}Tc$ reaction is well established, the reliable production of commercially viable amounts is dependent upon innovative targetry, giving due consideration to physical factors including melting point, thermal conductivity, thermal expansion, machinability, activation properties, mechanical strength, cost, and chemical inertness. Both Edmonton and TRIUMF achieve thermal power dissipation by use of an oblique target angle and collimation of the beam on a water-cooled target assembly. Target stations have been manufactured to allow mechanical or pneumatic transport to the hot cells for subsequent dissolution.

The target plates consist of approximately 80 mg of $^{100}Mo$ that is then sintered and pressed into the backing plate (Figure 8.3). Alberta uses an aluminium backing plate in a mechanical assembly fitted to the ACSI T24 cyclotron. TRIUMF use a $^{100}Mo$ target thickness of around 300um on a tungsten backing plate for use with both the ACSI TR19/TR24 and GE PET-TRACE cyclotrons.

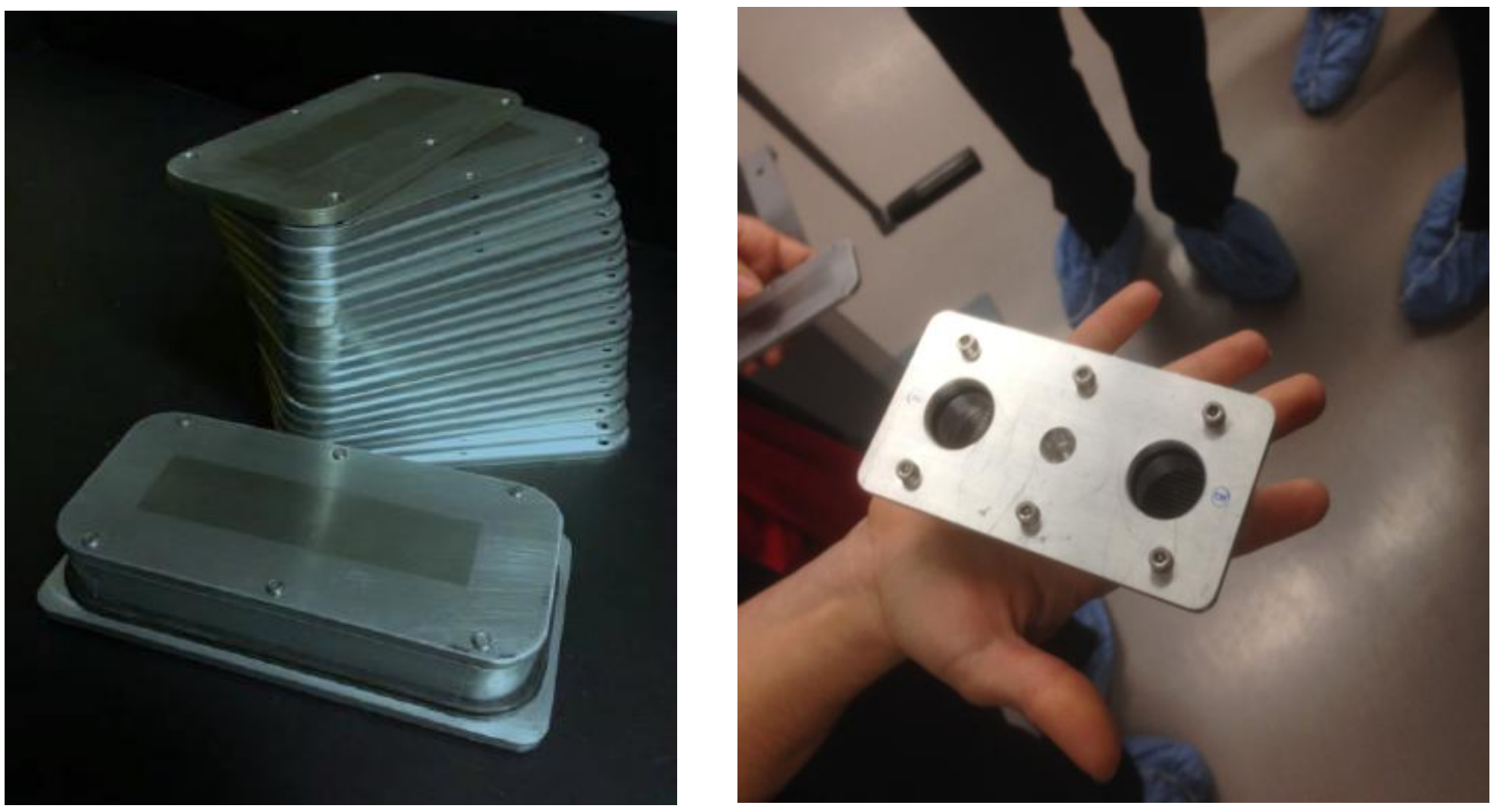

**Figure 8.3** Target and backing plate assembly. Left: $^{100}Mo$ deposited on aluminium plates with target assembly in foreground. Right: The underside of the target plate showing the entry and exit channels for water-cooling. Photographs courtesy of MICF, Edmonton

Enriched $^{100}Mo$ powder is supplied from Russia and obtained through ISOFLEX and URENCO Enrichment Services. Current stocks are of the order of 25kg, and are available at around US $500 per gram. Edmonton are currently considering target recycling through a central facility, however TRIUMF are not considering this option due to issues of impurities that may affect the administered radiation dose to the patient.

From experimental irradiation work a 6-hour bombardment is considered to be optimum for routine production. Experimental irradiations on $^{100}Mo$ targets running at 300 μA, 19 MeV, have produced ~340 GBq (TR19), ~174 GBq (PETtrace) with recovery of 93% as $Na^{99m}TcO_4$.

### *8.4.1 Target Dissolution and $^{99m}Tc/^{100}Mo$ Separation*

**(i) University of Alberta (MICF; Medical Isotope and Cyclotron Facility; Edmonton)**

The irradiated target was fitted onto the front of the dissolution chamber within the hot cell. The target was dissolved in ~30% $H_2O_2$, followed by the addition of 3 M ammonium carbonate.

The dissolved oxidized target solution was transferred under remote control to the $^{99m}Tc$ purification and $^{100}Mo$ recovery system. The extraction process was carried out on an automated synthesis unit (e.g. Neptis Mosaic LC) and the $^{99m}Tc/^{100}Mo$ solution passed through a solid phase extraction resin e.g. ABEC-2000 column (Aqueous Biphasic Extraction Chromatography) or Tentagel which retained the $^{99m}Tc$ pertechnetate while the $^{100}Mo$ molybdate was eluted. The column was washed with 3 M ammonium carbonate to maximise the $^{100}Mo$ recovery. Residual ammonia on the resin was removed by eluting with 1 M sodium carbonate. $^{99m}Tc$ pertechnetate was then eluted from the resin using distilled water and the solution was neutralised by passing through a cation exchange resin before transferring onto an alumina column. The $^{99m}Tc$ pertechnetate was immobilised on the alumina column, which was washed with water before eluting the $^{99m}Tc$ with saline into a collection vial [4, 5, 8].

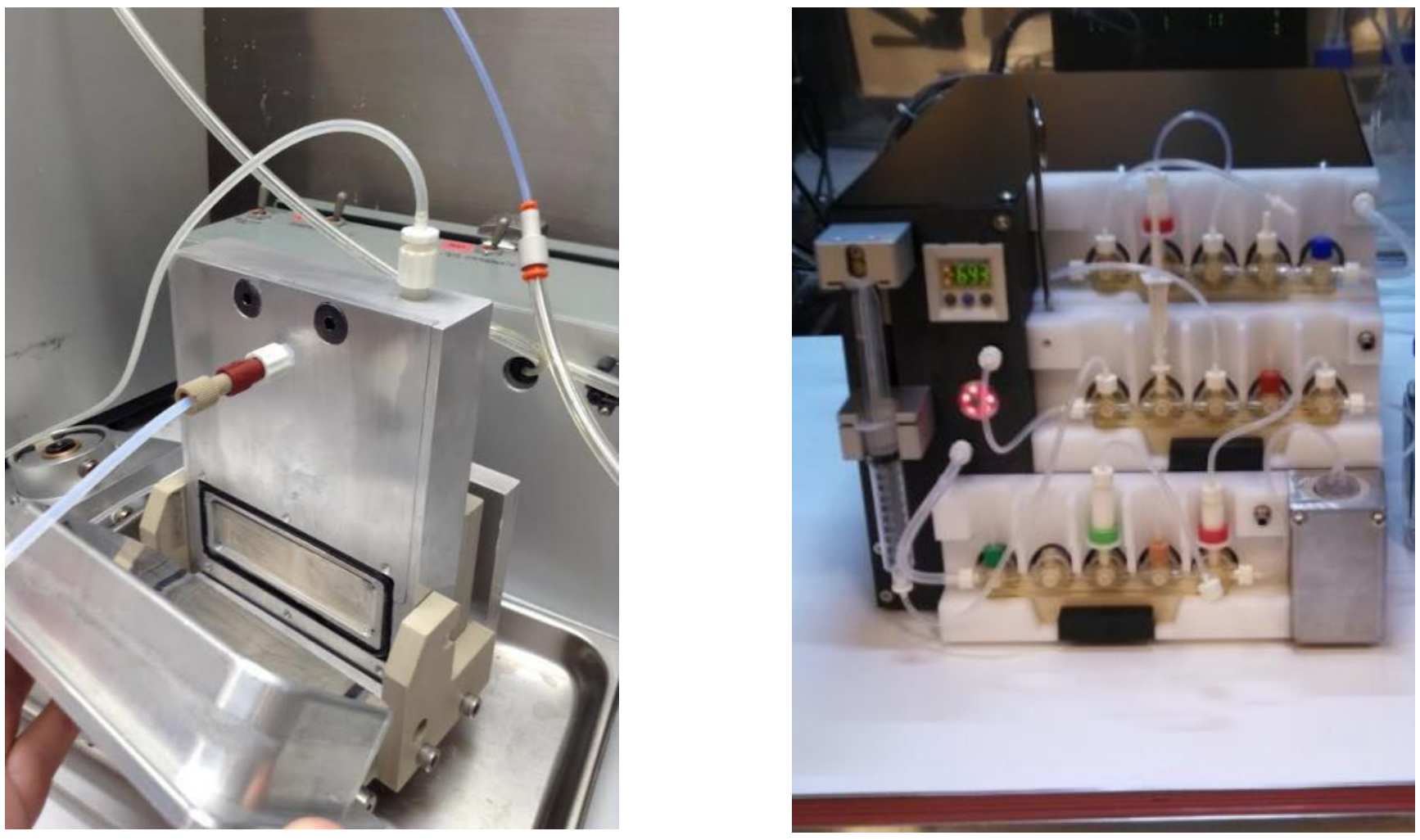

**Figure 8.4** Left: System for dissolution of the irradiated target.
Right: Automated separation synthesis unit.
Photographs courtesy of MICF, Edmonton.

The dissolution and separation processes (Figure 8.4) were carried out in Comecer hot cells (EC GMP Grade C). The final elution of $^{99m}Tc$ pertechnetate was terminally filtered using a 0.22 micron filter into the collection vial and this process was undertaken in an EC GMP Grade A environment.

**(ii) TRIUMF (Vancouver)**

The target dissolution and $^{99m}Tc$/molybdate separation followed a similar process to that described above for the University of Alberta. The target capsule was transferred into the hot cell and used directly as a dissolution vessel to extract $^{99m}Tc$ from the target plates. The target was dissolved in 30% $H_2O_2$ followed by evaporation of the target solution and dissolution in 4 N NaOH. The $^{99m}Tc$ pertechnetate purification process was carried out on an automated synthesis unit designed for use with disposable kits (e.g. miniAIO; Trasis). The $^{100}Mo$/$^{99m}Tc$ solution in 4 N NaOH was passed through a solid phase extraction cartridge containing a resin e.g. ABEC-2000, cross-linked PEG resins and the liquid that passed through the resin was collected to recover the molybdenum for recycling or reuse. The $^{99m}Tc$ remained trapped on the resin. The resin was washed with 4 N NaOH and the $^{99m}Tc$ eluted with water. The eluent was then passed through a cation exchange resin (On-guard II H; Dionex) and alumina cartridge (Sep-Pak Alumina A; Waters). After washing the cartridges with water, the alumina cartridge was eluted with 0.9% NaCl solution and the $^{99m}Tc$ pertechnetate solution was passed through a 0.22 micron filter into the final product vial [1, 2]. The approximate time for the dissolution and separation processes was 1 hour.

## 8.5 Preparation of $^{99m}$Tc Radiopharmaceuticals and Quality Control

A similar approach was undertaken from both the University of Alberta and TRIUMF. Health Canada has agreed that radiolabelling with representative kits (neutral, cationic, anionic) and showing that reconstituted kits meet the kit manufacturer specifications will be sufficient for approval of cyclotron produced $^{99m}$Tc. Kits for neutral ($^{99m}$Tc-exametazime; Ceretec; GE Healthcare), cationic ($^{99m}$Tc-tetrofosmin; GE Healthcare; $^{99m}$Tc-sestamibi; Cardiolite; Lantheus) and anionic ($^{99m}$Tc-MDP; Draximage) were reconstituted and tested according to the manufacturer's instructions (including maximum activity levels).

### *8.5.1 Radiochemical Purity*

All kits complied with the manufacturer's limits for radiochemical purity. Radiochemical purity was determined using standard thin layer chromatography. No significant radiochemical impurities were noted.

The radiochemical purity of $^{99m}$Tc pertechnetate was found to be > 95% from both centres. This complies with USP (United States Pharmacopoeia) and Ph. Eur (European Pharmacopoeia) and BP (British Pharmacopoeia) limits.

The expiry time of the $^{99m}$Tc pertechnetate has not yet been determined but it is envisaged that the shelf life of the cyclotron produced $^{99m}$Tc pertechnetate will be 18 or 24 h from production. This work is currently being undertaken.

The kits have not yet been stressed with maximum activity levels at the end of the expiry of cyclotron-produced $^{99m}$Tc for worst-case scenarios. This work is currently being undertaken in the two centres in Canada (November 2014).

### *8.5.2 Alumina Content*

The $^{99m}$Tc pertechnetate was tested for alumina content using a colorimetric test. In both centres the alumina content was found to be less than 5 ppm. This complies with USP, Ph. Eur and BP limits.

### *8.5.3 Radionuclidic Purity*

The radionuclidic purity of cyclotron-produced $^{99m}$Tc is dependent on the target material, irradiation conditions and the time after production at which the $^{99m}$Tc is used (see Chapter 5). The radionuclidic purity was found to be influenced by the isotopic composition of $^{100}$Mo. This was found to be more important than the absolute enrichment level. It was found that $^{94}$Mo, $^{95}$Mo, $^{96}$Mo and $^{97}$Mo are significant contributors to the production of radionuclide impurities. Both groups observed this. The composition of various batches is shown in Table 8.1.

**Table 8.1** Isotopic compositions of batches of cyclotron-produced 99mTc [1]

| | **Isotopic composition (%)** | | | | | | |
|---|---|---|---|---|---|---|---|
| **Batch** | **$^{92}$Mo** | **$^{94}$Mo** | **$^{95}$Mo** | **$^{96}$Mo** | **$^{97}$Mo** | **$^{98}$Mo** | **$^{100}$Mo** |
| *99.01* | 0.09 | 0.06 | 0.10 | 0.11 | 0.08 | 0.55 | 99.01 |
| *97.4* | 0.005 | 0.005 | 0.005 | 0.005 | 0.01 | 2.58 | 97.39 |
| *99.8* | 0.003 | 0.003 | 0.003 | 0.003 | 0.003 | 0.17 | 99.815 |

The $^{100}$Mo is currently sourced from Isoflex and can be obtained with the desired $^{100}$Mo specification for high quality 99.8% isotopic purity with $^{94-97}$Mo content below the detection limit. Another potential supplier (URENCO) is being investigated. The theoretical dosimetry estimations for radioisotopes produced by proton-induced reactions on natural and enriched molybdenum targets [6] are discussed in Chapter 5.
The reported radionuclidic purity from TRIUMF [2] for $^{99m}$Tc is 99.94% (decay corrected to end of beam). Only traces of other technetium radioisotopes ($^{97m}$Tc 0.003%, $^{96g}$Tc 0.002%, $^{95m}$Tc<0.0001%, $^{95g}$Tc 0.009%, $^{94m}$Tc 0.044%, $^{94g}$Tc 0.012%, $^{93g}$Tc 0.007%) were detected in the $^{99m}$Tc pertechnetate. Other radionuclides detected the waste recovery vessel but not the product vial were $^{99}$Mo (0.78%), $^{96}$Nb (0.06%) and $^{97}$Nb (4.0%). The University of Alberta has achieved a radionuclidic purity of $^{99m}$Tc pertechnetate of 99.98% (at end of beam).

### *8.5.4 Product Specifications*

There are currently no USP or Ph. Eur monographs for cyclotron produced $^{99m}$Tc pertechnetate but an EP monograph Working Group has been established to develop a specification (see Chapter 5). Both centres in Canada are developing their own product specifications for cyclotron-produced $^{99m}$Tc as part of their regulatory submission to Health Canada. Tables 8.2, 8.3 and 8.4 outline the product specification from the University of Alberta (MICF; Medical Isotope and Cyclotron Facility).

**Table 8.2** Specifications (USP [$^{98}$Mo]; MICF)

| **Test** | **USP ($^{99}$Mo from $^{98}$Mo)** | **MICF** |
|---|---|---|
| Appearance | clear, colourless solution | clear, colourless solution |
| pH | 4.5-7.5 | 4.5-7.5 |
| Radiochemical Purity | >95% (8/2 acetone/HCl) | >95% (acetone) |
| Radiochemical Identification | $R_f \approx 0.9$ (compare to known Tec) | $R_f$ >0.8 |
| Radionuclide Identity | photopeak @ 140 keV | photopeak @ 140 keV |
| Aluminium | ≤ 10 ppm | ≤ 5 ppm |
| Residual Solvents | n/a | <5000 ppm EtOH |
| Filter Integrity | n/a | ≥ 50 psi |
| Sterility | no growth after 14 days | no growth after 14 days |
| Endotoxins | <175 EU/V | <175 EU/V (< 5.8 EU/ml) |
| Radionuclidic Purity | ≤ 0.015% Mo-99<br>≤ 0.05% all other | see tables below |

**Table 8.3** Radionuclidic/Radioisotopic purity (Pre-release)

| Isotopes | MICF Lower Detection Limit (LDL) (pre-release) | TR-19 cyclotron data (End of bombardment (EoB) or 18h post, worst case) |
|---|---|---|
| $^{99}$Mo | 0.0048% | 0.0057% (0.0093%) (18h) |
| $^{96}$Nb | 0.0012% | < LDL (18h) |
| $^{97}$Nb | 0.0006% | < LDL (EoB) |
| $^{93m}$Tc | 0.0027% | < LDL (18h) |
| $^{93g}$Tc | 0.0009% | < LDL (18h) |
| $^{94m}$Tc | 0.0006% | < LDL (EoB) |
| $^{94}$Tc | 0.0006% | 0.0027% (0.0033%) (EoB) |
| $^{95}$Tc | 0.0006% | 0.0092% (0.012%) (18h) |
| $^{96}$Tc | 0.0007% | 0.0067% (0.009%) (18 h) |

**Table 8.4** Radionuclidic/Radioisotopic purity (Post-release)

| Isotope | MICF LDL Post release | TR19 Cyclotron data (18 h post EoB) |
|---|---|---|
| $^{99}$Mo | 0.0008% | 0.0057% (0.0093%) (18h) |
| $^{96}$Nb | 0.0001% | <LDL (18h) |
| $^{95m}$Tc | 0.00007% | 0.00011% (0.00014%) |
| $^{97m}$Tc | 0.014% | 0.0475% (0.0618%) |

Additional tests undertaken on the final product for chemical purity (colorimetric) are:

1. molybdenum (< 5 micrograms per ml)
2. peroxide (< 1 microgram per ml)

TRIUMF are also developing a similar specification for cyclotron-produced $^{99m}$Tc.

The approximate time for dissolution, separation and quality control tests to be completed (pre-release) was 2 h (post end of bombardment). At present, there has been no large-scale production of $^{99m}$Tc. It has been calculated that 1.5 TBq quantities of $^{99m}$Tc could be produced after a 6-hour bombardment time (500

□A) on a TR24 cyclotron. The University of Alberta is currently planning to undertake daily runs of high current large-scale production to demonstrate reliability over a 3-month period of continuous production.

## 8.6 Waste Products

At the current time the radioactive waste calculations are not complete. The long-lived waste products include $^{99}$Mo and $^{95}$Nb, these having multi-day half-lives.

## 8.7 Regulatory Aspects

Health Canada has two levels of approval and requires an Establishment Licence (for New Drug Submission (NDS)) to manufacture, supply and use cyclotron-produced $^{99m}$Tc pertechnetate for clinical investigations. These levels of approval are for:

1) Clinical Trials
2) New Drug Submission (NDS)

### *8.7.1 Clinical Trials*

Submission of a Clinical Trial Application (CTA) requires a 30-day review period and successful applications are granted an Issue of No Objection Letter (NOL) from Health Canada. CTA's require information (in ICH Common Technical Document (CTD) format) such as the clinical protocol, case report form, product monograph and a manufacturing file.

The University of Alberta completed a Phase 1 clinical trial in 2012 for cyclotron-produced $^{99m}$Tc using a TR19 cyclotron [7]. A clinical trial with the Centre Hospitalier Universitaire de Sherbrooke (CHUS) is being planned for $^{99m}$Tc produced from the TR24 cyclotron [9]. It is not yet known whether the clinical trial data obtained from the TR19 will have to be repeated using the TR24. TRIUMF are planning to submit a CTA in December 2014 and to initiate a clinical trial for cyclotron-produced $^{99m}$Tc in patients from January 2015.

### *8.7.2 New Drug Submission (NDS)*

A NDS requires a 45-day screening and 300-day review process (or 25 day screening and 180 day review process if granted priority review) from Health Canada. Successful submissions are granted an Issuance of Notice of Compliance (NOC) from Health Canada and this confers the ability to market the drug.

New Drug Submissions require information to be submitted such as product overview, supporting clinical/preclinical data, clinical indications, product monograph and a manufacturing file. The University of Alberta and TRIUMF both anticipate submitting a NDS for cyclotron-produced $^{99m}$Tc to Health Canada during 2015. It is expected that final approval from Health Canada will be obtained in late 2016.

## 8.8 Economic Considerations

The business plans for building, commissioning and operation are currently under consideration. The costs considered include capital expenditure, (cyclotron and radiopharmacy facilities), start-up (training, materials, regulatory submissions) variables (salaries, power, consumables, maintenance, etc.), administration, insurance, shipping and waste costs. According to the TRIUMF team the estimated price is comparable with that of reactor produced $^{99m}$Tc at < Ca \$1.00/mCi.

## 8.9 Conclusion

Two $^{99m}$Tc production methodologies using the $^{100}$Mo(p,2n)$^{99m}$Tc reaction on high powered cyclotrons have been accomplished. The key innovation for the implementation of this technology resides with the target plate technology. These methods have the potential for routine production of TBq amounts of GMP product and could be established in the UK along similar lines to those using regional cyclotrons for the production of positron emitting radiopharmaceuticals. The technology represents a valuable opportunity for UK science and technology, however the reliability of routine production and backup arrangements are still to be evaluated. Further commercial investment will be requited to develop operational stability and to secure the necessary regulatory approvals before this technology is suitable for routine medical use.

## Declarations of Interest

Participating authors were asked to declare any conflicts of interest in relation to their involvement in this report. The replies are noted below.

| | |
|---|---|
| **Brian Neilly**, | No conflicts of interest declared. |
| **Sarah Allen**, | No conflicts of interest declared. |
| **Jim Ballinger**, | No conflicts of interest declared. |
| **John Buscombe**, | No conflicts of interest declared. |
| **Rob Clarke**, | No conflicts of interest declared. |
| **Beverley Ellis**, | No conflicts of interest declared |
| **Glenn Flux**, | No conflicts of interest declared |
| **Louise Fraser:** | No conflicts of interest declared |
| **Adrian Hall**, | No conflicts of interest declared |
| **Hywel Owen**, | No conflicts of interest declared |
| **Audrey Paterson,** | No conflicts of interest declared |
| **Alan Perkins**, | No conflicts of interest declared |
| **Andrew Scarsbrook**, | No conflicts of interest declared |

December 2014